\documentclass[useAMS,usenatbib,usegraphicx,letterpaper]{mn2e}

\DeclareRobustCommand{\ion}[2]{%
  \textup{#1\,{\mdseries\textsc{#2}}}%
}

\renewcommand\vec[1]{\bmath{#1}}
\newcommand\mat[1]{\mathbf{#1}}

\newcommand{\nE}{N_{E}}
\newcommand{\nLz}{N_{L_{z}}}

\newcommand{\Rc}{R_{\rm c}}

\newcommand{\talp}{\alpha_{0}}
\newcommand{\Lz}{L_{z}}

\newcommand{\vecd}{\vec{d}}
\newcommand{\veceta}{\vec{\eta}}

\newcommand{\de}{\mathrm{d}}

\newcommand{\reg}{\mu}
\newcommand{\cauldron}{\textsc{cauldron}}

\newcommand{\slope}{\gamma'}

\newcommand{\qstar}{q_{\ast}}
\newcommand{\qspro}{q_{\mathrm{\ast, 2D}}}
\newcommand{\Reff}{R_{\mathrm{e}}}
\newcommand{\re}{r_{\mathrm{e}}}

\newcommand{\Mstar}{M_{\ast}}

\newcommand{\Mtot}{M_{\mathrm{tot,e}}}
\newcommand{\Mmb}{M_{\mathrm{\ast,e}}^{\mathrm{mb}}}
\newcommand{\Mchab}{M_{\mathrm{\ast,e}}^{\,\mathrm{Chab}}}
\newcommand{\Msalp}{M_{\mathrm{\ast,e}}^{\mathrm{Salp}}}
\newcommand{\REin}{R_{\mathrm{Einst}}}

\newcommand{\Rkin}{R_{\mathrm{kin}}}

\newcommand{\jz}{j_{z}}
\newcommand{\lamR}{\lambda_{R}}

\newcommand{\vos}{(v/\sigma, \epsilon)}
\newcommand{\epstar}{\epsilon_{\ast}}
\newcommand{\vphi}{\langle v_{\varphi} \rangle}
\newcommand{\fDM}{f_{\mathrm{DM}}}
\newcommand{\mslope}{\langle \slope \rangle}
\newcommand{\slopec}{\gamma'_{\mathrm{c}}}
\newcommand{\sigmams}{\sigma_{\gamma'}}
\newcommand{\sigmamsq}{{\sigma_{\gamma'}^{2}}}
\newcommand{\deslope}{\delta{\gamma'_{\mathrm{i}}}}
\newcommand{\deslopeq}{\delta{\gamma'_{\mathrm{i}}}^{2}}
\newcommand{\rhostar}{\rho_{\ast}}

\newcommand{\pr}{\mathrm{Pr}}

\usepackage{url}
\usepackage[varg]{txfonts}
\usepackage{color}
\usepackage{subfigure}
\usepackage{amssymb}
\usepackage{multirow}

\title[Two-dimensional kinematics of SLACS lenses -- III.]  
{Two-dimensional kinematics of SLACS lenses -- III. \\ Mass structure
and dynamics of early-type lens galaxies beyond~\boldmath{$z \simeq 0.1$}}

\author[M. Barnab\`e et al.]{%
  Matteo Barnab\`e$^{1,2}$\thanks{E-mail: mbarnabe@stanford.edu}, 
  Oliver Czoske$^{2,3}$,
  L\'eon V. E. Koopmans$^{2}$,
  Tommaso Treu$^{4}$
  \newauthor
  and Adam S. Bolton$^{5}$\\
  $^{1}$Kavli Institute for Particle Astrophysics and Cosmology,
  Stanford University, 452 Lomita Mall, Stanford, CA 94035-4085, USA\\
  $^{2}$Kapteyn Astronomical Institute, University of Groningen, 
  PO Box 800, 9700\,AV Groningen, the Netherlands\\
  $^{3}$Institut f\"ur Astronomie der Universit\"at Wien,
  T\"urkenschanzstra{\ss}e 17, 1180 Wien, Austria\\
  $^{4}$Department of Physics, University of California, Santa
  Barbara, CA 93101, USA\\
  $^{5}$Department of Physics and Astronomy, University of Utah, 115
  South 1400 East, Salt Lake City, UT 84112, USA
}

\begin{document}

\date{Accepted 2011 April 04. Received 2011 April 02; in original form
  2011 February 11}

\maketitle

\label{firstpage}

\begin{abstract}
We combine in a self-consistent way the constraints from both
gravitational lensing and stellar kinematics to perform a detailed
investigation of the internal mass distribution, amount of dark
matter, and dynamical structure of the sixteen early-type lens
galaxies from the SLACS Survey, at $z = 0.08 - 0.33$, for which both
\emph{HST}/ACS and NICMOS high-resolution imaging and VLT VIMOS
integral-field spectroscopy are available. Based on this data set, we
analyze the inner regions of the galaxies, i.e.\ typically within one
(three-dimensional) effective radius $\re$, under the assumption of
axial symmetry and by constructing dynamical models supported by
two-integral stellar distribution functions (DFs). For all systems,
the total mass density distribution is found to be well approximated
by a simple power-law $\rho_{\mathrm{tot}} \propto m^{-\slope}$ (with
$m$ being the ellipsoidal radius): this profile is on average slightly
super-isothermal, with a logarithmic slope $\mslope =
2.074^{+0.043}_{-0.041}$ (errors indicate the 68\% CL) and an
intrinsic scatter $\sigmams = 0.144^{+0.055}_{-0.014}$, and is fairly
round, with an average axial ratio $\langle q \rangle =
0.77\pm0.04$. The lower limit for the dark matter fraction ($\fDM$)
inside~$\re$ ranges, in individual systems, from nearly zero to almost
a half, with a median value of~12\%. By including stellar masses
derived from stellar population synthesis (SPS) models with a Salpeter
initial mass function (IMF), we obtain an average $\fDM = 31\%$, and
the corresponding stellar profiles are physically acceptable, with the
exception of two cases where they only marginally exceed the total
mass profile. The $\fDM$ rises to 61\% if, instead, a Chabrier IMF is
assumed. For both IMFs, the dark matter fraction increases with the
total mass of the galaxy (correlation significant at the 3-sigma
level). Based on the intrinsic angular momentum parameter calculated
from our models, we find that the galaxies can be divided into two
dynamically distinct groups, which are shown to correspond to the
usual classes of the (observationally defined) slow and fast
rotators. Overall, the SLACS systems are structurally and dynamically
very similar to their nearby counterparts, indicating that the inner
regions of early-type galaxies have undergone little, if any,
evolution since redshift $z \approx 0.35$.
\end{abstract}

\begin{keywords}
 galaxies: elliptical and lenticular, cD --- galaxies: kinematics and
 dynamics --- galaxies: structure --- gravitational lensing
\end{keywords}


\section{Introduction}
\label{sec:introduction}

Early-type galaxies are among the fundamental constituents of the
local observable Universe, contributing to most of the total stellar
mass \citep*{Fukugita1998, Renzini2006}, and it is therefore only
natural that these systems have become the focus of a vast amount of
observational and theoretical studies during recent years.

Understanding the formation and evolution mechanisms that can generate
such a remarkably homogeneous class of objects remains among the
most important open questions in present-day astrophysics: although
the standard scenario according to which massive ellipticals are
built-up via hierarchical merging of disk galaxies \citep[see
  e.g][]{Toomre1977, White-Frenk1991, Cole2000} has been steadily
gaining consensus throughout the last decades, only recently the
progress in high-resolution numerical simulations has allowed to start
investigating in detail the internal structural properties of
realistic systems (e.g.\ \citealt{Meza2003}, \citealt{Naab2007},
\citealt{Jesseit2007}, \citealt{Onorbe2007},
\citealt{GonzalesGarcia2009}, \citealt*{Johansson2009},
\citealt{Lackner-Ostriker2010}).

Clearly, in view of these developments, the availability of stringent
observational constraints has become even more crucial. Luckily, at
least as far as nearby ellipticals are concerned, the observational
and modelling effort has been very intense in the past few
years. Stellar dynamics represents the most widely used diagnostic
tool (see e.g.\ \citealt*{Saglia1992}, \citealt{Bertin1994},
\citealt{Franx1994}, \citealt{Carollo1995}, \citealt{Rix1997},
\citealt{Loewenstein-White1999}, \citealt{Gerhard2001},
\citealt{Borriello2003}, \citealt{vandenBosch2008},
\citealt{Thomas2007b}, \citeyear{Thomas2009b}, \citeyear{Thomas2011}
and in particular the SAURON project, \citealt{deZeeuw2002}, and the
Atlas3D project, \citealt{Cappellari2010}), although other studies
have also successfully employed\,---\,sometimes in combination with
stellar kinematics\,---\,different tracers such as planetary nebulae
and globular cluster kinematics \citep[e.g.][]{Cote2003,
  Romanowsky2003, Romanowsky2009, Coccato2009, deLorenzi2009,
  Napolitano2009, Rodionov-Athanassoula2011}, the occasional
$\ion{H}{I}$ gas disk or ring \citep[e.g.][]{Oosterloo2002,
  Weijmans2008}, and hot X-ray gas emission
\citep[e.g][]{Matsushita1998, Fukazawa2006, Humphrey-Buote2010,
  Das2010, Churazov2010}. In general, from these analyses there is
mounting evidence that early-type galaxies are embedded in dark
haloes, whose contribution is often significant already inside one
effective radius, and that their total density profiles are roughly
isothermal within the inner regions (but not necessarily at larger
radii, see \citealt{deLorenzi2009}, \citealt{Dutton2010}).

Ideally, if such studies could be extended to elliptical galaxies
beyond the local Universe, it would become possible to provide strong
constraints also to the intermediate steps of numerical simulations
throughout redshift, rather than just to their $z = 0$
end-products. The current observational limitations, unfortunately,
make it prohibitively difficult to apply the traditional techniques to
distant systems, i.e. at $z \approx 0.1$ and beyond. In particular,
the impossibility to reliably measure the higher-order velocity
moments needed to disentangle the mass--anisotropy degeneracy
\citep{Gerhard1993, vanderMarel-Franx1993} constitutes a serious
hindrance for dynamical studies. On the other hand, however, at higher
redshifts strong gravitational lensing \citep*[see
  e.g.][]{SaasFee2006} may become available as an additional probe,
opening up new possibilities to investigate the mass distribution of
galaxies.

The main obstacle with this approach, namely the difficulty in
discovering a suitable number of early-type galaxies that also act as
strong gravitational lenses, has been remedied by the Sloan Lens ACS
Survey, SLACS (\citealt{Bolton2006}, \citeyear{Bolton2008a},
\citeyear{Bolton2008b}, \citealt{Koopmans2006}, \citealt{Treu2006},
\citeyear{Treu2009}, \citealt{Gavazzi2007}, \citeyear{Gavazzi2008},
\citealt{Auger2009}, \citeyear{Auger2010}), which has yielded a large
sample of almost a hundred such objects. As highlighted by these
studies, as well as by a number of other works (see
\citealt{Treu-Koopmans2002a}, \citealt{Koopmans-Treu2003},
\citealt{Treu-Koopmans2004}, \citealt{Rusin-Kochanek2005},
\citealt{Jiang-Kochanek2007}, \citealt{Grillo2008},
\citealt{Koopmans2009}, \citealt{Trott2010}, \citealt{vandeVen2010},
\citealt{Dutton2011}, \citealt{Spiniello2011}), the combination of
gravitational lensing and stellar dynamics provides a particularly
powerful and robust method to determine the total mass density
distribution of distant galaxies.

A potential limitation in the joint analysis of the SLACS galaxies
mentioned above is that all the information about the dynamics is
extracted from a single data point, i.e. the average stellar velocity
dispersion measured within a 3~arcsec diameter aperture, obtained from
SDSS spectra. Moreover, the method is not fully self-consistent, in
the sense that axial symmetry is assumed for the lensing modelling,
but not for the dynamical modelling, which is based on spherical Jeans
equations. In order to address the concerns that these approximations
might lead to biased results, the SLACS data set has been
complemented, for about~30 lenses, with extended, two-dimensional
kinematic information, obtained either with VLT VIMOS integral field
unit (IFU) or from LRIS Keck long-slit spectroscopic observations
(with the slit positions offset along the galaxy minor axis in order
to mimic integral field capabilities). In addition, we have developed
a more rigorous, self-consistent modelling technique (the {\cauldron}
code, based on the superposition of two-integral distribution
functions, see Sect.~\ref{sec:analysis} and
\citealt{Barnabe-Koopmans2007}), aimed at taking full advantage of the
available data sets for each galaxy, namely the gravitationally lensed
image, the surface brightness distribution and the velocity moments
maps. This makes it possible to recover more accurate constraints on
the slope and the flattening of the total mass density distribution,
to derive information on the dark matter fraction within the inner
regions and to obtain insights on the global and local dynamical
status of the system, so that this analysis constitutes, to a good
extent, the higher redshift analogue of the detailed studies that are
routinely carried out on local ellipticals. We have applied this
method to half a dozen SLACS galaxies with VIMOS observations
\citep{Czoske2008, Barnabe2009}, as well as to one of the Keck systems
\citep{Barnabe2010}. In this paper, we extend this in-depth analysis
to the full sample of SLACS early-type lenses with available VIMOS IFU
spectroscopy (we refer to Czoske et al. 2011, in preparation, for a
comprehensive description of the complete data set), a total of~16
systems at redshift $z = 0.08$ to $0.35$, and covering a wide range in
mass and angular momentum. We also make use of stellar masses derived
from SPS models \citep{Auger2009} to impose further constraints on the
luminous distribution and estimate the contribution of the dark matter
component.

This paper is organized as follows: in Section~\ref{sec:data} we
provide an overview of the available data-sets. In
Section~\ref{sec:analysis}, after recalling the basic features of the
method for the combined analysis of the adopted model, we present the
main results of the study of the SLACS lenses mass density
profiles. The mass structure of the analyzed galaxies, in terms of
luminous and dark matter components, is described in
Section~\ref{sec:mass}, while Section~\ref{sec:dynamics} deals with
the recovered dynamical structure of the systems. We discuss our
findings in Section~\ref{sec:discussion} and we briefly summarize the
main conclusions in Section~\ref{sec:conclusions}.  Throughout this
paper we adopt a concordance $\Lambda$CDM model described by
$\Omega_{\mathrm{M}}=0.3$, $\Omega_{\Lambda} = 0.7$ and $H_{0} =
100\,h\,\mathrm{km\,s^{-1}\,Mpc^{-1}}$ with $h=0.7$, unless stated
otherwise.


\section{Observations}
\label{sec:data}

The joint lensing/dynamics analysis requires three types of input
data: (i) high-resolution images are used to trace the surface
brightness of the gravitationally lensed background galaxies; (ii)
near-infrared images provide the surface brightness distribution of
the stars in the lens galaxies; (iii) integral-field spectroscopic
observations are used to derive two-dimensional maps of systematic
velocity and velocity dispersion of the stars in the lens galaxies.

\subsection{Imaging}

High-resolution images of the lens systems in the sample were obtained
with Advanced Camera for Surveys (ACS) on the Hubble Space Telescope
(HST). We use images taken through the F814W filter. Deep images are
available for eight of the systems, while for the remaining nine only
single-exposure snapshot images are available. The pixel scale of the
images is $0.05\,\mathrm{arcsec}$. The structure of the
gravitationally lensed background images was isolated by subtracting
elliptical B-spline models of the lens galaxies' light distribution
\citep{Bolton2008a}. 

NICMOS F160W images were used to obtain the surface brightness of the
lens galaxies. In order to serve as luminosity weights to the
kinematic maps, the images were matched in resolution to the
spectroscopic observations ($0.8\,\mathrm{arcsec}$) and resampled from
the original pixel scale of $0.05\,\mathrm{arcsec}$ to the pixel grid
of the kinematic maps ($0.67\,\mathrm{arcsec}$). For three systems no
NICMOS observations were available. In these cases, the lens surface
brightness was obtained from the B-spline fits to the F814W images.

\subsection{Spectroscopy}

Two-dimensional spectroscopy of seventeen systems was obtained with
the integral-field unit of VIMOS on the VLT\footnote{One of the lens
  systems observed with VIMOS IFU, SDSS\,J1250$-$0135 (identified as a
  late-type galaxy with morphological type S0/SA), has been excluded
  from the sample examined in this work due to its spiral arms
  partially appearing in the lens-subtracted image, severely hindering
  an accurate lensing analysis.}. Three systems (J0037, J0912 and
J2321) were observed with the HR-Blue
grism($4000-6200$\,\AA)\footnote{ESO programme 075.B-0226,
  P.I.:~L.\ Koopmans} , while the remainder were observed with the
HR-Orange grism ($5000-7500$\,\AA)\footnote{ESO programme 177.B-0682,
  P.I.:~L.\ Koopmans}. The field of view of $27\arcsec\times27\arcsec$
was sampled with $40\times40$ spatial elements with a scale of
$0.67\,\mathrm{arcsec}$, the spectral resolution is $R\approx2500$
(corresponding to a velocity resolution of $\Delta
v\approx90-110\,\mathrm{km\,s^{-1}}$ FWHM in the rest frame of the
lens galaxies). Two-dimensional maps of systematic velocity (with
respect to the mean redshift of the lens galaxy) and velocity
dispersion were measured by fitting stellar templates convolved with
Gaussian line-of-sight velocity distributions to the spectra from
spatial elements with a sufficient signal-to-noise ratio ($S/N > 8$
per spectral element of $0.644$\,\AA).

A full description of the sample, observations, data reduction and the
resulting kinematic maps is given in a companion paper (Czoske et al.\
2011, in preparation).


\section{Combined lensing and dynamics analysis}
\label{sec:analysis}

In this Section\,---\,after recalling the adopted modelling assumptions
and the salient features of the method employed to self-consistently
combine lensing and dynamics constraints\,---\,we present the core results
of the study of the full sample of sixteen SLACS early-type lens galaxies
for which data sets including both high-resolution imaging and VIMOS
integral field spectroscopy are available.

The theoretical framework of the joint lensing and dynamics analysis
and the implementation of the method (i.e., the {\cauldron} code) are
described in full detail in \citet[][hereafter
  BK07]{Barnabe-Koopmans2007}, to which we refer the interested
reader.

\subsection{Method overview}
\label{sec:code}

We describe the total mass density distribution of the lens galaxy as
$\rho(\vec{x}; \veceta)$, where the vector $\vec{x}$ denotes the
spatial coordinates and $\veceta$ is a set of parameters that
characterizes the density profile. The corresponding total
gravitational potential, $\Phi(\vec{x})$, is calculated via the
Poisson equation and utilized for both the gravitational lensing and
the stellar dynamics modelling of the data set. By employing a
pixelated source reconstruction method (see
e.g.\ \citealt{Warren-Dye2003}, \citealt{Koopmans2005},
\citealt{Suyu2006}) for the lensing and some flavour of the
Schwarzschild orbit superposition method (\citealt{Schwarzschild1979};
see \citealt{Thomas2010} for a review of the developments of this
technique) for the dynamics, both these modelling problems can be
formally written in an analogous way as a set of regularized linear
equations, for which robust solving techniques are readily available
\citep[see e.g.][]{NR1992}.

Clearly, for any given set of observations, the ultimate goal of the
analysis is to determine the set of~$\veceta$ parameters that generate
the `best' density model, i.e.\ the most plausible in an Occam's
razor sense. This can be achieved by embedding the linear optimization
scheme within the framework of Bayesian inference, which provides an
objective way to quantify the probability of each model given the data
\citep[see e.g.][]{MacKay1999, MacKay2003}. The best model is thus the
maximum \emph{a posteriori} (MAP) model, i.e.\ the one that maximizes
the joint posterior probability density function (PDF). 

While this method is very general, allowing in principle for any
density distribution, in its current implementation\,---\,the
{\cauldron} algorithm\,---\,we make further assumptions in order to
achieve the computational efficiency that is needed for a thorough
exploration of the parameter space. Within the {\cauldron} code,
therefore, galaxies are modelled as axially symmetric systems,
i.e. $\rho(R,z)$, whose dynamics are described by a two-integral
distribution function (DF) $f = f(E, \Lz)$ which depends on the two
classical integrals of motions, the energy~$E$ and the angular
momentum along the rotation axis $\Lz$. This makes it possible to
construct dynamical models in a matter of seconds, by means of a fast
Monte Carlo numerical implementation developed by BK07 of the
two-integral Schwarzschild method originally introduced by
\citet{Cretton1999} and \citet{Verolme-deZeeuw2002}, which employs
two-integral components (TICs), rather than the commonly used stellar
orbits, as dynamical building blocks. TICs can be viewed as elementary
stellar systems completely specified by a particular choice of $E$ and
$\Lz$; they are characterized by analytic radial density distributions
and (unprojected) velocity moments, which makes them convenient and
(compared to orbits) computationally inexpensive building blocks.

Remarkably, the applicability of the code is not too severely limited
by the assumptions mentioned above: as shown in \citet{Barnabe2009},
{\cauldron} has been successfully tested for robustness by applying it
to a complex system obtained from a numerical N-body simulation of a
merger process, which obviously departs from the condition of
axisymmetry. Nevertheless, several important global properties of the
system (particularly the total density slope and, when rotation is
visible in the kinematics maps, the angular momentum and the ratio of
ordered to random motions) are recovered quite reliably, typically
within~10 to~25 per cent of the true value.

\subsection{The galaxy model}
\label{sec:model}

We model the total mass density profile of the lens galaxy with an
axially symmetric power-law distribution,
\begin{equation}
  \label{eq:rho}
  \rho(m) = \frac{\rho_{0}}{m^{\slope}} ,
\end{equation}
where $\rho_{0}$ is a density scale, $\slope$ is the logarithmic slope
of the profile, and the ellipsoidal radius~$m$ is defined as
\begin{equation}
  \label{eq:m}
  m^2 \equiv \frac{R^2}{a_0^2} + \frac{z^2}{a_0^2 q^2} ;
\end{equation}
here $a_{0}$ denotes the arbitrary length-scale and $q$ is the axial
ratio ($q = 1$ for a spherical galaxy, while oblate systems have $0
\le q < 1$). The corresponding gravitational potential is obtained by
applying the \citet{Chandrasekhar1969} formula for homoeoidal density
distributions and can be written as a rather simple expression
\citep[cf.][]{Barnabe2009} that just requires the computation of a
one-dimensional integral.

Despite its simplicity, the power-law model (typically with a slope
close to the isothermal case, i.e. $\slope = 2$) provides a good
description of the total mass distribution of early-type galaxies over
a large radial range, as reported by previous analyses of the SLACS
systems \citep{Koopmans2006, Gavazzi2007, Czoske2008, Barnabe2009,
  Koopmans2009, Auger2010}, as well as a number of stellar dynamics
\citep[e.g.][]{Gerhard2001}, gravitational lensing
\citep[e.g.][]{Dye2008, Tu2009} and X-ray studies
\citep[e.g.][]{Humphrey-Buote2010} of ellipticals.

Three physical parameters (the non-linear parameters~$\veceta$, in the
notation of Sect.~\ref{sec:code}) characterize the power-law model:
the logarithmic density slope~$\slope$, the axial ratio~$q$ and the
lens strength~$\talp$, a dimensionless quantity directly related to
the normalization of the potential well (see Appendix~B of BK07).
Furthermore, four `geometrical' parameters are needed to describe the
observed configuration of the lens galaxy in the sky, i.e. the
inclination angle~$i$ (with $i = 90^{\circ}$ being an edge-on system),
the position angle and the sky-coordinates of the centre of the
system. Luckily, the last three parameters are well constrained by the
surface brightness distribution of the lens galaxy and of the lensed
images. Therefore, after determining their values by means of fast
preliminary explorations, we maintain them fixed during the
computationally expensive optimization and error analysis runs.

The level of regularization, which controls the smoothness in the
reconstructed background source and TIC weight maps, is set through
three additional non-linear parameters, usually known as
hyper-parameters (see \citealt{Suyu2006} and BK07 for a more technical
explanation). The optimal values of the hyper-parameters are also
found via maximization of the joint posterior PDF.

\subsection{Bayesian inference and uncertainties}
\label{sec:unc}

For our analysis we follow the standard Bayesian framework described
by, e.g., \citet{MacKay2003} involving multiple levels of inference;
here we summarize the main steps involved in this approach, referring
the interested reader to~BK07 for a detailed explanation of the
method.

Let us denote the combined data set as $\vecd$, the noise in the data
as $\vec{n}$ and the considered hypothesis (i.e. the adopted model) as
$\mathcal{H}$($\veceta$), where the non-linear parameters $\veceta$
are the physical and geometrical quantities (such as the density
slope, axial ratio, lens strength and inclination angle) that
characterize the model. At the \emph{first level of inference}, the
model is assumed to be true (i.e. we fix one choice for the set of
$\veceta$ parameters) and we aim to solve for $\vec{w}$ the linear
problem
\begin{equation}
  \label{eq:linear_problem}
  \vec{d} = \mat{A}\vec{w} + \vec{n} \, ,
\end{equation}
where~$\vec{w}$ are the linear parameters\,---\,namely, the surface
brightness distribution of the lensed background source and the
weights of the dynamical building blocks\,---\,being mapped onto the
observables by the operator~$\mat{A} =
\mat{A}[\mathcal{H}(\veceta)]$. The construction of the mapping
operator~$\mat{A}$ for both the lensing and dynamics modelling
constitutes the core of the {\cauldron} code, and is described
in~BK07. From Bayes' theorem, the posterior probability of the
parameters~$\vec{w}$ given the data is written as
\begin{equation}
  \label{eq:infer01}
  \pr(\vec{w} \, | \, \vec{d}, \reg, \veceta) = 
  \frac{
    \pr(\vec{d} \, | \, \vec{w}, \veceta) \,
    \pr(\vec{w} \, | \, \reg)
  }
  {Z_{1}} \, ,
\end{equation}
where the probability $\pr(\vec{d} \, | \, \vec{w}, \veceta)$ is the
likelihood and $\pr(\vec{d} \, | \, \vec{w}, \veceta)$ represents the
prior. Here, we adopt as prior a curvature regularization function
that describes the degree of smoothness that we expect to find in the
solution; the amount of regularization to be applied is set by the
value of the so-called `hyper-parameter' $\reg$. The evidence~$Z_{1} =
\pr(\vec{d} \, | \, \reg, \veceta) = \int \pr(\vec{d} \, | \, \vec{w},
\veceta) \, \pr(\vec{w} \, | \, \reg) \, \de \vec{w}$ is simply a
normalization constant in Eq.~(\ref{eq:infer01}), but it becomes
important at the upper level of inference, where it appears as the
likelihood. The set of parameters $\vec{w}_{\mathrm{MAP}}$ that
maximizes the posterior can be obtained by means of standard linear
optimization techniques.

Analogously, at the \emph{second level of inference} the $\veceta$
parameters are still maintained fixed and the posterior probability of
the hyper-parameter is
\begin{equation}
  \label{eq:infer02}
  \pr(\reg \, | \, \vec{d}, \veceta) = 
  \frac{
    \pr(\vec{d} \, | \, \reg , \veceta) \, \pr(\reg)
  }
  {Z_{2}} \, ;
\end{equation}
here we adopt a scale invariant prior $\pr(\reg) \propto 1/\reg$
(uniform in log) to formalize our ignorance of the order of magnitude
of the hyper-parameter.

Finally, at the \emph{third level of inference} we can compare the
different models by studying the posterior probability distribution of
the set of non-linear parameters~$\veceta$,
\begin{equation}
  \label{eq:infer03}
  \pr(\veceta \, | \, \vec{d}) =
  \frac{
  \pr(\vec{d} \, | \, \veceta) \, 
  \pr(\veceta)
  }
  {Z_{3}} \, ,
\end{equation}
and infer the parameters~$\vec{\eta}_{\mathrm{MAP}}$ for which the
posterior is maximized. For each galaxy, the corresponding MAP model
constitutes the `best model', meaning the most plausible joint set of
model parameters given the data, and therefore throughout the paper we
present the reconstructed observables, background source, TIC weights
maps and inferred quantities that are obtained with the choice
$\veceta = \vec{\eta}_{\mathrm{MAP}}$. As prior function
$\pr(\veceta)$, we initially adopt very broad uninformative uniform
priors: for the logarithmic slope we consider the interval $\gamma \in
[1, 3]$; for the lens strength $\talp \in [0,1]$; the inclination can
assume all the values $i \in [0^{\circ}, 90^{\circ}]$; finally, the
axial ratio $q \in [0, 1.5]$ can account for all the flattenings
between a thin disk and a very prolate distribution. Fast preliminary
exploration runs then permit us to select, for each system, narrower
priors that remain, however, generously non-restrictive (i.e. they
contain the bulk of the posterior and a wide margin of safety all
around it).

The uncertainties on each individual parameter $\eta_{k}$ are obtained
by marginalizing the posterior PDF over all other parameters. We refer
to the parameter value that corresponds to the maximum of the $k$-th
one-dimensional marginalized distribution as $\eta_{k,
  \mathrm{max}}$. In this context, it is important to note that, in
general, $\veceta_{\mathrm{MAP}}$ and $\veceta_{\mathrm{max}}$ do not
coincide and can actually differ significantly. In other words, the
MAP parameters \emph{individually} are not guaranteed to be probable:
although the MAP solution corresponds by definition to the highest
value of the joint posterior PDF, it does not necessarily occupy a
large volume in the multivariate parameter space and could easily lie
outside of the bulk of the posterior PDF if the distribution is
strongly non-symmetric.

The computationally expensive task of exploring the posterior
distribution $\pr(\veceta \, | \, \vec{d})$ is accomplished by making
use of the very efficient and robust \textsc{MultiNest} algorithm
\citep{Feroz-Hobson2008, Feroz2009}, which implements the `nested
sampling' Monte Carlo technique \citep{Skilling2004, SS2006}, and can
provide reliable posterior inferences even in presence of multi-modal
and degenerate multivariate distributions. For the analysis of the
SLACS lenses, we launch \textsc{MultiNest} with~200 live points (the
live or active points are the initial samples, drawn from the prior
distribution, from which the posterior exploration is started) to
obtain the individual marginalized posterior PDFs of the non-linear
parameters~$i$, $\talp$, $\gamma$ and~$q$. In the following, unless
stated otherwise, we quote the~99 per cent confidence intervals
(calculated such that the probability of being below the interval is
the same as being above it) obtained from these distributions as our
uncertainty on the parameters.

\begin{table*}
  \centering
  \caption{Recovered parameters for the sixteen sample SLACS lens galaxies.}
  \smallskip
  \begin{tabular}{ l c c c c c c @{} p{2em} @{} c c c c }
    \hline
    \noalign{\smallskip}
    \multirow{3}{*}{Galaxy name} & \multirow{3}{*}{$z_{\mathrm{lens}}$} & \multirow{3}{*}{$z_{\mathrm{src}}$} & \multicolumn{4}{c}{$\eta_{\mathrm{MAP}}$} & & \multicolumn{4}{c}{$\eta_{\mathrm{max}}$}\\
    \noalign{\smallskip}
    \cline{4-7} \cline{9-12}
    \noalign{\smallskip}
     & & & $\slope$ & $\talp$ & $q$ & $i$ & & $\slope$ & $\talp$ & $q$ & $i$ \\
    \noalign{\smallskip}
    \hline
    \noalign{\smallskip}
    SDSS\,J0037$-$0942 & 0.1955 & 0.6322 & 1.968 & 0.434 & 0.693 & 65.5 & & $1.969^{+0.014}_{-0.051}$ & $0.433^{+0.018}_{-0.009}$ & $0.694^{+0.047}_{-0.041}$ & $65.4^{+2.6}_{-0.9}$ \\ 
    \noalign{\smallskip}                                                                                                                                                             
    SDSS\,J0216$-$0813 & 0.3317 & 0.5235 & 1.973 & 0.344 & 0.816 & 70.0 & & $1.972^{+0.035}_{-0.077}$ & $0.344^{+0.018}_{-0.012}$ & $0.818^{+0.065}_{-0.061}$ & $69.5^{+6.3}_{-2.2}$ \\ 
    \noalign{\smallskip}                                                                                                                                                             
    SDSS\,J0912$+$0029 & 0.1642 & 0.3240 & 1.877 & 0.412 & 0.672 & 87.8 & & $1.880^{+0.035}_{-0.016}$ & $0.413^{+0.005}_{-0.004}$ & $0.670^{+0.021}_{-0.026}$ & $87.6^{+0.4}_{-1.6}$ \\ 
    \noalign{\smallskip}                                                                                                                                                             
    SDSS\,J0935$-$0003 & 0.3475	& 0.4670 & 2.285 & 0.225 & 0.982 & 34.8 & & $2.280^{+0.072}_{-0.056}$ & $0.228^{+0.013}_{-0.019}$ & $0.982^{+0.033}_{-0.124}$ & $34.0^{+2.3}_{-2.8}$ \\ 
    \noalign{\smallskip}                                                                                                                                                             
    SDSS\,J0959$+$0410 & 0.1260 & 0.5349 & 1.873 & 0.323 & 0.930 & 80.4 & & $1.869^{+0.013}_{-0.024}$ & $0.325^{+0.009}_{-0.008}$ & $0.920^{+0.029}_{-0.035}$ & $80.8^{+1.9}_{-1.5}$ \\ 
    \noalign{\smallskip}                                                                                                                                                             
    SDSS\,J1204$+$0358 & 0.1644 & 0.6307 & 2.226 & 0.399 & 1.000 & 65.2 & & $2.223^{+0.057}_{-0.074}$ & $0.382^{+0.020}_{-0.011}$ & $0.986^{+0.028}_{-0.104}$ & $65.8^{+6.3}_{-3.4}$ \\
    \noalign{\smallskip}                                                                                                                                                             
    SDSS\,J1250$+$0523 & 0.2318 & 0.7953 & 2.131 & 0.339 & 0.771 & 29.1 & & $2.134^{+0.025}_{-0.074}$ & $0.336^{+0.021}_{-0.011}$ & $0.772^{+0.054}_{-0.133}$ & $28.4^{+3.9}_{-6.0}$ \\ 
    \noalign{\smallskip}                                                                                                                                                             
    SDSS\,J1251$-$0208 & 0.2243	& 0.7843 & 1.925 & 0.244 & 0.796 & 85.6 & & $1.924^{+0.009}_{-0.030}$ & $0.245^{+0.005}_{-0.006}$ & $0.805^{+0.018}_{-0.046}$ & $85.4^{+1.6}_{-2.9}$ \\ 
    \noalign{\smallskip}                                                                                                                                                             
    SDSS\,J1330$-$0148 & 0.0808	& 0.7115 & 2.276 & 0.155 & 0.414 & 74.3 & & $2.280^{+0.039}_{-0.037}$ & $0.154^{+0.008}_{-0.012}$ & $0.414^{+0.018}_{-0.073}$ & $75.3^{+1.9}_{-1.7}$ \\ 
    \noalign{\smallskip}                                                                                                                                                             
    SDSS\,J1443$+$0304 & 0.1338	& 0.4187 & 2.435 & 0.149 & 0.698 & 71.5 & & $2.431^{+0.069}_{-0.076}$ & $0.150^{+0.020}_{-0.021}$ & $0.682^{+0.042}_{-0.085}$ & $71.8^{+2.5}_{-1.0}$ \\ 
    \noalign{\smallskip}                                                                                                                                                             
    SDSS\,J1451$-$0239 & 0.1254	& 0.5203 & 2.053 & 0.332 & 0.999 & 40.8 & & $2.028^{+0.030}_{-0.019}$ & $0.337^{+0.003}_{-0.007}$ & $0.997^{+0.012}_{-0.055}$ & $40.8^{+1.9}_{-3.0}$ \\ 
    \noalign{\smallskip}                                                                                                                                                             
    SDSS\,J1627$-$0053 & 0.2076 & 0.5241 & 2.122 & 0.369 & 0.851 & 56.4 & & $2.123^{+0.029}_{-0.016}$ & $0.369^{+0.004}_{-0.011}$ & $0.853^{+0.014}_{-0.040}$ & $55.8^{+1.1}_{-1.2}$ \\ 
    \noalign{\smallskip}                                                                                                                                                             
    SDSS\,J2238$-$0754 & 0.1371	& 0.7126 & 2.088 & 0.362 & 0.781 & 79.8 & & $2.098^{+0.022}_{-0.053}$ & $0.358^{+0.011}_{-0.005}$ & $0.767^{+0.042}_{-0.067}$ & $79.1^{+1.1}_{-2.2}$ \\ 
    \noalign{\smallskip}                                                                                                                                                             
    SDSS\,J2300$+$0022 & 0.2285	& 0.4635 & 1.921 & 0.349 & 0.642 & 59.4 & & $1.923^{+0.033}_{-0.043}$ & $0.344^{+0.007}_{-0.004}$ & $0.622^{+0.034}_{-0.012}$ & $58.7^{+4.0}_{-1.9}$ \\ 
    \noalign{\smallskip}                                                                                                                                                             
    SDSS\,J2303$+$1422 & 0.1553	& 0.5170 & 2.102 & 0.436 & 0.642 & 89.1 & & $2.098^{+0.030}_{-0.004}$ & $0.437^{+0.003}_{-0.007}$ & $0.644^{+0.004}_{-0.033}$ & $85.9^{+3.5}_{-0.8}$ \\ 
    \noalign{\smallskip}                                                                                                                                                             
    SDSS\,J2321$-$0939 & 0.0819 & 0.5324 & 2.058 & 0.467 & 0.744 & 67.3 & & $2.059^{+0.013}_{-0.066}$ & $0.468^{+0.009}_{-0.003}$ & $0.744^{+0.029}_{-0.011}$ & $67.5^{+2.1}_{-2.0}$ \\ 
    \noalign{\smallskip}
    \hline
    \noalign{\smallskip}
  \end{tabular}

  \begin{minipage}{1.00\hsize}
    \textit{Note.} The non-linear $\eta$~parameters are: the
    logarithmic slope $\slope$; the dimensionless lens strength
    $\talp$; the axial ratio $q$ and the inclination $i$ (in degrees)
    with respect to the line-of-sight. Columns~4 to~7 list the MAP
    parameters, i.e. the parameters that maximize the joint posterior
    distribution (`best model' parameters). Columns~8 to~11
    encapsulate a description of the one-dimensional marginalized
    posterior PDFs (shown as histograms in Appendix~\ref{app:unc} in
    the online version of the journal): $\eta_{\mathrm{max}}$ is the
    maximum of that distribution and the indicated errors represent
    the lower and upper limits of the~99 per cent confidence interval.
  \end{minipage}
  \label{tab:eta}
\end{table*}

\subsection{Results}
\label{sec:results}

The {\cauldron} code has been applied to the analysis of all sixteen
lens galaxies for which VIMOS integral-field spectroscopic data are
available. These objects are all early-type galaxies, with the
exception of J1251$-$0208, which is morphologically classified by
\citet{Auger2009} as a late-type galaxy. We include the latter in the
E/S0 sample studied in this paper, since it is a bulge-dominated
system which was spectroscopically selected from the SDSS using the
same criteria\,---\,optimized to detect bright early-type lens
galaxies\,---\,adopted for the other objects \citep[see][]{Bolton2005,
  Bolton2006}. For six of the systems, the results of the combined
lensing and dynamics study were reported in previous publications
\citep{Czoske2008, Barnabe2009}. However, updated and extended
kinematic maps are available for J2321 and therefore we have
re-analyzed that galaxy.

Table~\ref{tab:eta} lists, for each galaxy, the recovered non-linear
parameters for the MAP model, i.e. the inclination~$i$, the lens
strength~$\talp$, the logarithmic slope~$\slope$ and the axial
ratio~$q$ of the total density distribution. The reconstructed
observables corresponding to this model are presented and compared to
the data sets in Appendix~\ref{app:models} (available in the online
version of the journal), where also the recovered unlensed background
source is shown; the MAP model reconstruction of the weighted DF is
presented in Sect.~\ref{sec:dynamics}, where we study the dynamical
structure of the lens galaxies. None of the galaxies requires external
shear to be added to the model.

The marginalized posterior PDFs for the non-linear model
parameters\,---\,which quantify the statistical uncertainties on those
parameters (see Sect.~\ref{sec:unc})\,---\,are presented in
Appendix~\ref{app:unc} (available in the online version of the
journal) for all sixteen galaxies in the sample\footnote{The error
  analysis for the galaxies studied in \citet{Barnabe2009} was
  performed using a less robust implementation of the nested sampling
  technique. For consistency, the posterior distributions for those
  systems have been re-analyzed here using the more robust
  \textsc{MultiNest} algorithm. The revised confidence intervals are
  found to be wider, as a consequence of the improved parameter space
  exploration.}. In Table~\ref{tab:eta} we also indicate, as a compact
description of the errors, the parameter values which correspond to
the maximum of the marginalized posterior distributions (subscript
`max') and the limits of the~99 per cent confidence intervals.

All the observables are, in general, accurately reproduced within the
noise level. In particular, the lensed image residual maps do not
exhibit the prominent structured patterns that can be seen when the
method is applied to complex simulated systems, as in the
\citet{Barnabe2008} tests, thus indicating that the underlying density
distribution of real ellipticals is fairly smooth and well-behaved,
and can be satisfactorily modelled by the axisymmetric power-law
profile of Eq.~(\ref{eq:rho}).

The reconstructed background sources generally appear as simple and
smooth systems dominated by a single component, but a few of them
display a more complex morphology or fainter secondary components. To
understand this, we recall that the SLACS candidates are
spectroscopically selected from the SDSS database by identifying those
systems which have composite spectra consisting of both an
absorption-dominated continuum (due to the foreground galaxy) and
multiple nebular emission lines at a higher redshift (due to the
background object), which directly trace star formation
\citep[see][]{Bolton2006}. Therefore, the background objects tend to
be actively star-forming galaxies which, in the typical redshift range
where the sources are located, i.e. $0.4 \lesssim z \lesssim 0.8$, can
sometimes be late-type systems characterized by clumpy, irregular and
flocculent appearances \citep[cf.\ e.g.][]{Zamojski2007,
  Elmegreen2009}.

The complicated and patchy morphology of certain sources (as in the
case of SDSS\,J2303 or, more conspicuously, of the SLACS lens
SDSS\,J0728 analyzed in \citealt{Barnabe2010}) as well as the presence
of multiple sources almost adjacent to each other (e.g. SDSS\,J1251,
SDSS\,J2321) can thus be interpreted as distinct star-forming regions
within the same galaxy. Analogously, the small clumps observed in
SDSS\,J1250 can likely be identified as bright blue star-forming
regions that are observed within an interacting system with a tidal
tale.

\begin{figure}
  \centering
  \resizebox{1.00\hsize}{!}{\includegraphics[angle=-90]
            {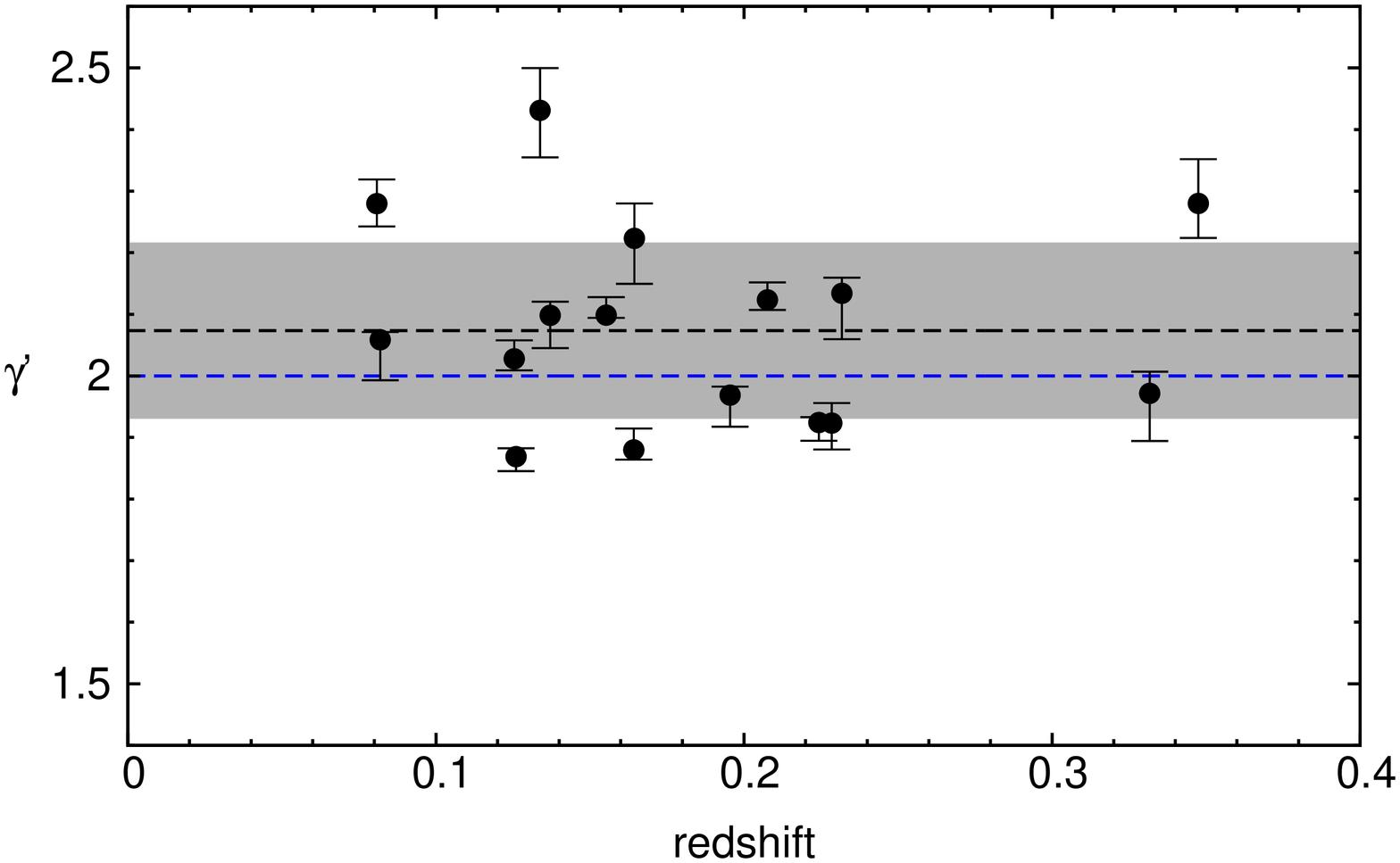}}
  \caption{The logarithmic slope of the total density profile plotted
    against redshift for the sixteen early-type lens galaxies in the
    ensemble. The error bars are calculated by considering the region
    of the marginalized posterior PDF for $\slope$ (see
    Figure~\ref{fig:NSerrors}) which contains 99 per cent of the
    probability. The blue line indicates the slope $\slope = 2$,
    corresponding to the isothermal profile, while the dashed black
    line denotes the average slope $\mslope = 2.07$ for the
    sample. The grey band region visualizes the intrinsic scatter of
    about~7 per cent in $\slope$ (see text for details).}
  \label{fig:gamma}
\end{figure}

\begin{figure}
  \centering
  \resizebox{1.00\hsize}{!}{\includegraphics[angle=-90]
            {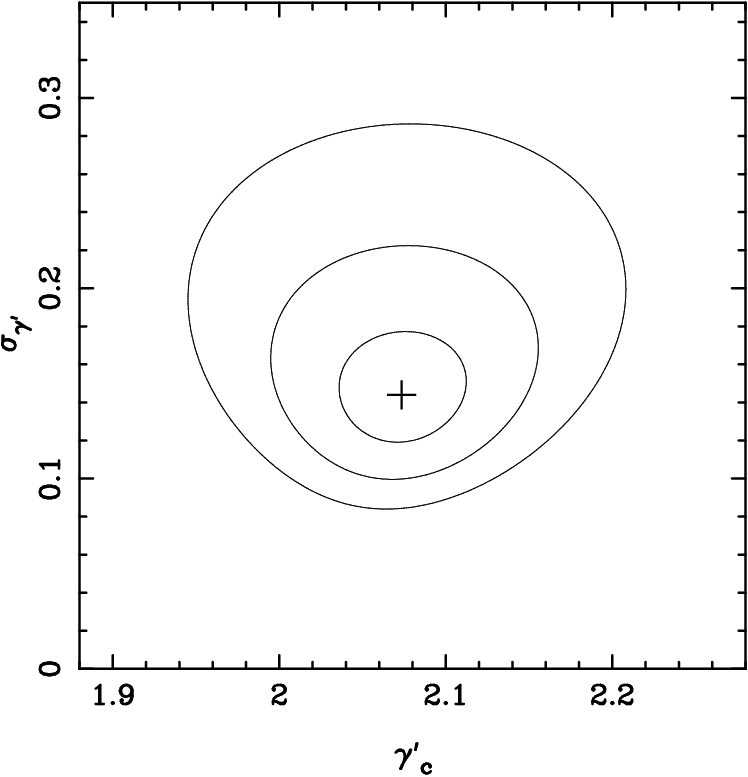}}
  \caption{Joint posterior probability for $\slopec$ and $\sigmams$
    (see Eq.~[\ref{eq:jointP}]). The cross marks the position of the
    joint maximum. The three contours correspond to the posterior
    ratios $\mathcal{P}/\mathcal{P_{\mathrm{max}}} = e^{- \Delta
      \tilde{\chi}^{2} / 2}$, with $\Delta \tilde{\chi}^{2} = 1, 4,
    9$. The contours are only for indication: formally, they can be
    interpreted as 1, 2 and 3\,$\sigma$ contours only for a
    (multivariate) Gaussian distribution, in which case
    $\tilde{\chi}^{2}$ represents the usual chi-square.}
  \label{fig:jointP}
\end{figure}

\subsubsection{The slope of the density profile}
\label{ssec:slope}

By examining the recovered values for the logarithmic slope of the
total density distribution, we find no evidence of evolution
of~$\slope$ with redshift within the probed range ($z = 0.08$
to~$0.35$), confirming the results of \citet{Koopmans2009} and
\citet{Auger2010} on the full SLACS sample (see
Figure~\ref{fig:gamma}). Moreover, from the same Figure, we also see
that, while almost all systems have a nearly (i.e. within 10 per cent)
isothermal profile, very few of them exhibit a slope that is actually
consistent with $\slope = 2$ within the uncertainties.

We now proceed to study in a more quantitative way the distribution of
density slopes for the ensemble at hand. Under the assumption that the
slope $\slope_{\mathrm{i}}$ (with a symmetrized 1\,$\sigma$ error
$\deslope$) of each lens system is drawn from a parent Gaussian
distribution of centre $\slopec$ and standard deviation $\sigmams$,
the joint posterior probability $\mathcal{P}$ of these two parameters
for the sample is given by
\begin{equation}
\label{eq:jointP}
  \mathcal{P} = \pr(\slopec, \sigmams \, | \, \{ \slope_{\mathrm{i}}\}) 
  \propto
  p(\slopec, \sigmams) \, \prod_{\mathrm{i}}
  \frac{
  \exp{\left[ - \frac{(\slope_{\mathrm{i}} - \slopec)^{2}}
  {2 (\sigmamsq + \deslopeq)} \right]}
  }
  {\sqrt{2 \pi (\sigmamsq + \deslopeq)}} \, ,
\end{equation}
where $p(\slopec, \sigmams)$ indicates the prior, for which we adopt a
uniform distribution\footnote{Since the results are essentially data
  driven, the specific choice of the prior distribution is largely
  irrelevant. If, for instance, we adopt a scale-invariant Jeffreys
  prior $p \propto 1/\sigmams$ (which indicates the absence of \emph{a
    priori} information on the scale of $\sigmams$) the recovered MAP
  value for $\sigmams$ differs less than~4 per cent from the value
  obtained above with $p = \mathrm{constant}$.}. This probability
function, visualized in Figure~\ref{fig:jointP}, yields the average
logarithmic density slope
\begin{displaymath}
\label{eq:mslope}
\mslope = 2.074^{+0.043}_{-0.041} \quad (68\% \; \mathrm{CL}) ,
\end{displaymath}
which is slightly super-isothermal. This result is consistent with
the value of $\mslope$ determined by the \citet{Koopmans2009} and
\citet{Auger2010} combined lensing and dynamics analysis of the full
SLACS sample (which does not include two-dimensional kinematic
information), using spherical Jeans equations and assuming an
isotropic stellar velocity dispersion tensor.

The \emph{intrinsic} scatter around the average slope, also derived
from the posterior distribution of Eq.~(\ref{eq:jointP}), is 
\begin{displaymath}
\sigmams = 0.144^{+0.055}_{-0.014} \quad (68\% \; \mathrm{CL}) ,
\end{displaymath}
which corresponds to $6.9^{+2.7}_{-0.6}$ per cent of~$\mslope$, and is
displayed in Figure~\ref{fig:gamma} as a grey band. This value is only
marginally lower than the intrinsic scatter of about~10 per cent
obtained by \citet{Koopmans2009} and \citet{Auger2010}, and the two
$\sigmams$ are consistent within the errors.

\begin{figure}
  \centering
  \resizebox{0.95\hsize}{!}{\includegraphics[angle=-90]
            {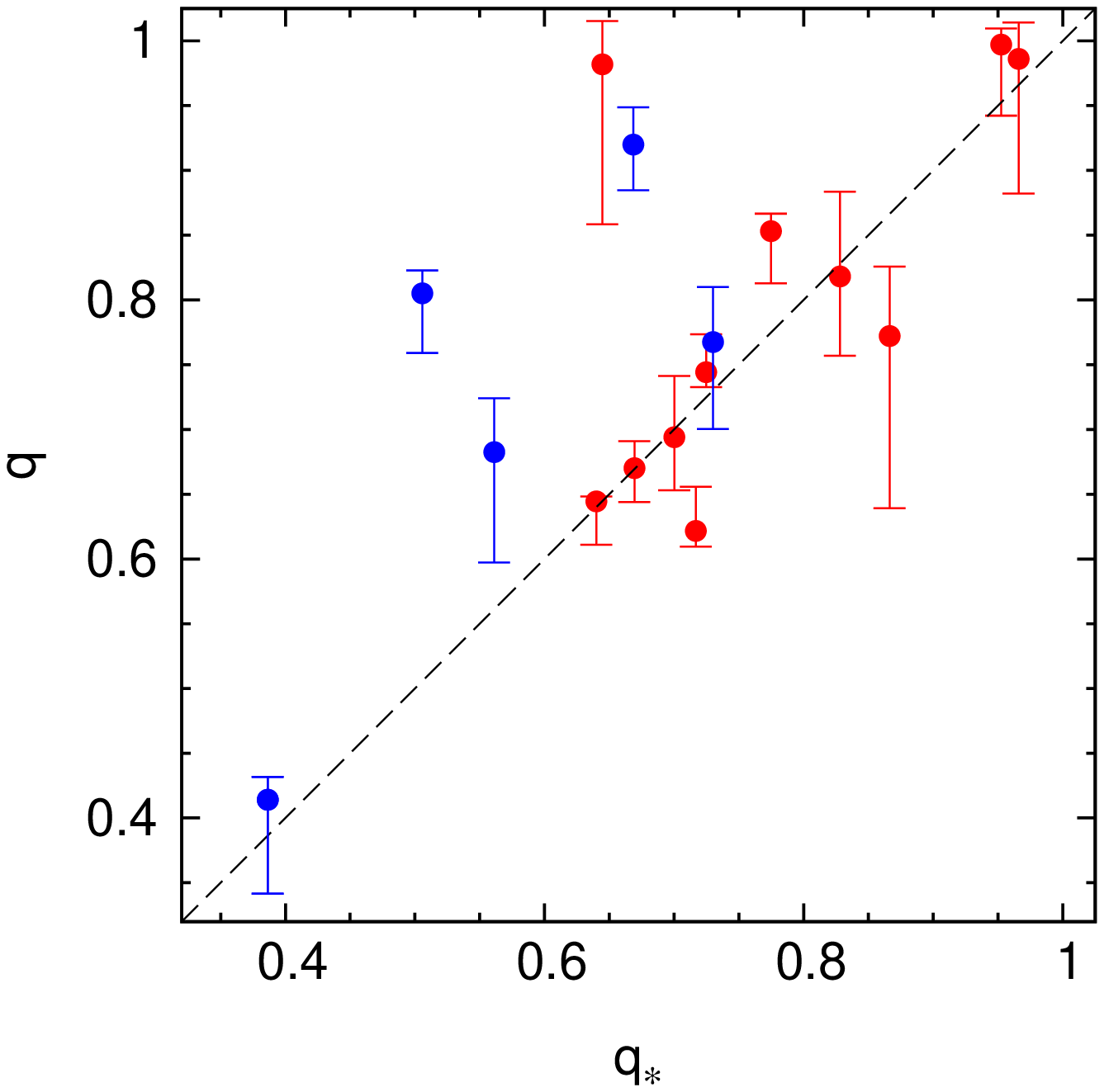}}
  \caption{Axial ratio $q$ of the total density distribution plotted
    against the intrinsic axial ratio $\qstar$ of the luminous
    distribution. For each galaxy, $\qstar$ is calculated by
    deprojecting the observed isophotal axis ratio $\qspro$, using the
    recovered MAP value of the inclination. The dashed line
    corresponds to $q = \qstar$. Red and blue symbols denote,
    respectively, slow and fast rotators.}
  \label{fig:q}
\end{figure}

\subsubsection{The flattening of the density profile}
\label{ssec:axialratio}

All the analyzed galaxies\,---\,with the exception of
SDSS\,J1330\,---\,are found to be in the range between moderately
flattened and perfectly spherical, with a total density axial ratio
$0.6 < q \le 1.0$. The ensemble-average axial ratio, obtained from the
joint posterior probability as described in Sect.~\ref{ssec:slope}, is
\begin{displaymath}
\langle q \rangle = 0.770^{+0.042}_{-0.041} \quad (68\% \; \mathrm{CL}) ,
\end{displaymath}
with a quite substantial intrinsic scatter of about 19 per cent around
this value:
\begin{displaymath}
\sigma_{q} = 0.148^{+0.049}_{-0.019} \quad (68\% \; \mathrm{CL}) .
\end{displaymath}

The three roundest galaxies in the sample, and the only ones
consistent with being spherical, are all slow rotators (see
\citealt{Emsellem2007}, and Sect.~\ref{ssec:jz} for our definition of
slow and fast rotating systems based on the specific angular
momentum). The system SDSS\,J1330, which is peculiar for its
remarkably flattened axial ratio $q \simeq 0.4$, also bears the
distinction of being the least massive object and one of the only five
fast-rotating galaxies in the sample. Fast rotators, however, are not
necessarily very flattened: SDSS\,J0959, for instance, is found to be
almost spherical, with $q = 0.9$.

In Figure~\ref{fig:q}, $q$~is compared to the intrinsic
(i.e. three-dimensional) axial ratio of the luminous distributions,
$\qstar$, calculated as
\begin{equation}
  \label{eq:qintr}
  \qstar = \sqrt{1 - (1 - q^{2}_{\mathrm{\ast, 2D}})/{\sin}^{2} i} 
           \: \textrm{,}
\end{equation}
where $\qspro$ is the observed isophotal axial ratio and $i$ indicates
the recovered MAP value for the inclination. For most systems, the
flattening of the luminous distribution is consistent or very similar
(i.e. within $\sim 10$ per cent) to the flattening of the total
distribution; we note that this does not necessarily imply that mass
follows light as the total and luminous radial density profiles could
be very different. Only three galaxies exhibit a significant
discrepancy, all of them in the sense of having a total axial ratio
much rounder than the luminous one, i.e. $q/\qstar \gtrsim 1.4$. Of
these systems, two (SDSS\,J0959 and SDSS\,J1251) are fast rotators,
whereas the third one, SDSS\,J0935, is a very massive slow rotator
(and the most massive galaxy in the whole sample,
cf.\ Sect.~\ref{sec:mass}). Out of the remaining fast rotating
galaxies, SDSS\,J1443 shows $q/\qstar \simeq 1.2$, while SDSS\,J1330
and SDSS\,J2238 are slightly rounder in the total distribution but
still consistent with $q = \qstar$. In conclusion, from the small
sample at hand, we find that fast rotators tend to be more flattened
in the stellar mass distribution than in the total density
distribution (i.e., $q$ is at least as round as~$\qstar$, and
generally rounder).


\section{Mass structure}
\label{sec:mass}

\subsection{Total mass}

\begin{figure*}
  \begin{center}
    \subfigure{\label{fig:combo-1330}\includegraphics[angle=-90,width=0.24\textwidth]{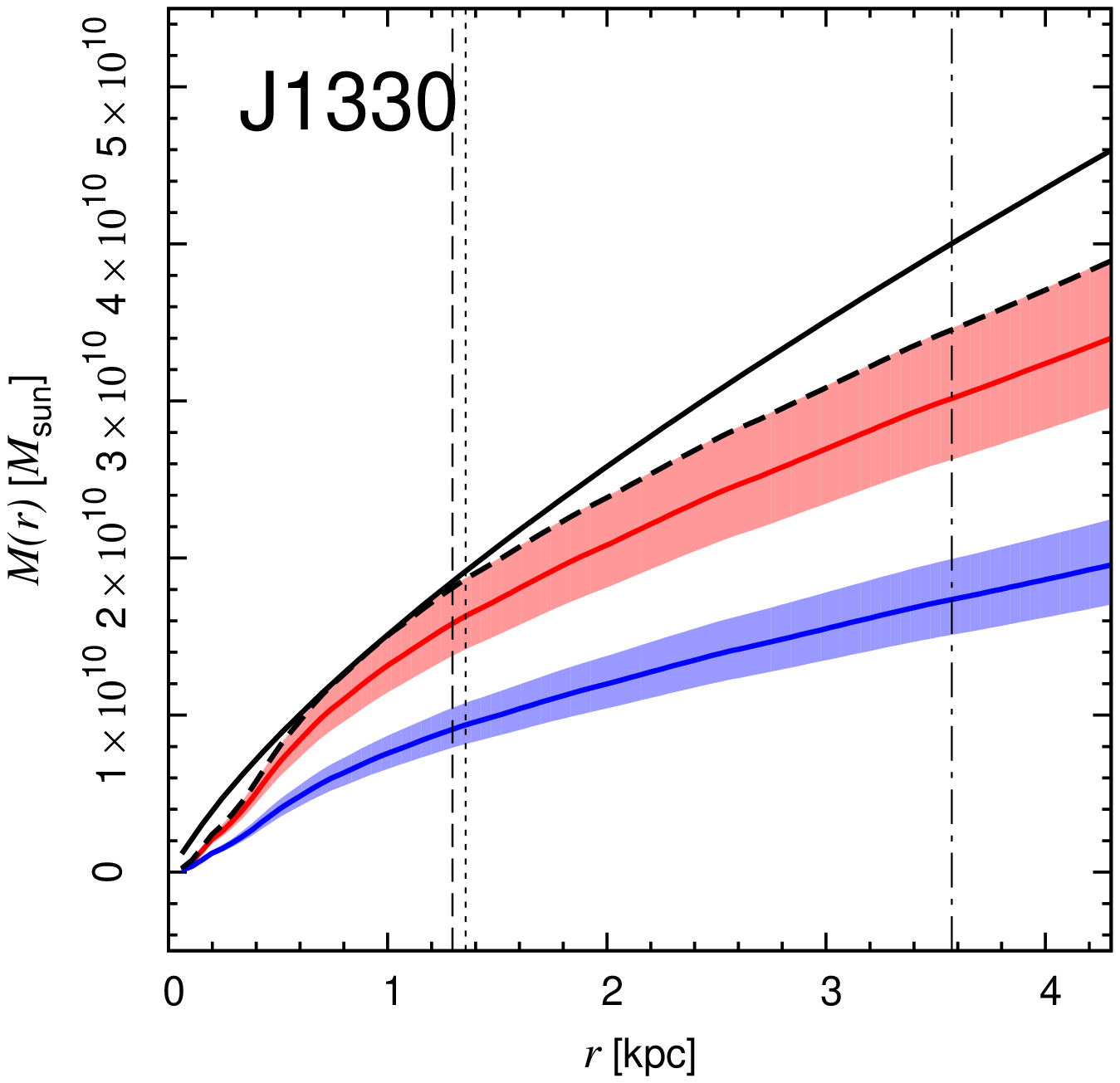}}
    \subfigure{\label{fig:combo-1443}\includegraphics[angle=-90,width=0.24\textwidth]{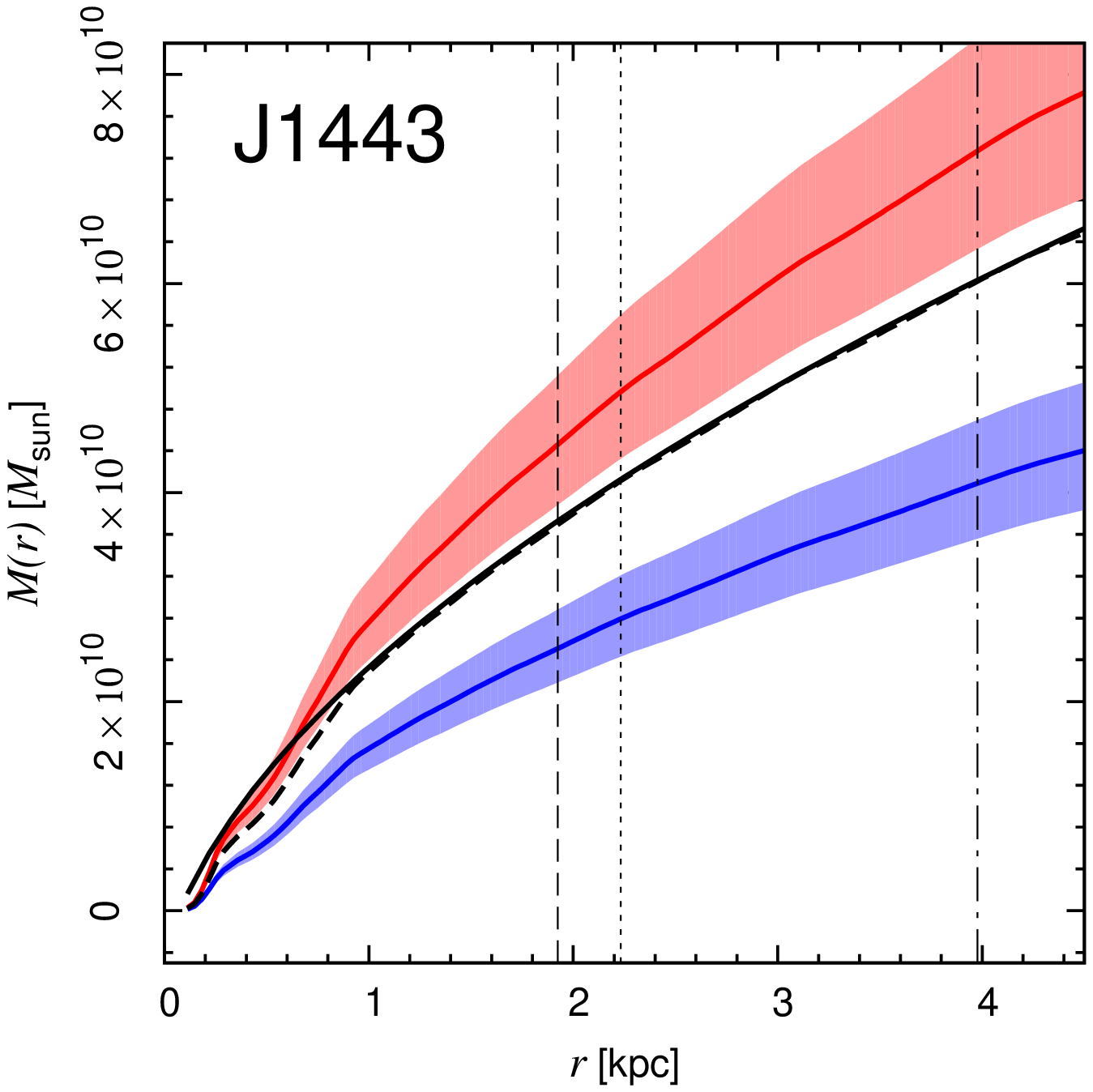}}
    \subfigure{\label{fig:combo-0959}\includegraphics[angle=-90,width=0.24\textwidth]{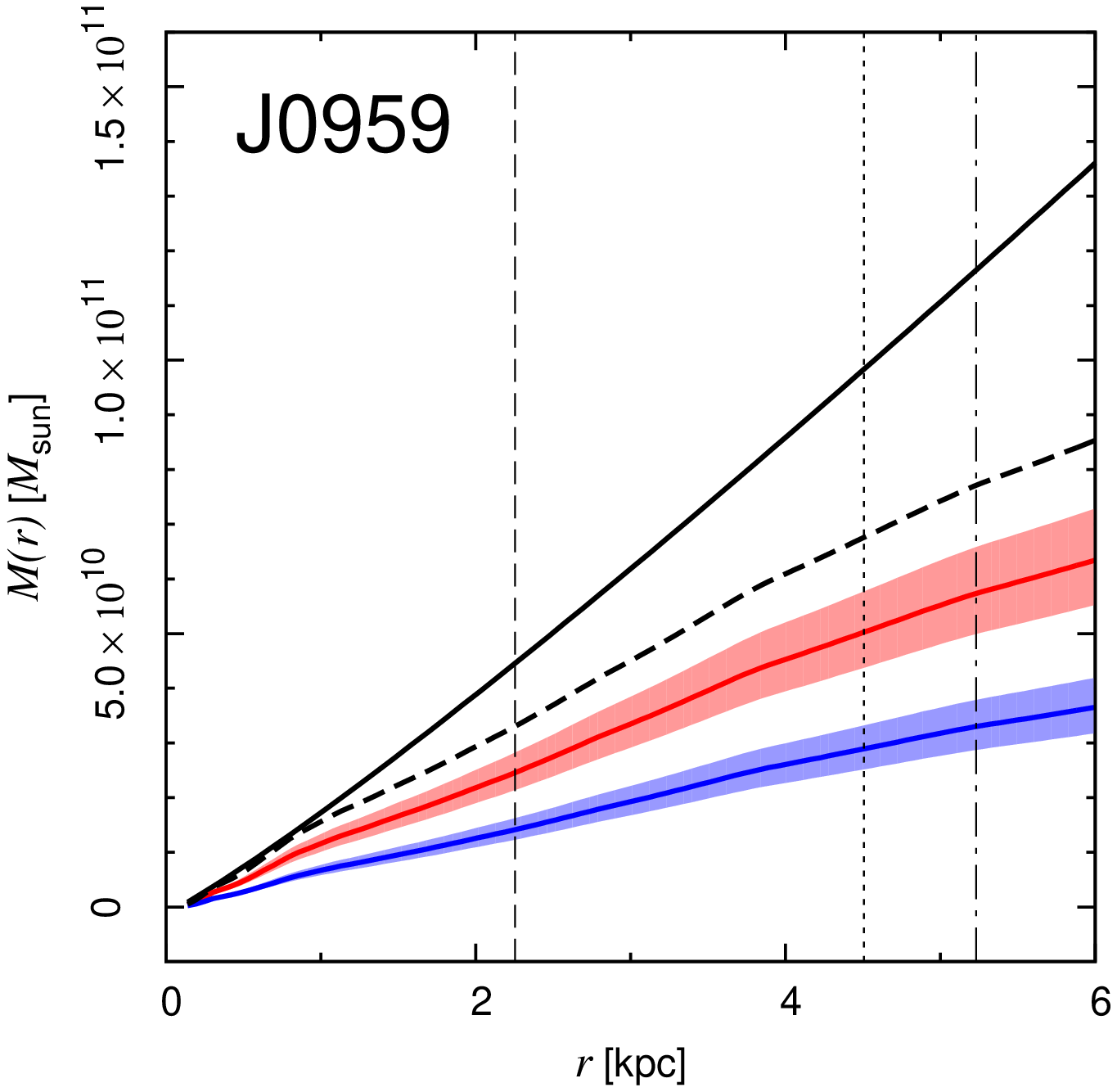}}
    \subfigure{\label{fig:combo-1451}\includegraphics[angle=-90,width=0.24\textwidth]{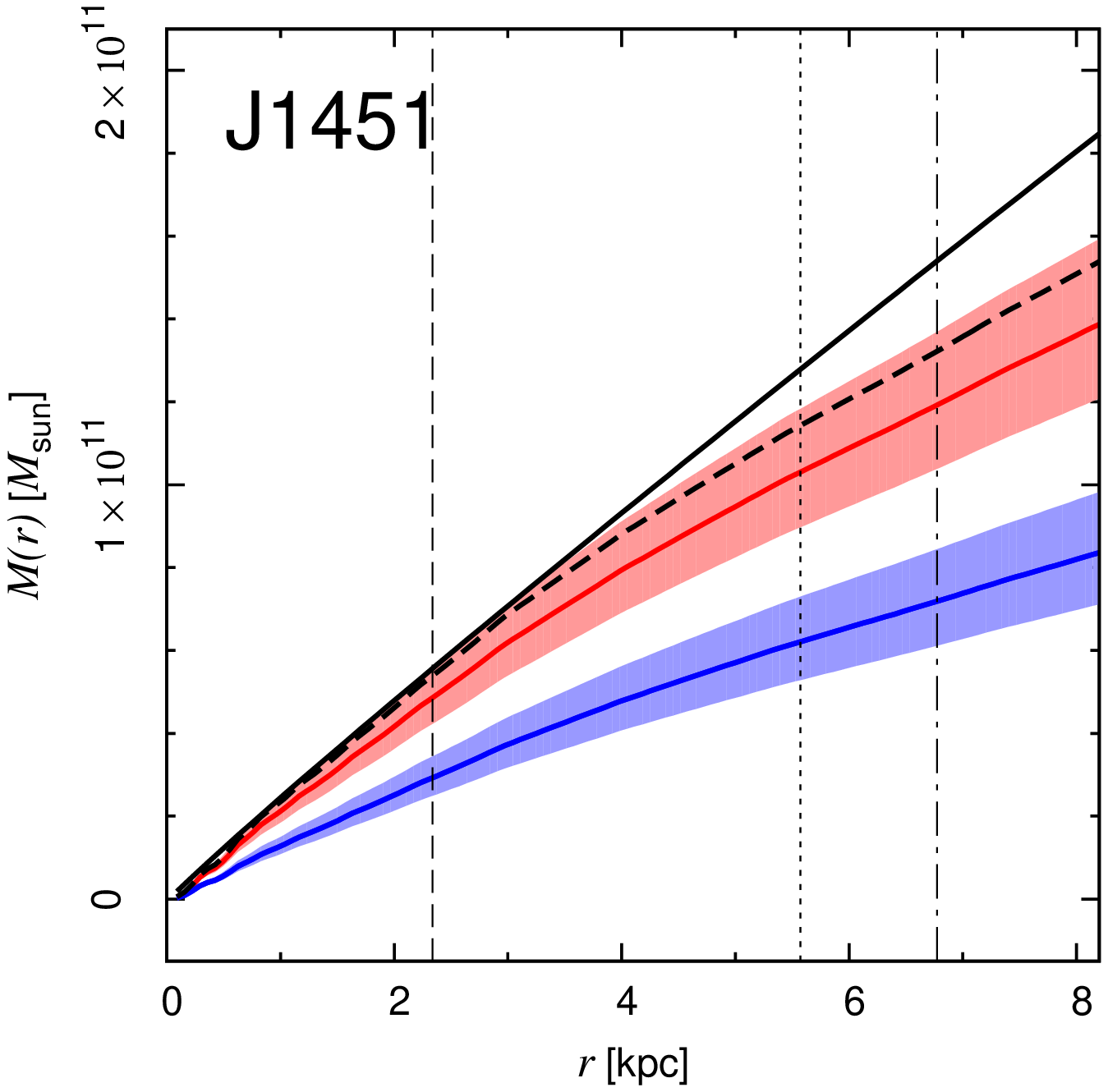}}
  \end{center}
  \vspace{-0.5cm}

  \begin{center}
    \subfigure{\label{fig:combo-1204}\includegraphics[angle=-90,width=0.24\textwidth]{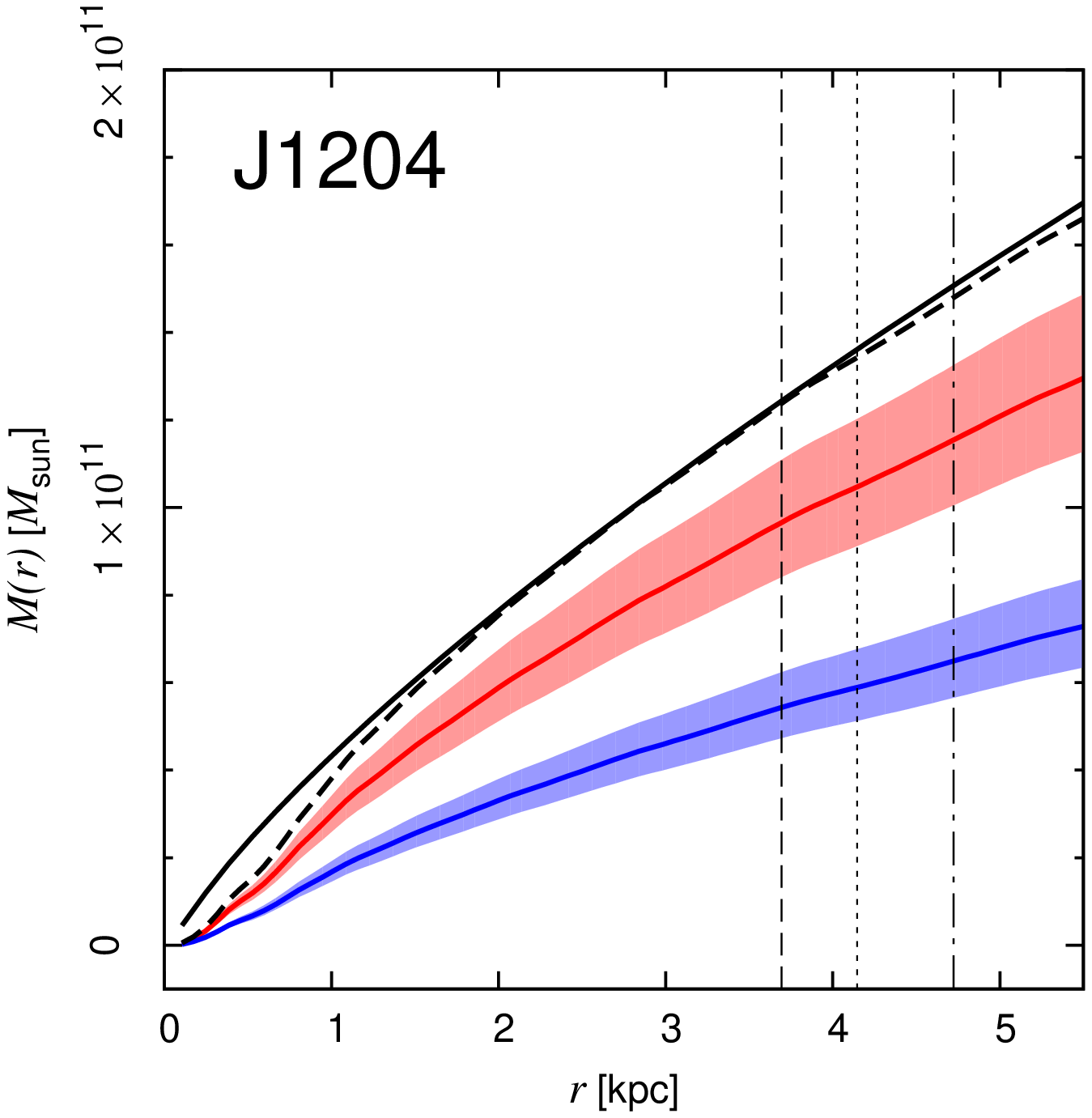}}
    \subfigure{\label{fig:combo-2238}\includegraphics[angle=-90,width=0.24\textwidth]{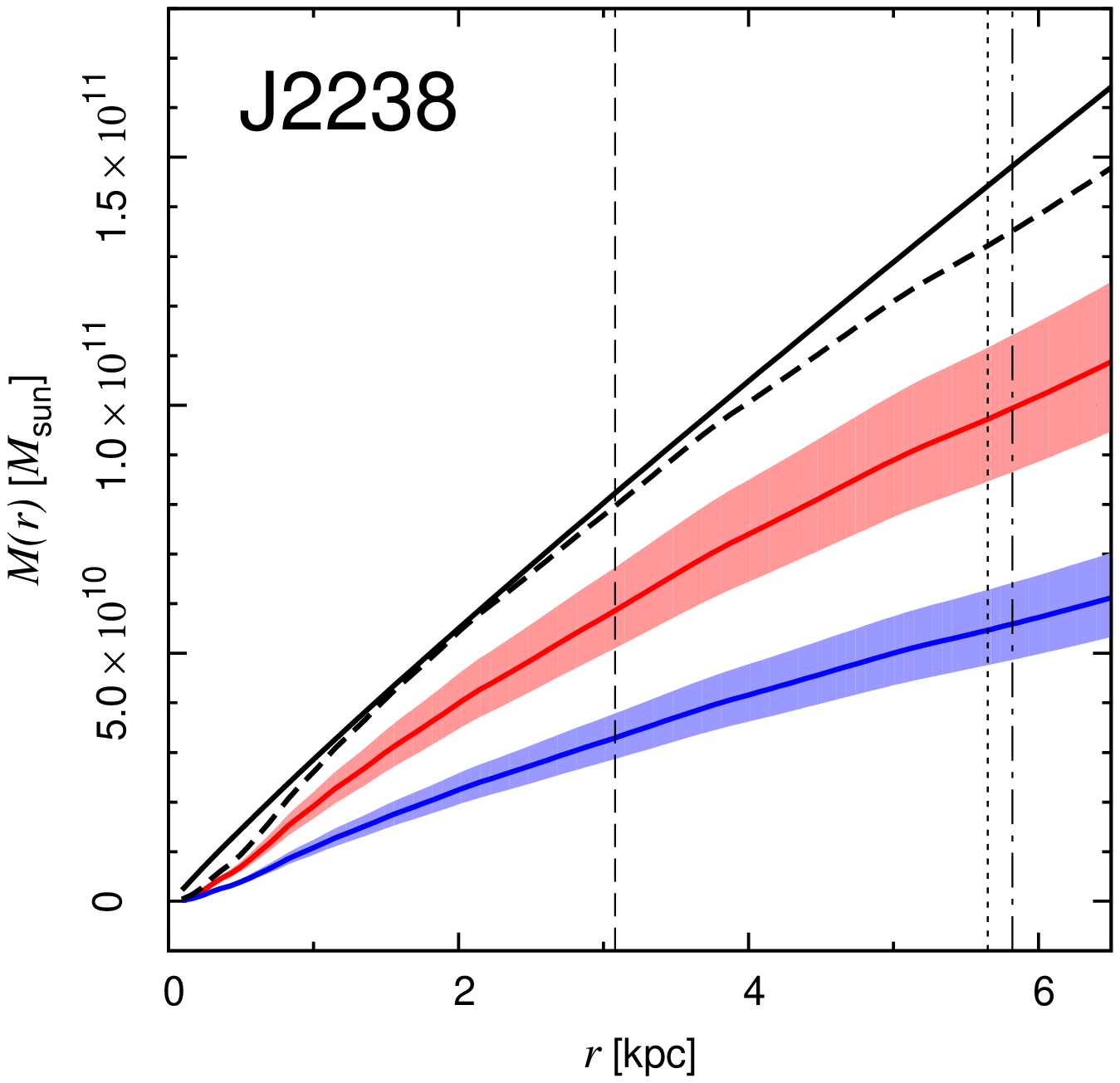}}
    \subfigure{\label{fig:combo-1251}\includegraphics[angle=-90,width=0.24\textwidth]{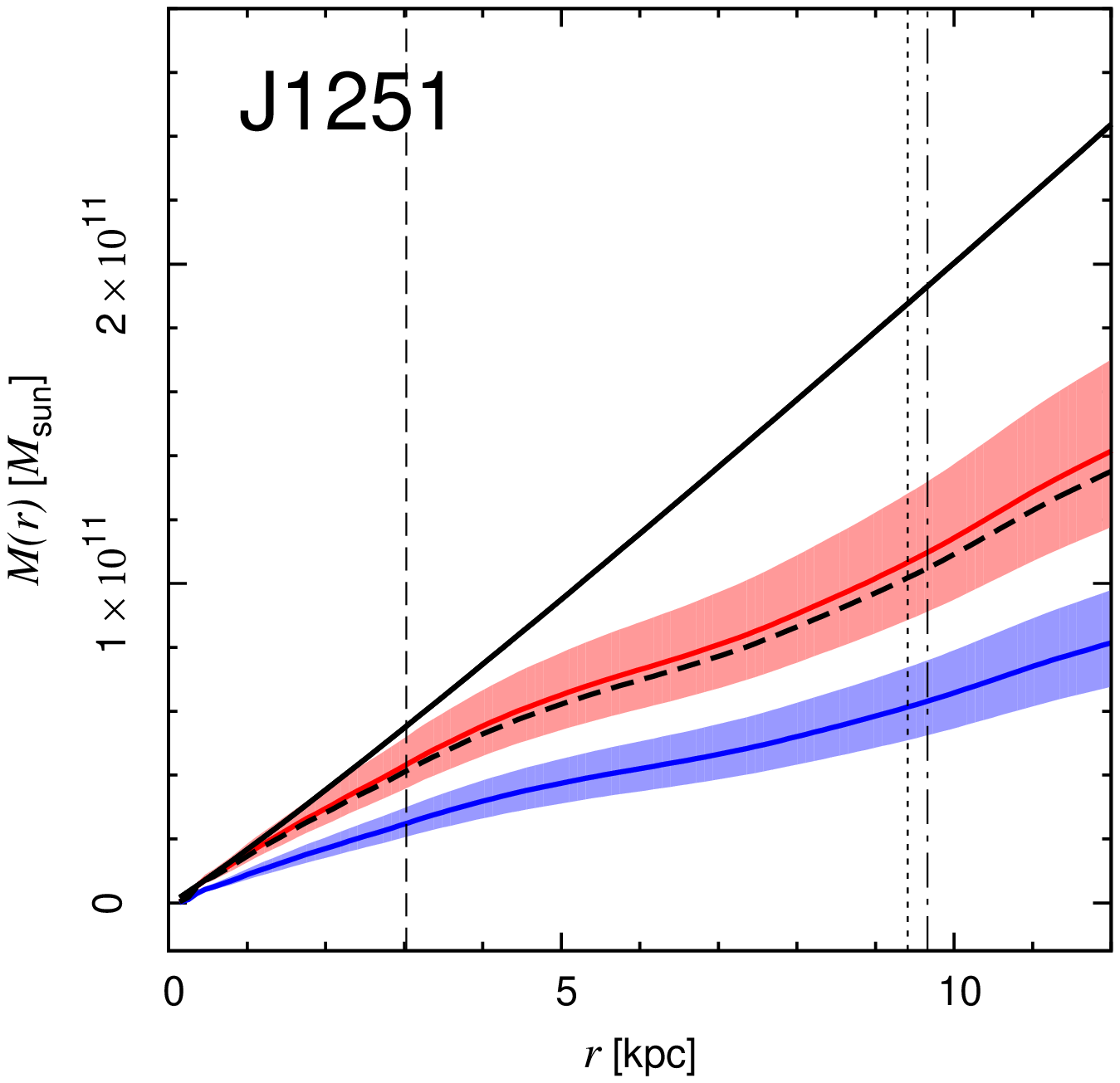}}
    \subfigure{\label{fig:combo-2321}\includegraphics[angle=-90,width=0.24\textwidth]{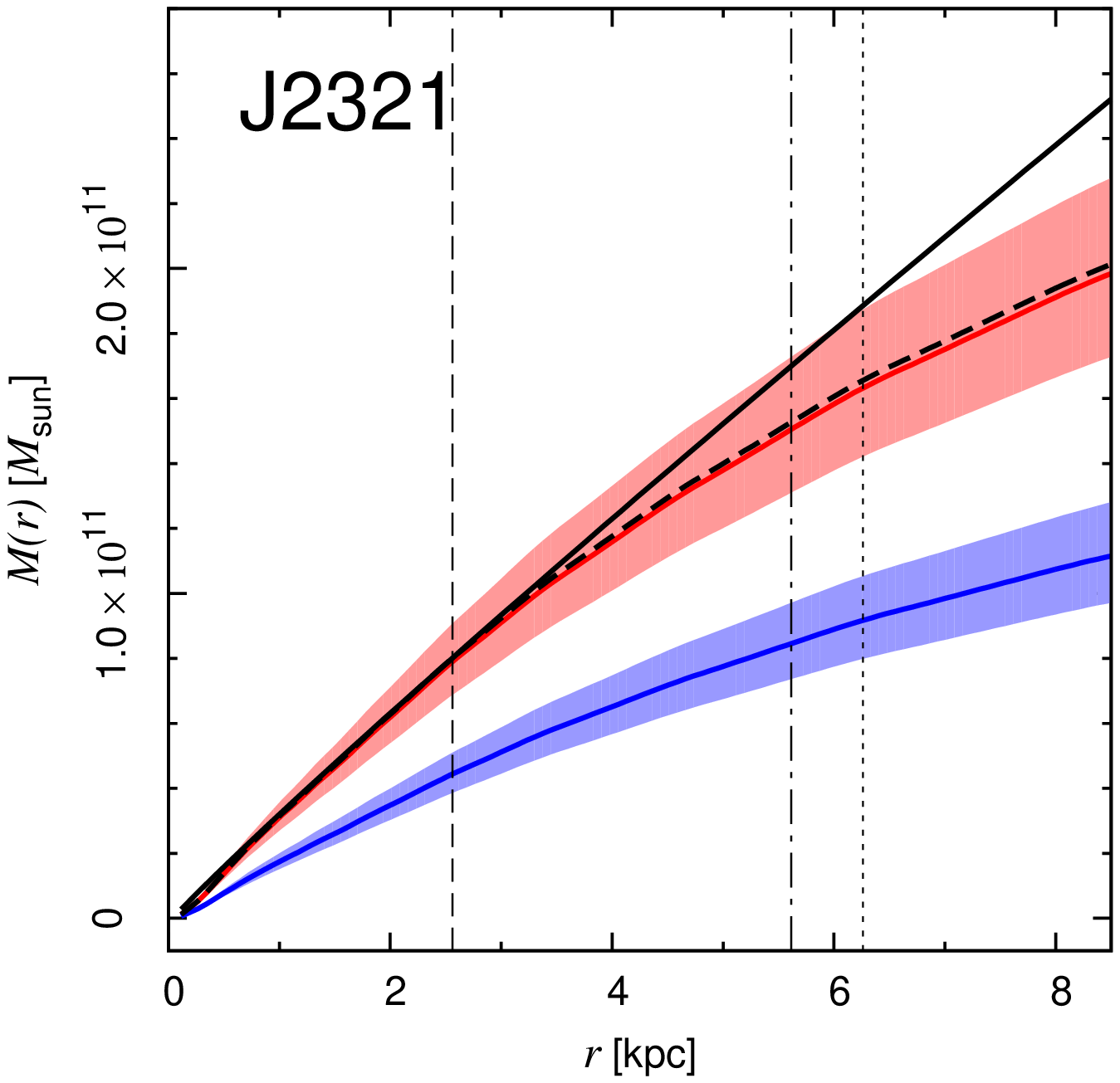}}
  \end{center}
  \vspace{-0.5cm}

  \begin{center}
    \subfigure{\label{fig:combo-1250}\includegraphics[angle=-90,width=0.24\textwidth]{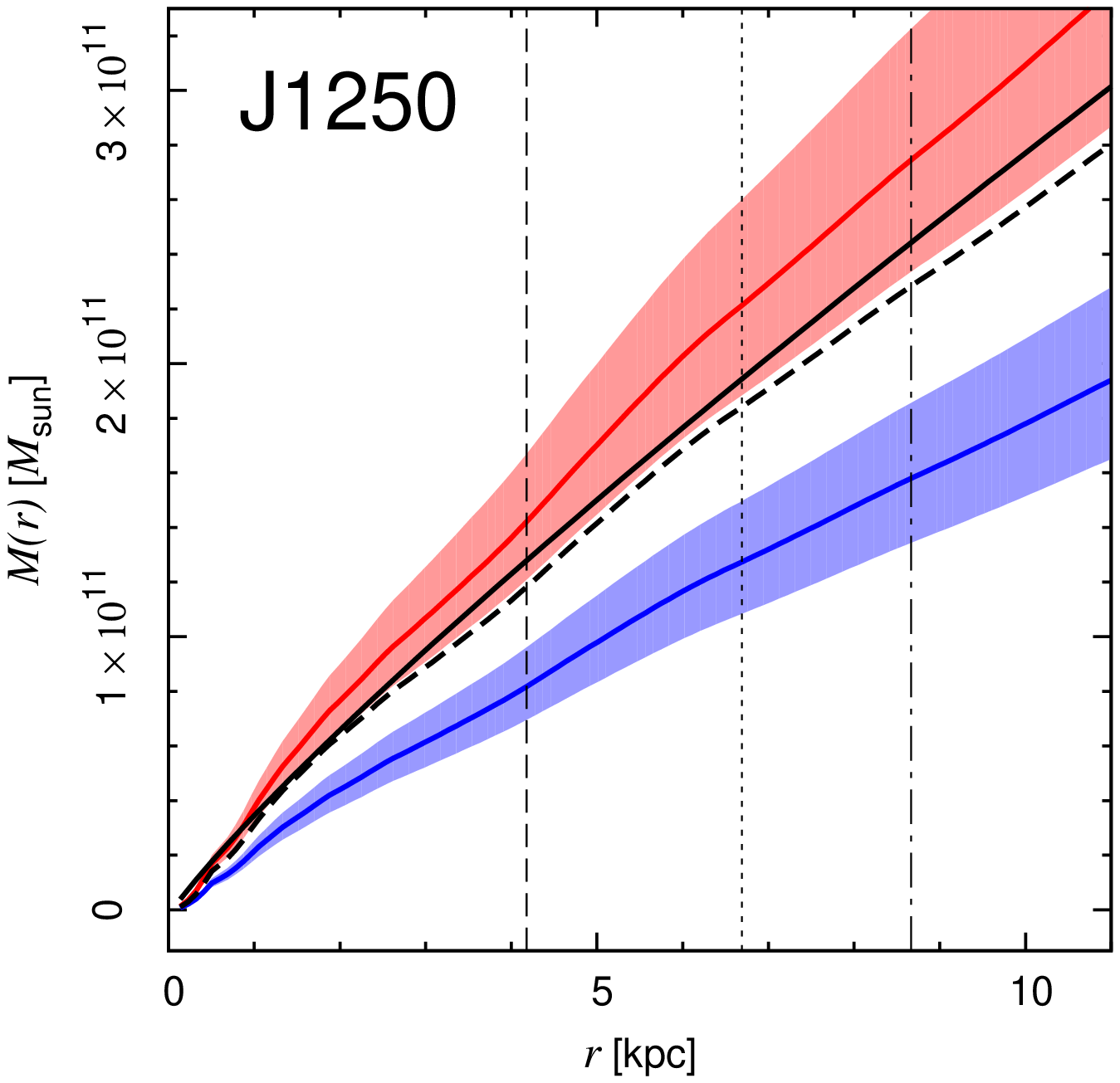}}
    \subfigure{\label{fig:combo-1627}\includegraphics[angle=-90,width=0.24\textwidth]{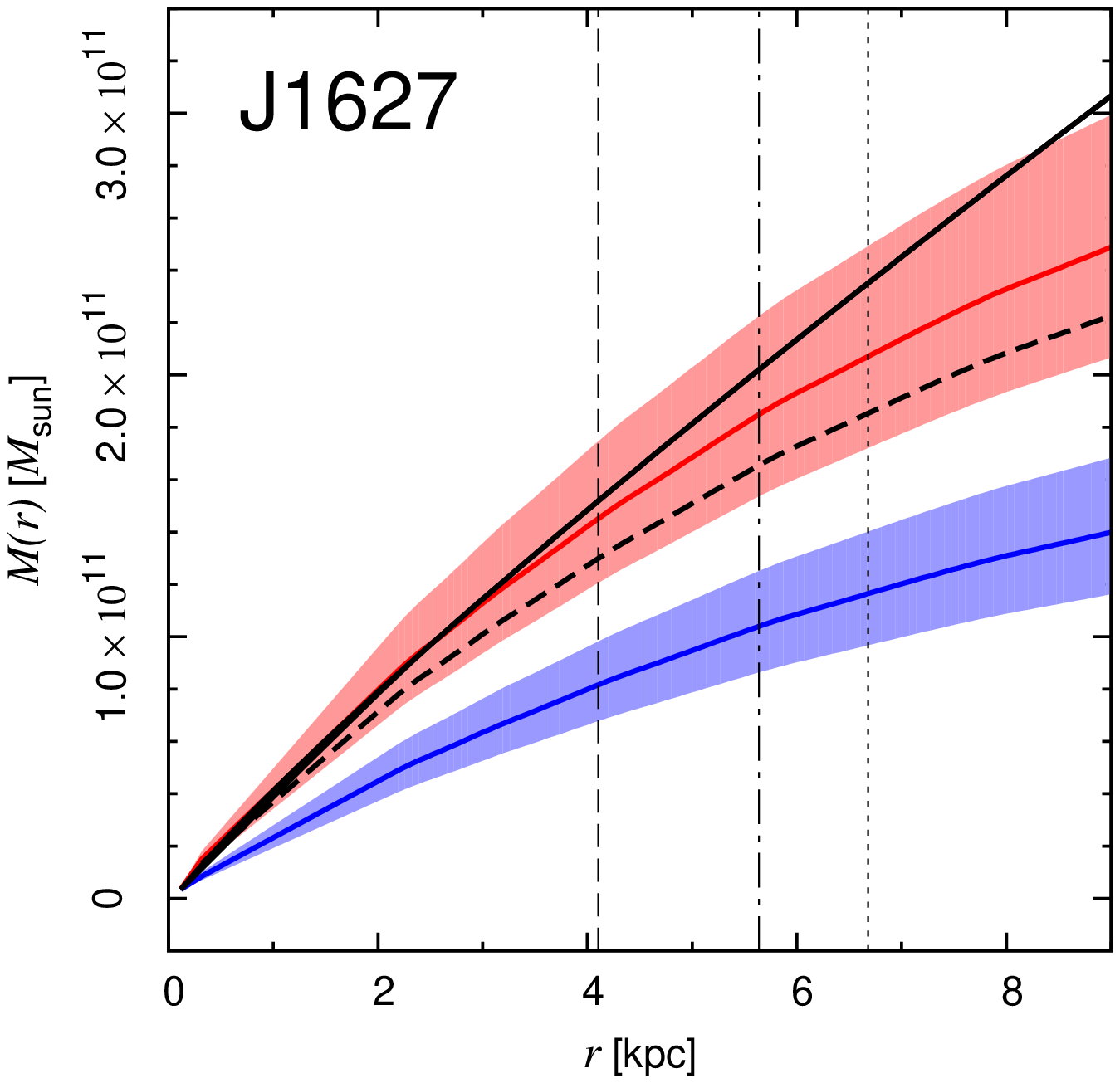}}
    \subfigure{\label{fig:combo-2300}\includegraphics[angle=-90,width=0.24\textwidth]{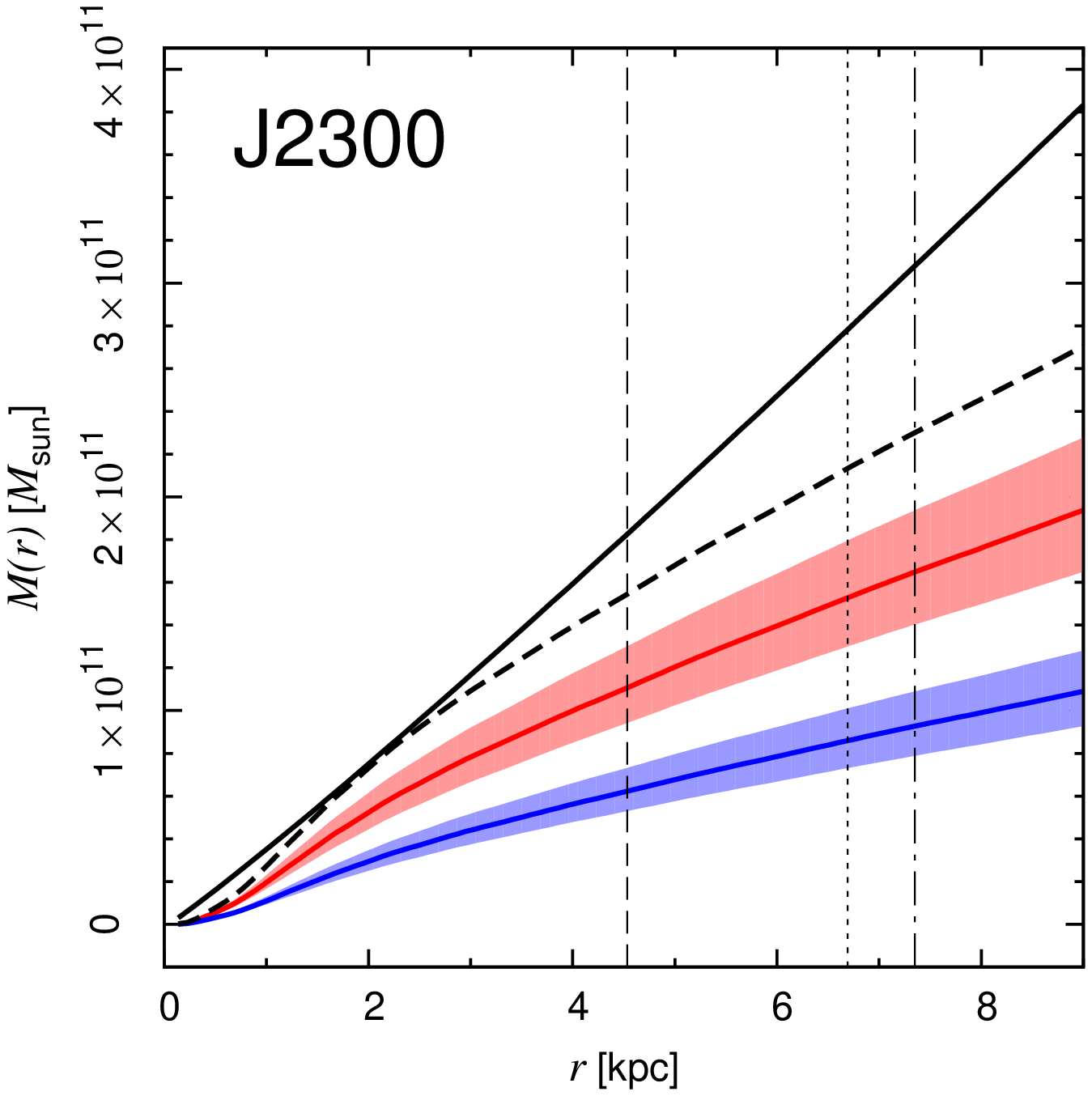}}
    \subfigure{\label{fig:combo-0037}\includegraphics[angle=-90,width=0.24\textwidth]{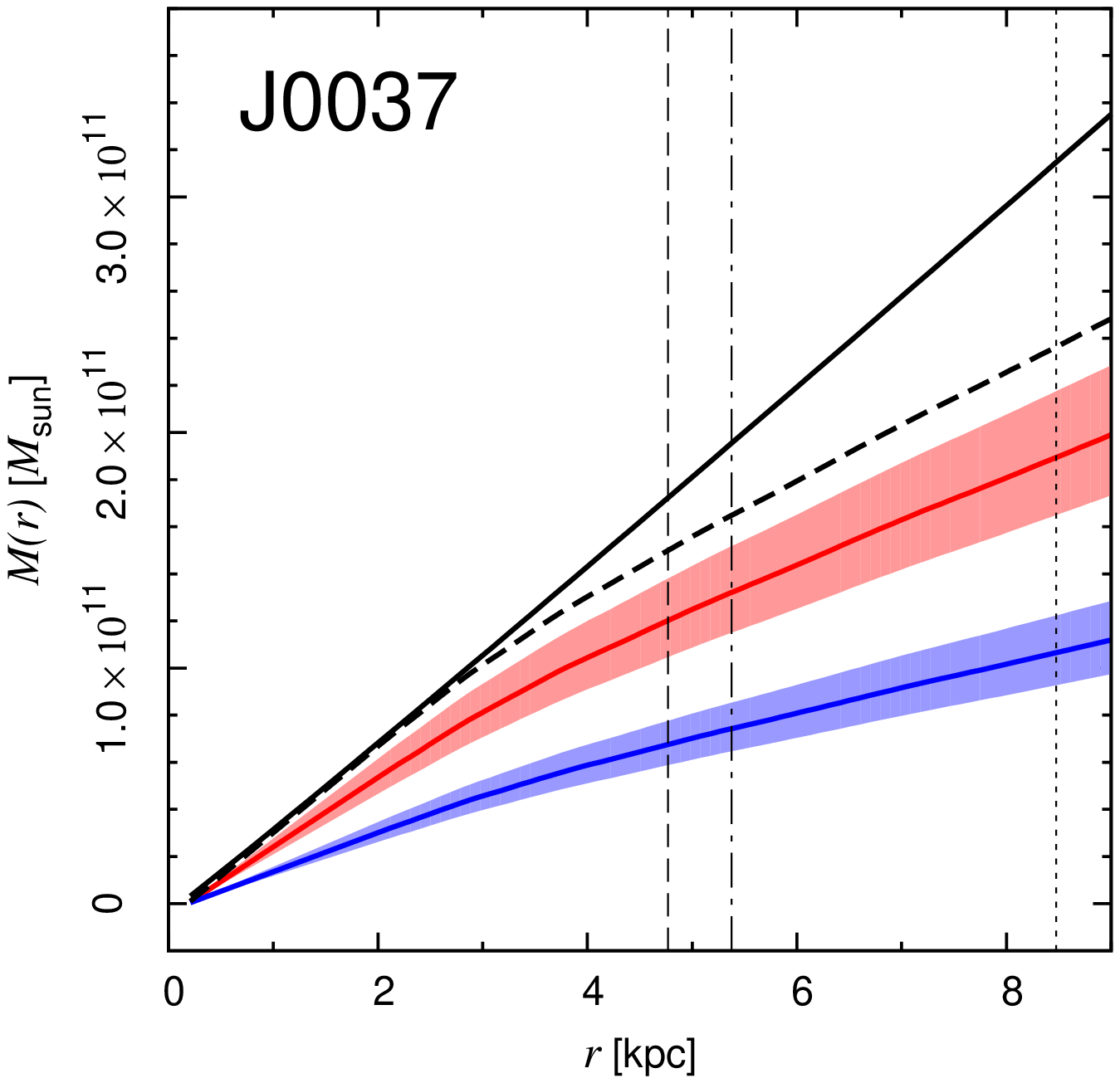}}
  \end{center}
  \vspace{-0.5cm}

  \begin{center}
    \subfigure{\label{fig:combo-2303}\includegraphics[angle=-90,width=0.24\textwidth]{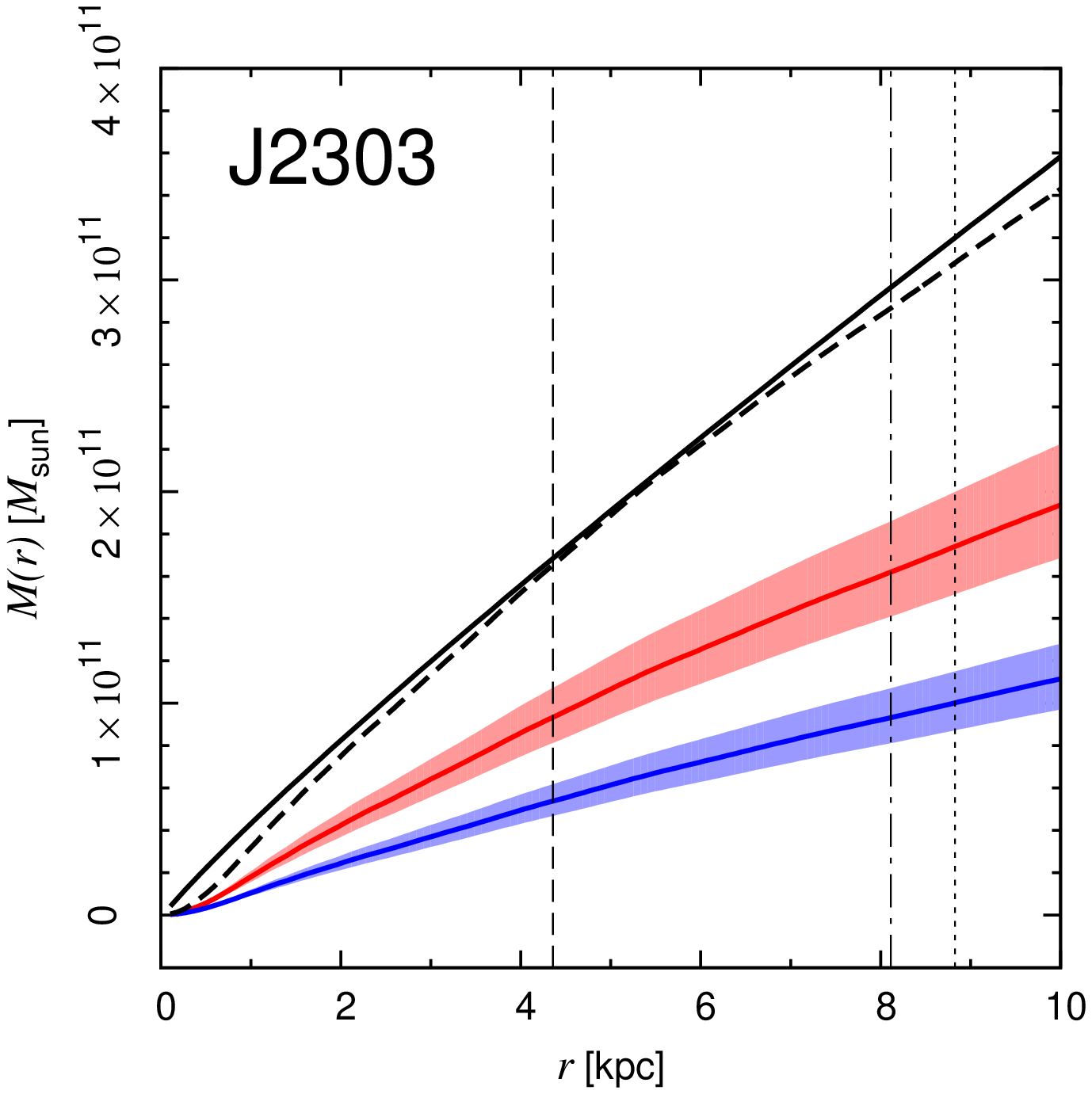}}
    \subfigure{\label{fig:combo-0912}\includegraphics[angle=-90,width=0.24\textwidth]{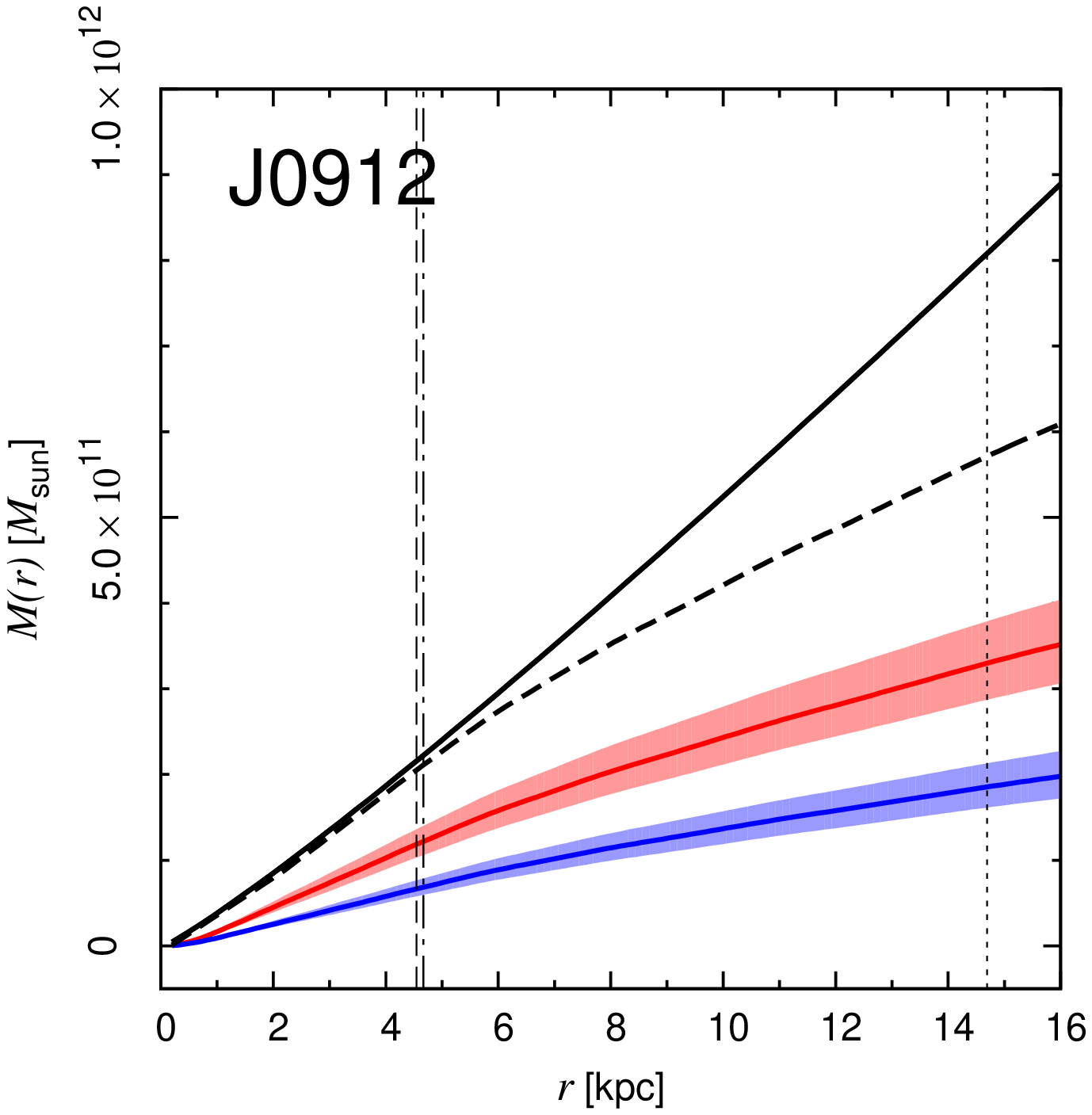}}
    \subfigure{\label{fig:combo-0216}\includegraphics[angle=-90,width=0.24\textwidth]{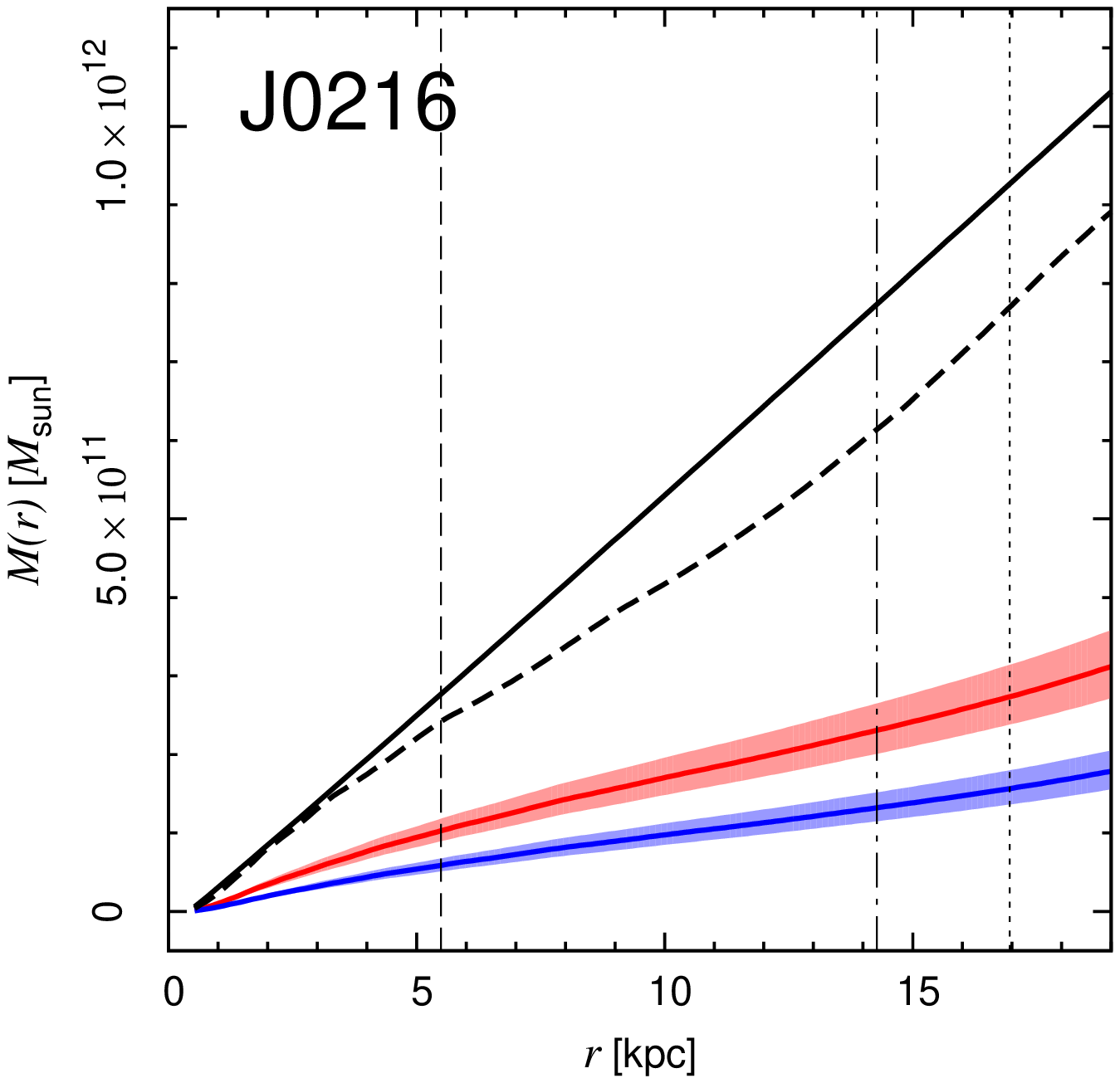}}
    \subfigure{\label{fig:combo-0935}\includegraphics[angle=-90,width=0.24\textwidth]{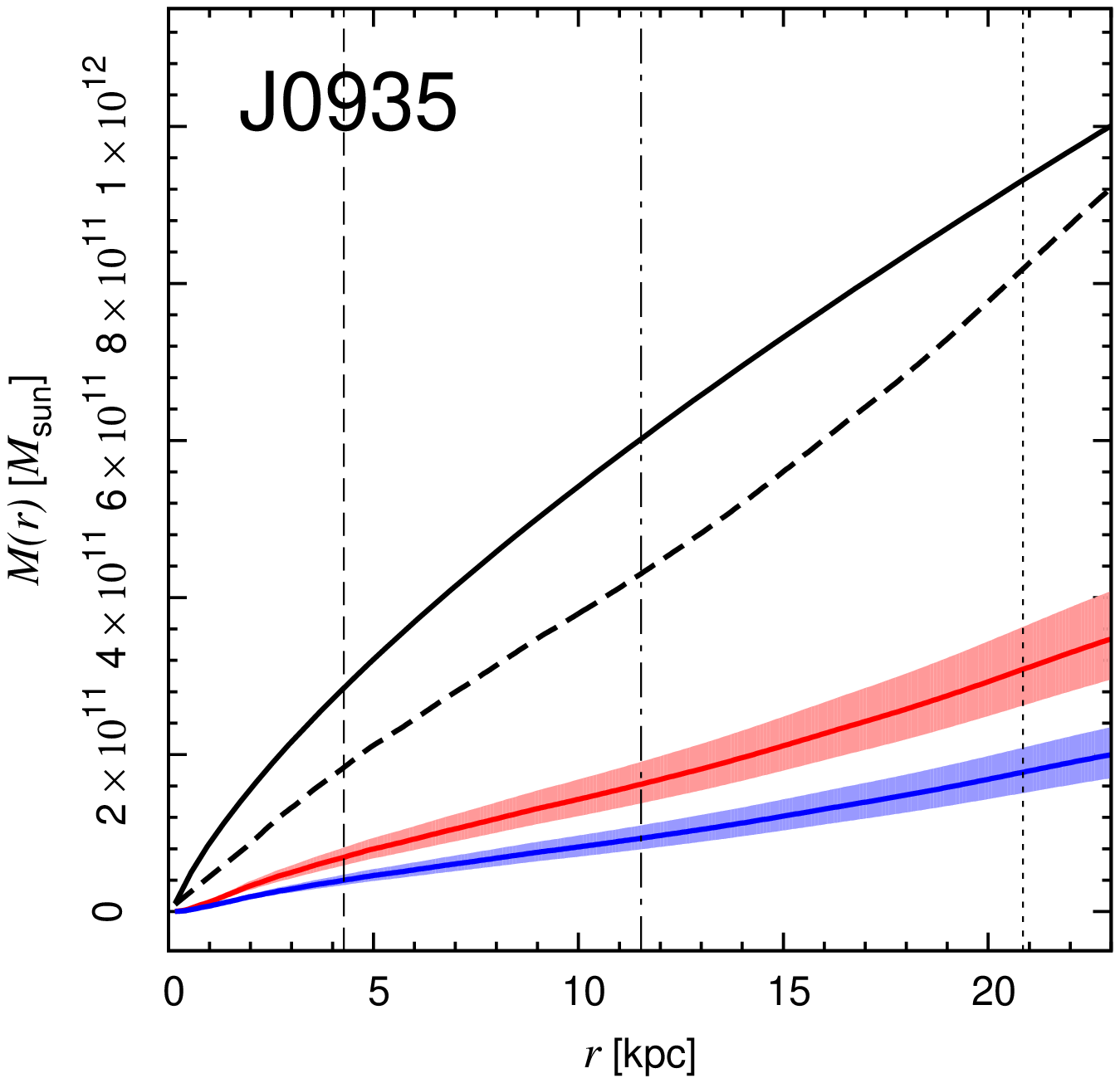}}
  \end{center}
  \vspace{-0.5cm}

  \caption{Compilation of spherically averaged mass distributions for
    the sample galaxies, sorted by increasing $\Mtot$. In each panel,
    the solid black line shows the total mass profile obtained from
    the best reconstructed model, and the dashed black line indicates
    the maximal luminous mass profile calculated under the maximum
    bulge (i.e. minimum halo) hypothesis (see
    Sect.~\ref{ssec:max-bulge}). The solid red line shows the luminous
    mass profile rescaled using the value for stellar mass determined
    from SPS analysis \citep{Auger2009}, adopting a Salpeter IMF. The
    red-shaded region represents the corresponding uncertainty (i.e.,
    the standard deviation of the marginalized posterior for the
    stellar mass). The blue curves and regions have the same meaning,
    but are obtained with a Chabrier IMF. The vertical lines indicate
    the location of the effective radius (dotted line), the Einstein
    radius (dashed line) and the outermost boundary of the kinematic
    data (dash-dotted line).  }

  \label{fig:Mprof.combo}
\end{figure*}

For each galaxy, the (spherically averaged) total mass profile
corresponding to the best reconstructed model is calculated from the
density distribution of Eq.~(\ref{eq:rho}), and shown as the solid black
curve in Figure~\ref{fig:Mprof.combo}. The vertical lines indicate,
for reference, the location of the (two-dimensional) effective
radius~$\Reff$, the Einstein radius~$\REin$ and the outermost radius
$\Rkin$ probed by the kinematic data set. For most systems the
kinematic maps extend up to and beyond the half-light radius,
providing strong constraints within this region, and in all cases
(with the exception of SDSS\,J0912) $\Rkin$ exceeds at least
$\Reff/2$.

While $\REin$ and $\Rkin$ provide useful yardsticks to compare the
data coverage in different systems, it should be noted that the
observational constraints to the mass models are not strictly limited
to these radial values since, as discussed in \citet{Czoske2008}, the
data sets also include the contribution of more distant galaxy regions
that are located along the line-of-sight.

The analyzed sample spans almost two orders of magnitude in mass: the
total mass $\Mtot$ enclosed within the three-dimensional radius $r =
\Reff$, hereafter denoted as $\re$ for brevity, ranges from a few
$10^{10} M_{\sun}$ for the two smallest systems SDSS\,J1330 and
SDSS\,J1443 to the almost $10^{12} M_{\sun}$ of SDSS\,J0216 and
SDSS\,J0935, with most of the galaxies falling in the range $1 - 3
\times 10^{11} M_{\sun}$ (see Table~\ref{tab:dyn}).

\begin{figure}
  \centering
  \resizebox{1.00\hsize}{!}{\includegraphics[angle=-90]
            {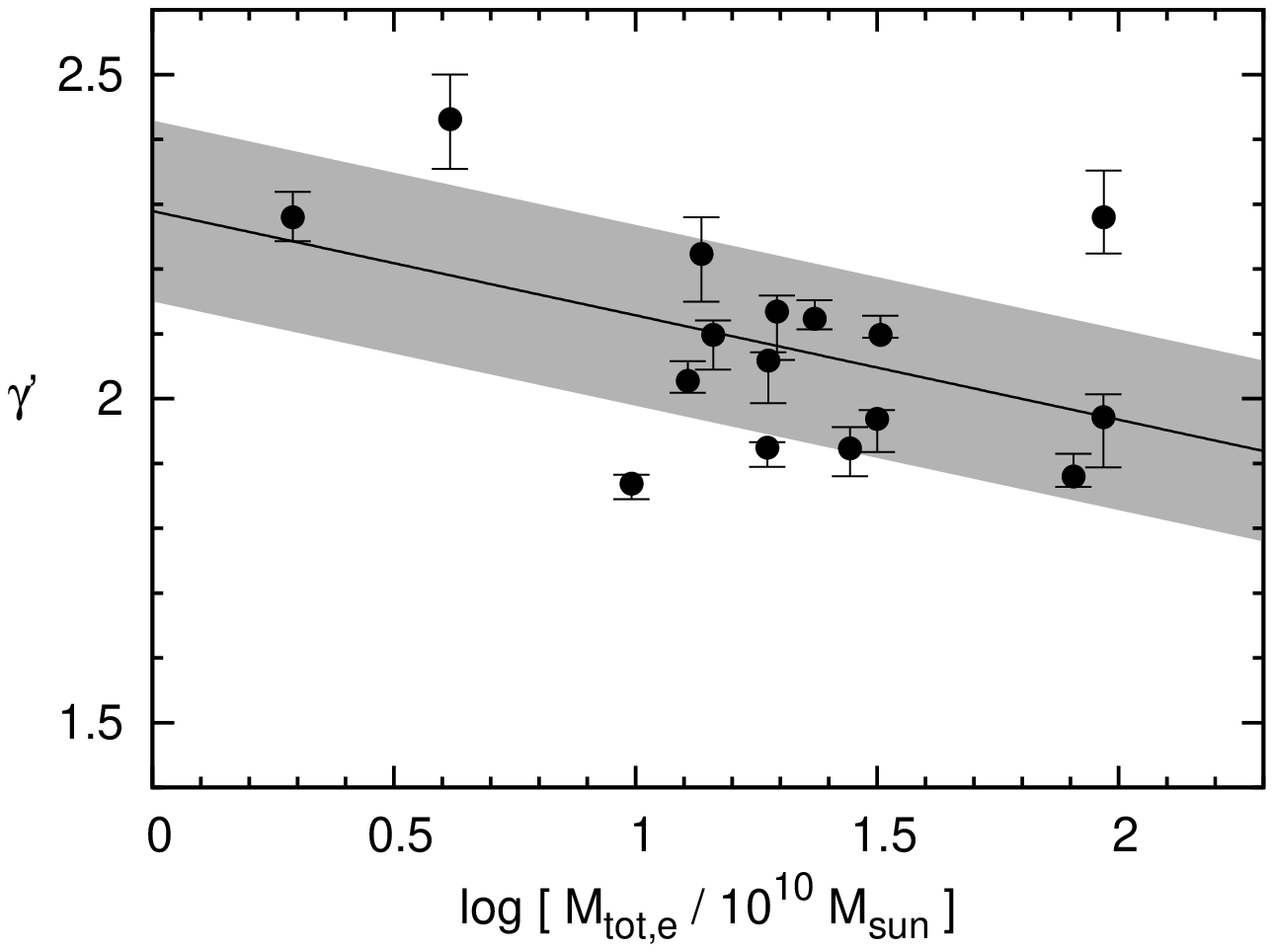}}

  \vspace{0.0cm}

  \resizebox{1.00\hsize}{!}{\includegraphics[angle=-90]
            {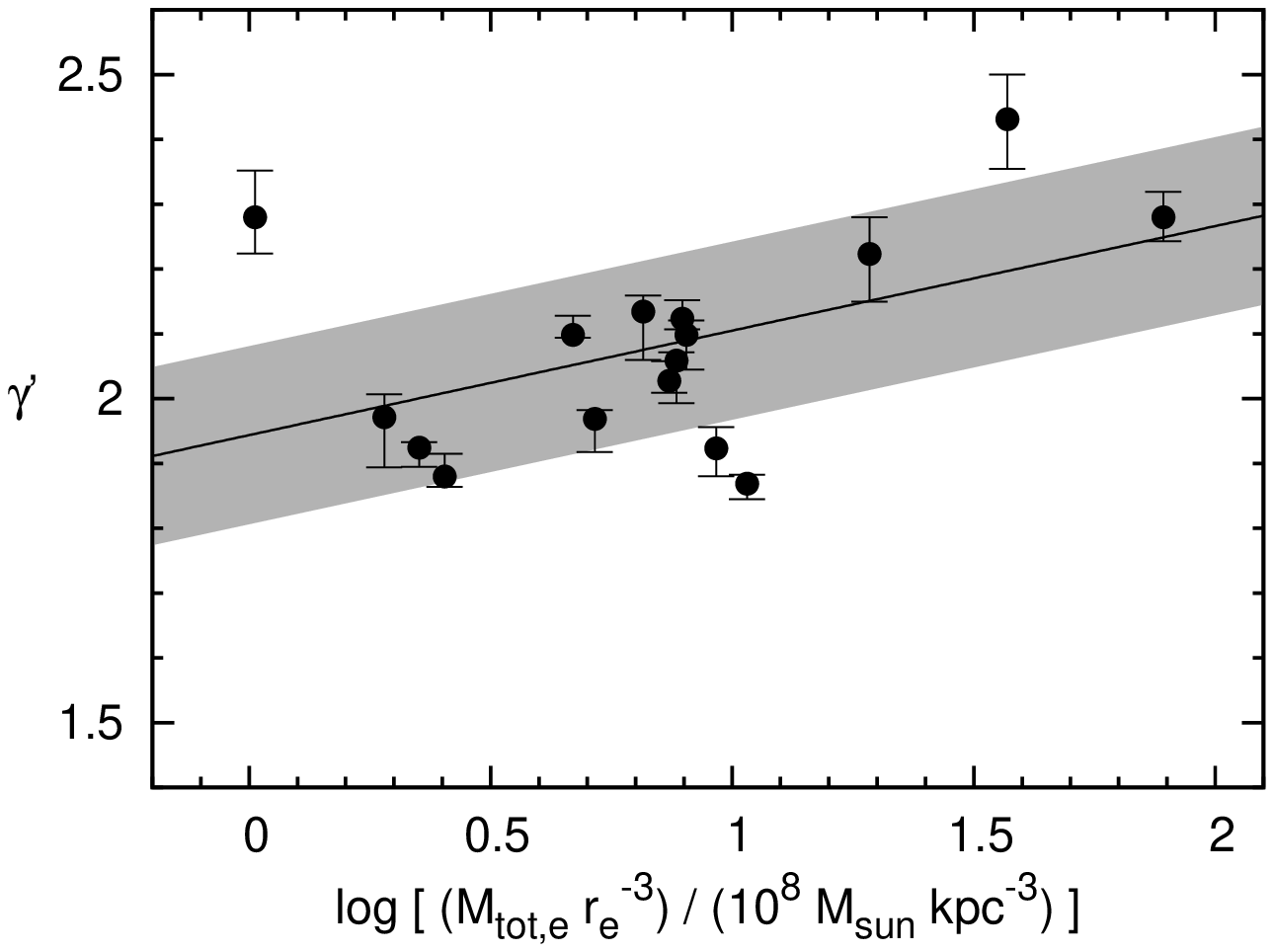}}

  \caption{Logarithmic slope of the total density profile $\slope$
    against the total mass enclosed within the three-dimensional
    radius $\re$ (upper panel) and the average total mass density
    inside $\re$ (lower panel). The solid black line is a linear fit
    to the relation (including scatter), and the gray band indicates
    the intrinsic scatter.}
  \label{fig:Mtot_vs_gamma}
\end{figure}

There is some evidence (non-zero slope with about $2$-sigma
significance) for a slight anti-correlation between the total mass
$\Mtot$ and $\slope$ (see Figure~\ref{fig:Mtot_vs_gamma}, upper panel,
and Table~\ref{tab:fits}), i.e. the total density slope tends to be
steeper for the least massive systems. The most notable outlier is
SDSS\,J0935, which, despite being the most massive galaxy in the
sample, has a $\slope$ comparable to that of the two systems with
$\Mtot < 10^{11} M_{\sun}$. We also find a correlation (with similar
significance) between $\slope$ and the quantity $\log
(\Mtot/\re^{3})$, which is a proxy for the average total mass density
inside the three-dimensional radius~$\re$
(Figure~\ref{fig:Mtot_vs_gamma}, lower panel). This trend is in
agreement with the findings of \citet{Auger2010}, who observe a
relatively tight correlation between~$\slope$ and the central surface
mass density of the full SLACS sample.

We note that the covariance between $\slope$ and $\Mtot$ is generally
very small. In fact, for a given Einstein mass and radius, the
dependence on the slope of the total mass $M_{\mathrm{tot}}(r)$
enclosed inside the three-dimensional radius $r$ is least in the
region where $r/\REin \simeq 2$. For most of the galaxies in our
sample, the ratio $\Reff/\REin$ is not far from that value, so that
the 1-sigma uncertainty on $\slope$ usually translates into a mass
change of less than~1 per cent (and even in the case of J0935, which
has $\Reff/\REin \simeq 5$, it only reaches about 4 per cent,
i.e. less than 0.02 in dex).

\begin{table*}
  \centering
  \caption{Recovered masses, dark matter fractions and dynamical
    quantities for the sixteen sample SLACS lens galaxies.}
  \smallskip
  \begin{tabular}{ l c c c c c c c c c c }
    \hline
    \noalign{\smallskip}
    Galaxy name & $\log \frac{\displaystyle \Mtot}{\displaystyle M_{\sun}}$ & $\Mstar/L_{\sun, B}$ & $f_{\mathrm{DM}}^{\mathrm{mb\phantom{p}}}$ & ${f}_{\mathrm{DM}}^{\mathrm{Chab}\phantom{p}}$ & $f^{\mathrm{Salp}}_{\mathrm{DM}}$ & $v/\sigma$ & $\jz$ & $\lamR$ & $\delta$ & $\gamma$ \\
    \noalign{\smallskip}
    \hline
    \noalign{\smallskip}
    SDSS\,J0037$-$0942  & 11.50 & 4.85 & 0.23 & $0.66^{+0.04}_{-0.05}$ & $\phantom{-}0.40^{+0.08}_{-0.09}$ & 0.08 & 0.25 & 0.05 & $\phantom{-}0.16$  & $-0.37$ \\ 
    \noalign{\smallskip}                                                                                   
    SDSS\,J0216$-$0813  & 11.97 & 9.80 & 0.17 & $0.83^{+0.03}_{-0.03}$ & $\phantom{-}0.70^{+0.04}_{-0.05}$ & 0.04 & 0.12 & 0.02 & $\phantom{-}0.08$  & $-0.17$ \\ 
    \noalign{\smallskip}                                                                                   
    SDSS\,J0912$+$0029  & 11.91 & 9.00 & 0.30 & $0.77^{+0.03}_{-0.04}$ & $\phantom{-}0.59^{+0.06}_{-0.07}$ & 0.12 & 0.23 & 0.13 & $\phantom{-}0.07$  & $-0.15$ \\ 
    \noalign{\smallskip}                                                                                   
    SDSS\,J0935$-$0003  & 11.97 & 7.17 & 0.12 & $0.81^{+0.03}_{-0.03}$ & $\phantom{-}0.67^{+0.05}_{-0.06}$ & 0.10 & 0.17 & 0.03 & $\phantom{-}0.24$  & $-0.64$ \\ 
    \noalign{\smallskip}                                                                                   
    SDSS\,J0959$+$0410  & 10.99 & 7.63 & 0.30 & $0.71^{+0.04}_{-0.05}$ & $\phantom{-}0.49^{+0.07}_{-0.08}$ & 0.49 & 0.65 & 0.42 & $-0.16$            & $\phantom{-}0.27$  \\ 
    \noalign{\smallskip}                                                                                   
    SDSS\,J1204$+$0358  & 11.14 & 7.55 & 0.02 & $0.57^{+0.06}_{-0.08}$ & $\phantom{-}0.23^{+0.10}_{-0.11}$ & 0.03 & 0.07 & 0.03 & $\phantom{-}0.09$  & $-0.19$ \\ 
    \noalign{\smallskip}                                                                                   
    SDSS\,J1250$+$0523  & 11.29 & 3.55 & 0.05 & $0.35^{+0.10}_{-0.11}$ & $-0.14^{+0.17}_{-0.20}$           & 0.15 & 0.16 & 0.09 & $\phantom{-}0.22$  & $-0.56$ \\
    \noalign{\smallskip}                                                                                   
    SDSS\,J1251$-$0208  & 11.27 & 3.56 & 0.46 & $0.67^{+0.05}_{-0.07}$ & $\phantom{-}0.43^{+0.10}_{-0.11}$ & 0.82 & 0.81 & 0.66 & $-0.36$            & $\phantom{-}0.53$  \\ 
    \noalign{\smallskip}                                                                                   
    SDSS\,J1330$-$0148  & 10.29 & 5.55 & 0.05 & $0.51^{+0.06}_{-0.07}$ & $\phantom{-}0.15^{+0.11}_{-0.12}$ & 0.52 & 0.49 & 0.48 & $-0.05$            & $\phantom{-}0.09$  \\ 
    \noalign{\smallskip}                                                                                   
    SDSS\,J1443$+$0304  & 10.62 & 4.04 & 0.01 & $0.32^{+0.09}_{-0.10}$ & $-0.20^{+0.16}_{-0.18}$           & 0.36 & 0.60 & 0.37 & $\phantom{-}0.13$  & $-0.30$ \\
    \noalign{\smallskip}                                                                                   
    SDSS\,J1451$-$0239  & 11.11 & 4.04 & 0.12 & $0.51^{+0.07}_{-0.09}$ & $\phantom{-}0.19^{+0.10}_{-0.12}$ & 0.10 & 0.31 & 0.05 & $\phantom{-}0.15$  & $-0.35$ \\ 
    \noalign{\smallskip}                                                                                   
    SDSS\,J1627$-$0053  & 11.37 & 4.85 & 0.21 & $0.50^{+0.09}_{-0.11}$ & $\phantom{-}0.12^{+0.16}_{-0.20}$ & 0.07 & 0.18 & 0.06 & $\phantom{-}0.16$  & $-0.38$ \\ 
    \noalign{\smallskip}                                                                                   
    SDSS\,J2238$-$0754  & 11.16 & 5.91 & 0.08 & $0.62^{+0.05}_{-0.06}$ & $\phantom{-}0.33^{+0.09}_{-0.10}$ & 0.43 & 0.66 & 0.42 & $-0.18$            & $\phantom{-}0.30$  \\ 
    \noalign{\smallskip}                                                                                   
    SDSS\,J2300$+$0022  & 11.44 & 7.22 & 0.23 & $0.69^{+0.05}_{-0.05}$ & $\phantom{-}0.45^{+0.08}_{-0.10}$ & 0.03 & 0.15 & 0.02 & $\phantom{-}0.08$  & $-0.17$ \\ 
    \noalign{\smallskip}                                                                                   
    SDSS\,J2303$+$1422  & 11.51 & 7.48 & 0.04 & $0.69^{+0.04}_{-0.05}$ & $\phantom{-}0.46^{+0.07}_{-0.08}$ & 0.05 & 0.19 & 0.04 & $\phantom{-}0.14$  & $-0.31$ \\ 
    \noalign{\smallskip}                                                                                   
    SDSS\,J2321$-$0939  & 11.27 & 4.81 & 0.12 & $0.51^{+0.08}_{-0.10}$ & $\phantom{-}0.13^{+0.15}_{-0.18}$ & 0.06 & 0.08 & 0.06 & $\phantom{-}0.15$  & $-0.36$ \\ 
    \noalign{\smallskip}
    \hline
    \noalign{\smallskip}
  \end{tabular}

  \begin{minipage}{1.00\hsize}
    \textit{Note.} For each galaxy we list: the logarithm of the total
    mass $\Mtot$ enclosed within the three-dimensional radius $\re$;
    the stellar mass-to-light ratio (in the $B$-band) corresponding to
    the maximal luminous profile; the fraction of dark over total mass
    within $\re$ for the maximum bulge assumption
    ($f_{\mathrm{DM}}^{\mathrm{mb\phantom{p}}}$) and for the Chabrier
    and Salpeter IMFs (${f}_{\mathrm{DM}}^{\mathrm{Chab}\phantom{p}}$
    and $f^{\mathrm{Salp}}_{\mathrm{DM}}$, respectively); the
    inclination-corrected $v/\sigma$ ratio; the dimensionless angular
    momentum~$\jz$; the Emsellem kinematic parameter~$\lamR$; the
    global anisotropy parameters $\delta$ and $\gamma$.
  \end{minipage}
  \label{tab:dyn}
\end{table*}

\subsection{Luminous mass and dark matter content}

The radial profile for the luminous component is obtained from the
best reconstructed stellar DF. However, since the adopted modelling
approach treats the stellar distribution as a tracer embedded in the
total potential, the normalization of the luminous profile remains
unconstrained, and further assumptions (\S~\ref{ssec:max-bulge}) or
additional information (\S~\ref{ssec:stellar-pop}) are therefore
necessary to set its scale and determine the contribution of the dark
matter component at different radii.

\subsubsection{Dark matter fraction: constraints from the maximum bulge 
approach}
\label{ssec:max-bulge}

When additional information on the normalization is unavailable, the
most straightforward approach consists in maximally rescaling the
luminous profile under the physical constraint that it does not exceed
the total mass distribution at any point, and the assumption that the
stellar mass-to-light ratio is fairly position-independent.  This
procedure is known as the minimum halo or maximum bulge hypothesis,
and is the analogue for early-type galaxies of the classical maximum
disk hypothesis \citep{vanAlbada-Sancisi1986} routinely adopted in the
study of late-type galaxies. It is frequently employed (see
e.g.\ \citealt{Barnabe2009} and \citealt{Weijmans2009} for recent
applications) since it provides a consistent and robust method to
determine a conservative lower limit for the dark matter fraction in
the galaxy, within the considered region.

The (spherically averaged) luminous mass profiles obtained under the
maximum bulge hypothesis are presented as dashed black lines in
Figure~\ref{fig:Mprof.combo}. Let $\Mmb$ denote the corresponding
stellar mass enclosed inside the spherical radius $\re$. We see that
the lower limit for the (three-dimensional) fraction of dark over
total mass within that radius, defined as $\fDM \equiv 1 -
\Mmb/\Mtot$, varies very significantly from system to system
(cf.\ Table~\ref{tab:dyn}): three of the galaxies, SDSS\,J1443,
SDSS\,J1204 and SDSS\,J2303, show a dark matter lower limit $\fDM < 5$
per cent within the probed region, whereas\,---\,on the opposite
end\,---\,several galaxies require at least one fifth of the total
mass enclosed within $\re$ to be dark, with a peak of 46 per cent for
SDSS\,J1251. The average and median of the dark matter fraction lower
limits for the sample, obtained under the maximum bulge approach, are,
respectively, $\fDM = 16$ and $12$ per cent.

The stellar mass-to-light ratios (in the $B$-band) corresponding to
the maximal luminous profiles are in the range $3 < \Mstar/L_{\sun, B}
< 10$, fully consistent with the values determined for local
early-type galaxies (see e.g.\ \citealt{Kronawitter2000}
\citealt{Gerhard2001}, \citealt*{Trujillo2004},
\citealt{Mamon-Lokas2005}). However, when taking into account passive
stellar evolution, galaxies at higher redshift are expected to have
smaller $\Mstar/L_{\sun, B}$ than their local counterparts (an effect
that becomes significant already at $z \gtrsim 0.1$, see
\citealt{Treu2002}).

\subsubsection{Dark matter fraction: constraints from stellar mass estimates}
\label{ssec:stellar-pop}

While calculating a lower limit for the dark matter content is useful,
galaxies need not have maximal bulges. We therefore set the scale for
the luminous mass profiles by adopting, for each system, the stellar
mass inferred from stellar population synthesis models, assuming
either a \citet{Chabrier2003} or a \citet{Salpeter1955} IMF. We use
stellar mass values taken from the stellar population analysis
performed by \citet{Auger2009} on a data set constituted by deep,
high-resolution, multi-band \emph{HST} observations of the SLACS
lenses.

\begin{figure}
  \centering
  \resizebox{0.99\hsize}{!}{\includegraphics[angle=-90]
            {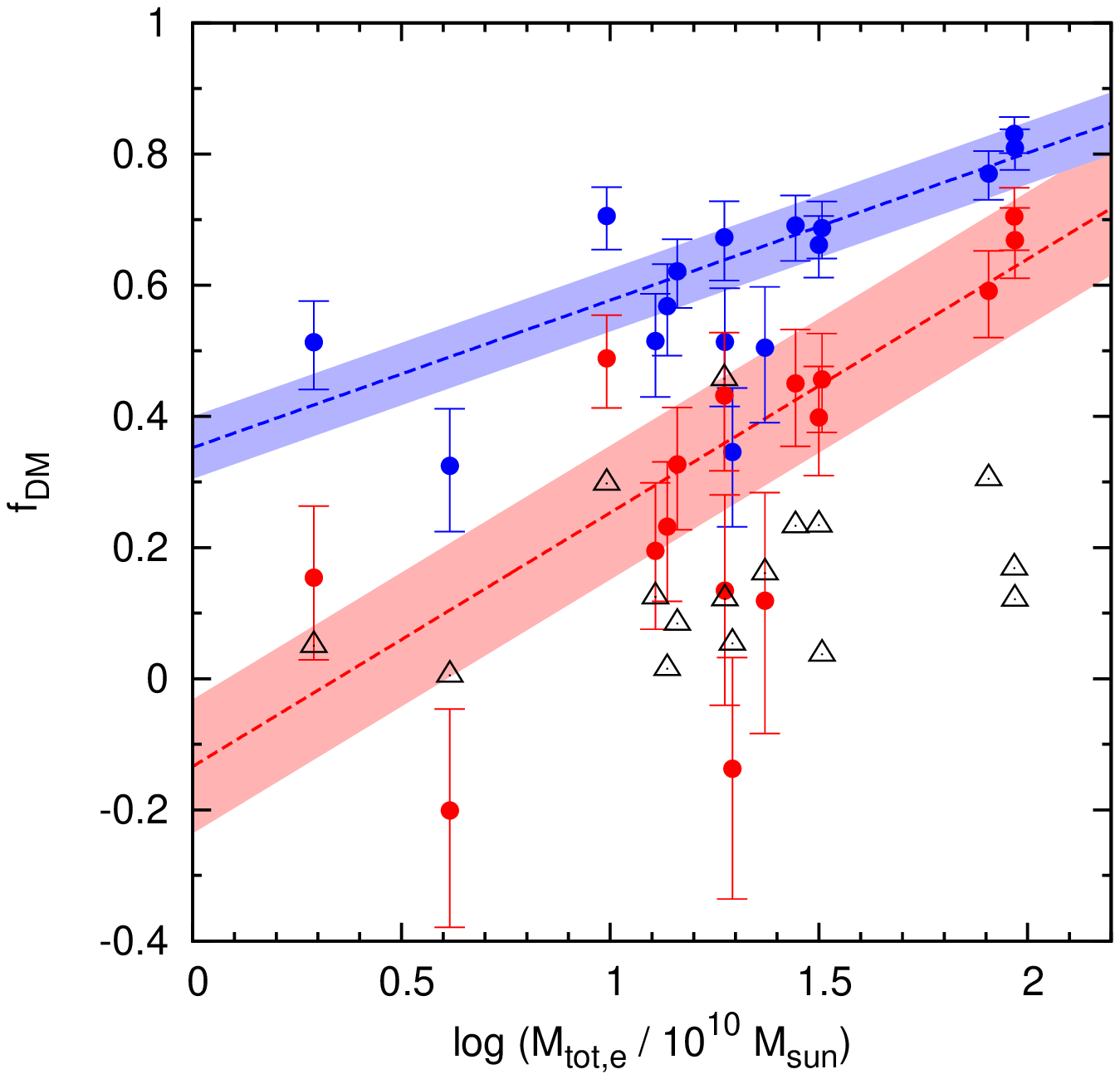}}
  \caption{Dark matter fractions versus the total mass enclosed within
    the three-dimensional radius $\re$. The upwards black triangles
    indicate the dark matter fraction lower limits calculated from the
    maximum bulge approach. The full circles show the dark matter
    fractions obtained when luminous masses are determined via SPS
    analysis, assuming Salpeter (red) or Chabrier (blue) IMFs. The
    dashed lines are linear fits to the relation, calculated including
    intrinsic scatter (which is indicated by the coloured bands).}
  \label{fig:fDM:IMF}
\end{figure}

The spherically averaged luminous mass profiles obtained in this way
are presented in Figure~\ref{fig:Mprof.combo}.
Figure~\ref{fig:fDM:IMF} shows the corresponding dark matter
fractions within $\re$, following the definition for $\fDM$ given
above in \S~\ref{ssec:max-bulge} and replacing $\Mmb$ with the
stellar masses $\Mchab$ and $\Msalp$ for Chabrier and Salpeter IMFs,
respectively. 

We find that, independently of the specific choice of the IMF, there
is a significant correlation (greater than 3-sigma) between the dark
matter fraction within $\re$ and the total mass (the parameters of the
fit are given in Table~\ref{tab:fits}). Such a correlation was first
observed, in the SLACS sample, by \citet{Auger2010}. Their result is
confirmed and strengthened by the present study, where we conduct a
more rigorous and detailed joint analysis on a sub-sample of SLACS
galaxies, taking advantage of the extended kinematic information
(rather than an estimate of the stellar velocity dispersion
$\sigma_{\mathrm{SDSS}}$ within a single 3~arcsec diameter aperture),
and self-consistently including axial symmetry and stellar anisotropy
in the model, and we reach analogous conclusions.

The normalization based on the stellar masses inferred from Salpeter
IMF produces a luminous profile that, when compared to Chabrier, is
always closer to the one determined under the maximum bulge
assumption, and in several cases is fully consistent with
it.\footnote{The SLACS lens galaxy SDSS\,J0728 (analyzed in
  \citealt{Barnabe2010}), which has kinematic maps obtained from Keck
  spectroscopy, represents, however, an exception to this trend, in
  the sense that for that galaxy the luminous mass profile obtained
  with a Chabrier IMF almost coincides with the maximum bulge
  determined one. We note that this systems lies close to the low end
  of the probed range in velocity dispersion.} For two systems,
SDSS\,J1250 and in particular SDSS\,J1443, this luminous profile
(including the error band) locally exceeds the maximum bulge curve,
indicating that it unphysically overshoots the total mass in places,
thus producing the two negative $\fDM$ points in
Figure~\ref{fig:fDM:IMF}. Taken at face value, this would lead to the
conclusion that the Salpeter relation constitutes an inadequate IMF
for these two galaxies (and therefore must be rejected for all systems
if we assume the IMF to be universal), whereas the Chabrier
IMF\,---\,which gives lower stellar masses by almost a factor
of~2\,---\,never fails to produce physical results. However, it must
be noted that in both cases the lower limit of the error band is only
slightly higher ($\lesssim 5$ per cent) than the maximal luminous
curve, and the adopted uncertainties represent 1-$\sigma$ errors (see
\citealt{Auger2009} for the details of how these uncertainties are
calculated within a Bayesian framework). Moreover, several systematic
effects (such as the circularization of the profile and the
discreteness of the TIC building blocks used to construct the model
galaxy) might interfere at the few per cent level with the minute
details of the reconstructed luminous profile. Therefore, we
conservatively conclude that all the analyzed systems are consistent
with having stellar masses determined from a Salpeter-like
IMF.\footnote{The situation, i.e. a recovered luminous profile
  slightly overshooting the physical limit of the maximum bulge, and
  the conclusions are analogous for the Keck system SDSS\,J0728
  mentioned above.}

With a Salpeter IMF, the recovered dark matter fraction for systems
having total masses $\log (\Mtot/M_{\sun}) \lesssim 11.5$ spans from~0
to~$\sim 50$ per cent, whereas the three most massive galaxies are
already dominated by dark matter ($\fDM$ is approximately~60 to~70 per
cent) inside 1~$\re$. The average and median dark matter fractions
are, respectively, $f^{\mathrm{Salp}}_{\mathrm{DM}} = 31$ and $37$ per
cent. The importance of the dark component becomes even more extreme
if a Chabrier IMF is considered (average
$f^{\mathrm{Chab}}_{\mathrm{DM}} = 61$ per cent, median
$f^{\mathrm{Chab}}_{\mathrm{DM}} = 64$ per cent), with the luminous
component contributing less than~50 per cent to $\Mtot$ in all systems
but two.

It has been suggested that the IMF is not universal but becomes
systematically `heavier' for more massive galaxies (see
\citealt{Treu2009}, \citealt{Auger2010imf},
\citealt{vanDokkum-Conroy2010}, \citealt{Thomas2011}; but see also
\citealt*{Napolitano2010} for a different conclusion). The data
presented here are consistent with this fact and cannot resolve the
degeneracy between varying IMF and varying dark matter fraction, as we
discuss in Section~\ref{sec:discussion}.

\begin{table}
  \centering
  \caption{Summary of linear relations derived for the SLACS lenses.}
  \smallskip
  \begin{tabular}{ l c c c c }
    \hline
    \noalign{\smallskip}
    \multicolumn{1}{c}{X} & Y & Slope & Intercept & Scatter\\
    \noalign{\smallskip}
    \hline
    \noalign{\smallskip}
    $\log(\Mtot)$ & $\slope$ 
    & $-0.16 \pm 0.09$ & $\phantom{-}2.29 \pm 0.13$ & $0.14^{+0.06}_{-0.01}$ \\
    \noalign{\smallskip}
    $\log(\Mtot/\re^{3})$ & $\slope$ 
    & $\phantom{-}0.16 \pm 0.09$ & $\phantom{-}1.94 \pm 0.08$ & $0.14^{+0.06}_{-0.01}$ \\
    \noalign{\smallskip}
    $\log(\Mtot)$ & ${f}_{\mathrm{DM}}^{\mathrm{Chab}}$ 
    & $\phantom{-}0.22 \pm 0.04$ & $\phantom{-}0.35 \pm 0.06$ & $0.05^{+0.04}_{-0.03}$ \\
    \noalign{\smallskip}
    $\log(\Mtot)$ & ${f}_{\mathrm{DM}}^{\mathrm{Salp}}$ 
    & $\phantom{-}0.39 \pm 0.08$ & $-0.13 \pm 0.12$ & $0.10^{+0.06}_{-0.04}$ \\
    \noalign{\smallskip}
    \hline
    \noalign{\smallskip}
  \end{tabular}

  \begin{minipage}{1.00\hsize}
    \textit{Note.} The mass $\Mtot$ is in units of $10^{10} M_{\sun}$,
    while the density $\Mtot/\re^{3}$ is in units of $10^{8} M_{\sun}$
    kpc$^{-3}$.
  \end{minipage}
  \label{tab:fits}
\end{table}


\begin{figure*}
  \begin{center}
    \subfigure[J1204]{\label{fig:DF-1204}\includegraphics[angle=-90,width=0.24\textwidth]{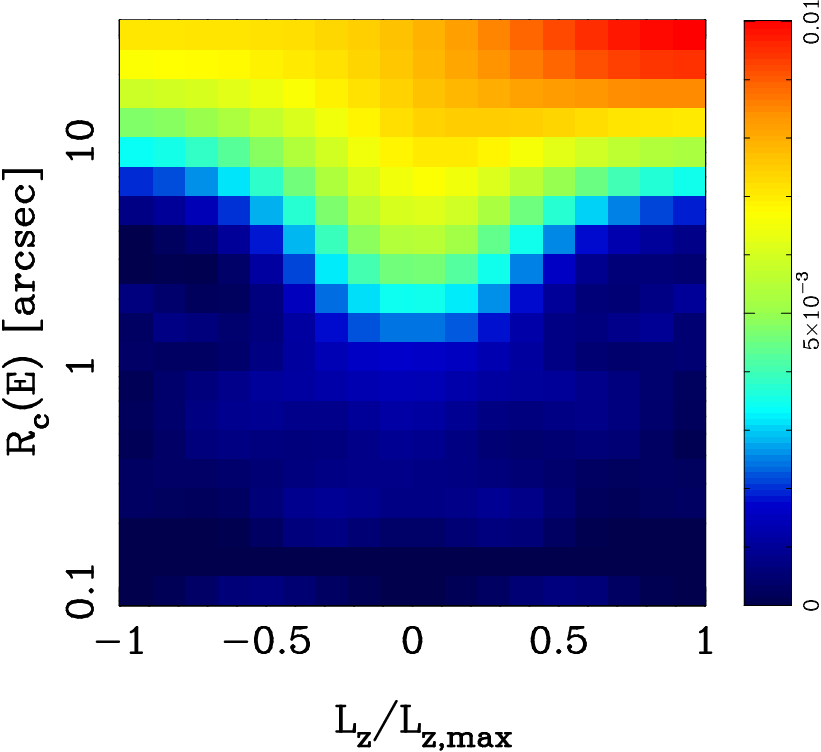}}
    \subfigure[J2321]{\label{fig:DF-2321}\includegraphics[angle=-90,width=0.24\textwidth]{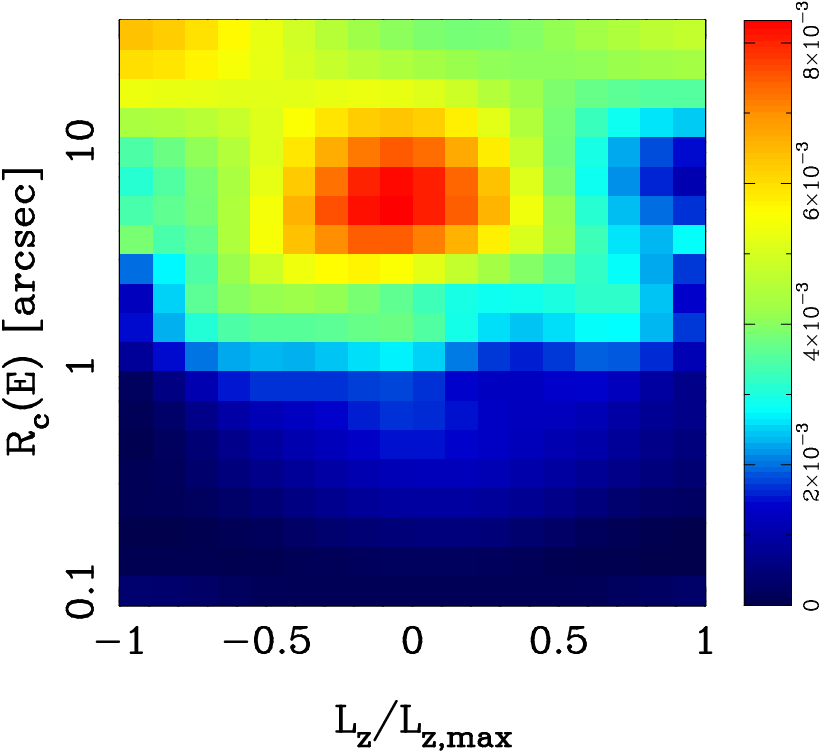}}
    \subfigure[J0216]{\label{fig:DF-0216}\includegraphics[angle=-90,width=0.24\textwidth]{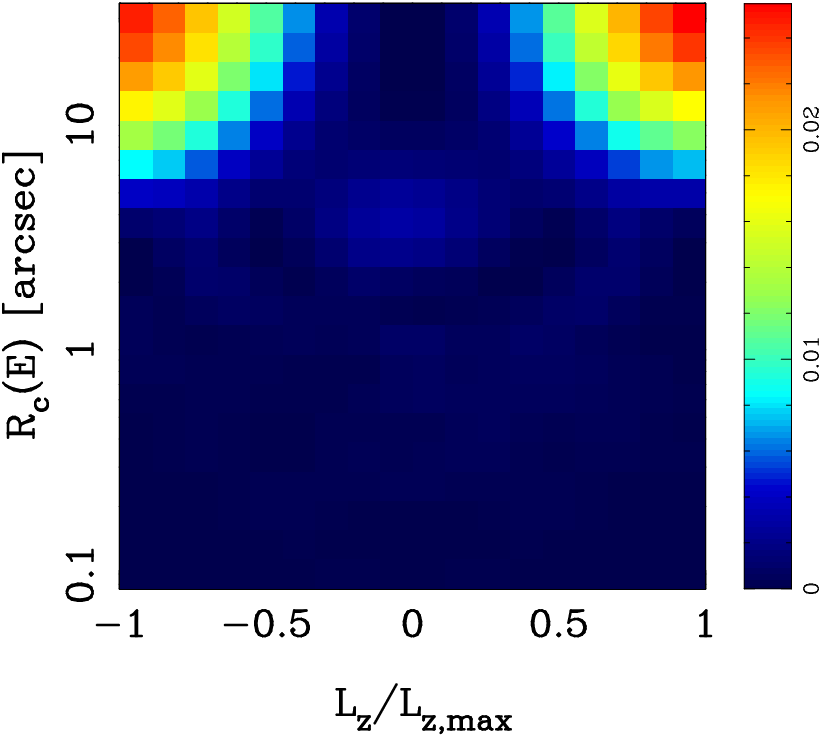}}
    \subfigure[J1250]{\label{fig:DF-1250}\includegraphics[angle=-90,width=0.24\textwidth]{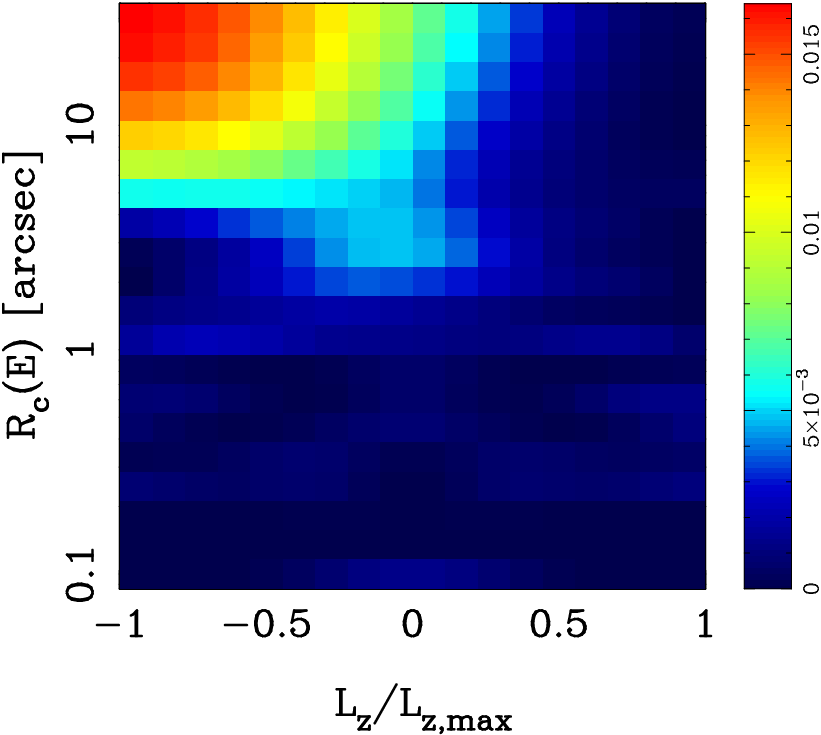}}
  \end{center}
  \vspace{-0.1cm}

  \begin{center}
    \subfigure[J2300]{\label{fig:DF-2300}\includegraphics[angle=-90,width=0.24\textwidth]{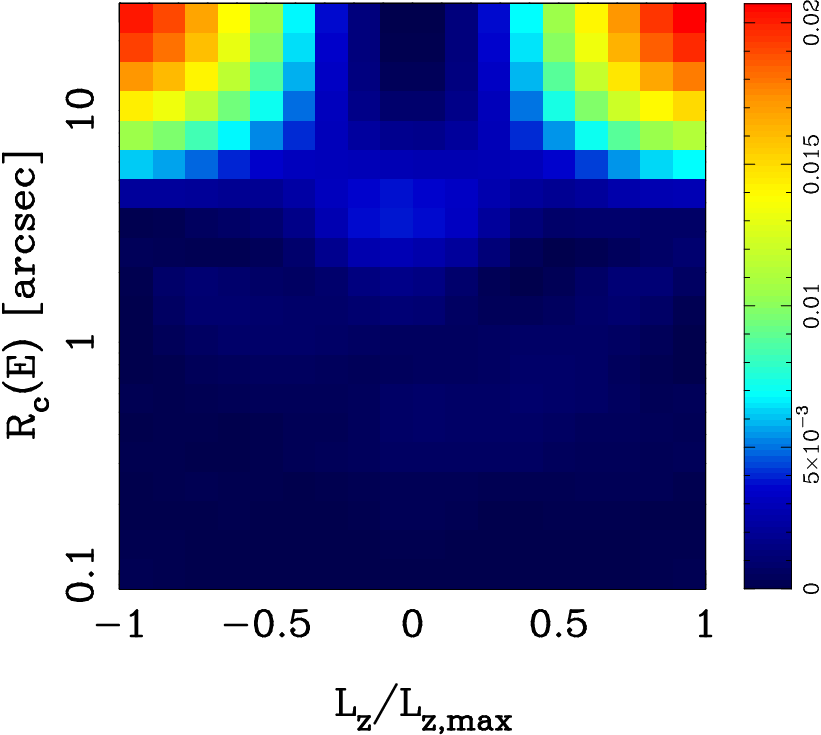}}
    \subfigure[J0935]{\label{fig:DF-0935}\includegraphics[angle=-90,width=0.24\textwidth]{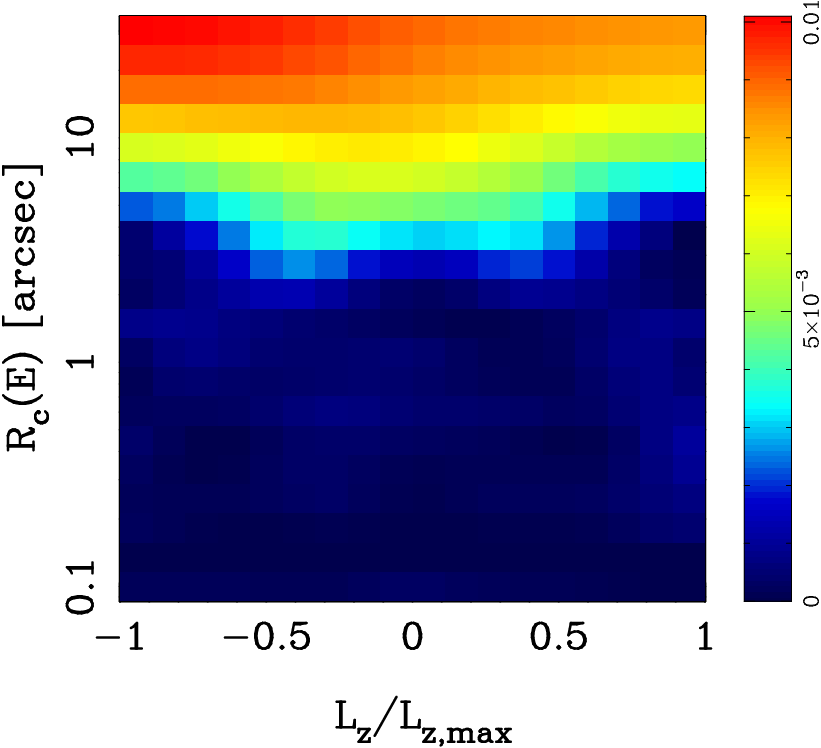}}
    \subfigure[J1627]{\label{fig:DF-1627}\includegraphics[angle=-90,width=0.24\textwidth]{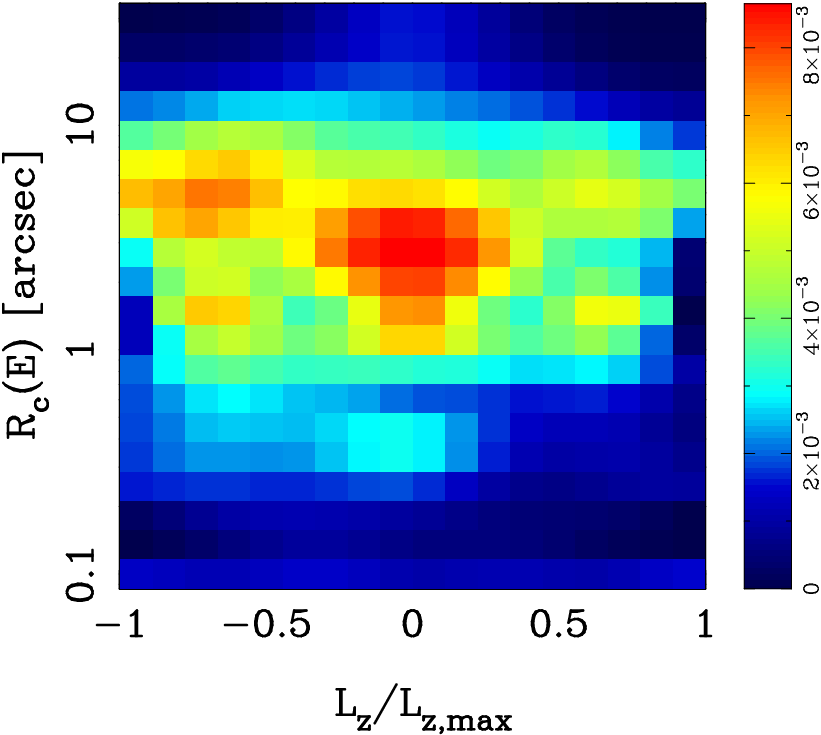}}
    \subfigure[J2303]{\label{fig:DF-2303}\includegraphics[angle=-90,width=0.24\textwidth]{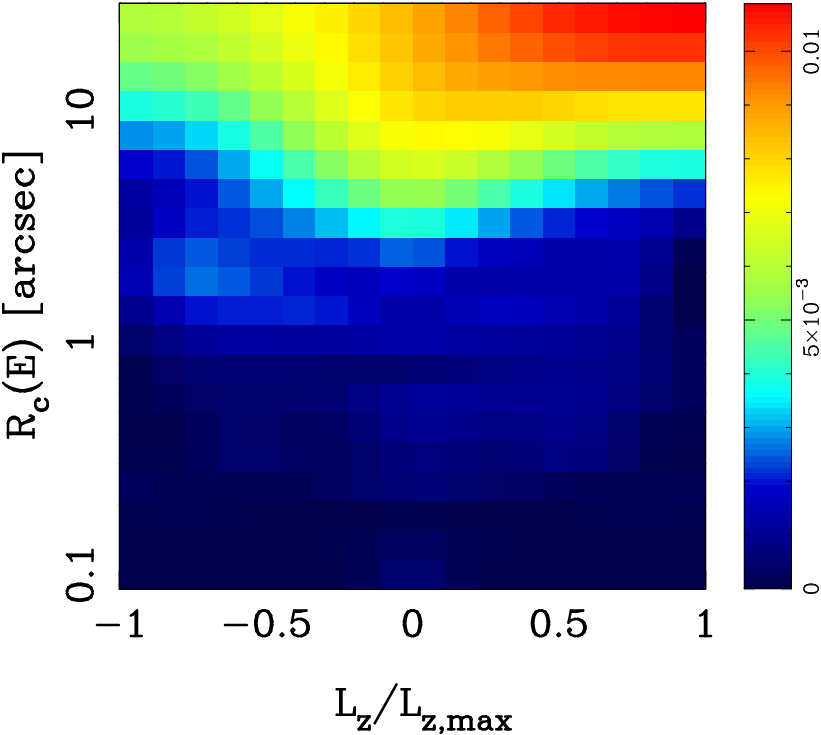}}
  \end{center}
  \vspace{-0.1cm}

  \begin{center}
    \subfigure[J0912]{\label{fig:DF-0912}\includegraphics[angle=-90,width=0.24\textwidth]{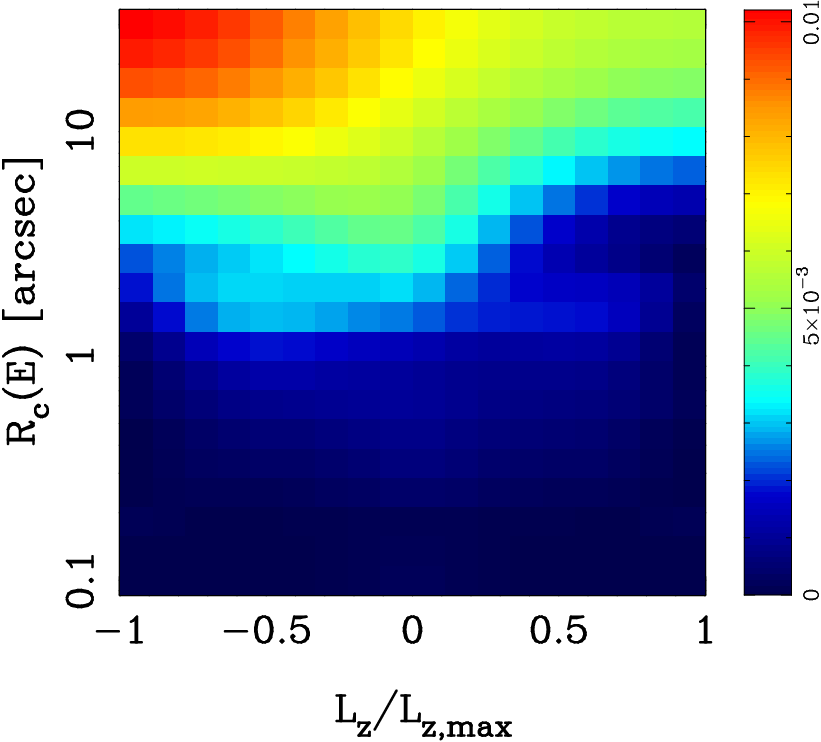}}
    \subfigure[J0037]{\label{fig:DF-0037}\includegraphics[angle=-90,width=0.24\textwidth]{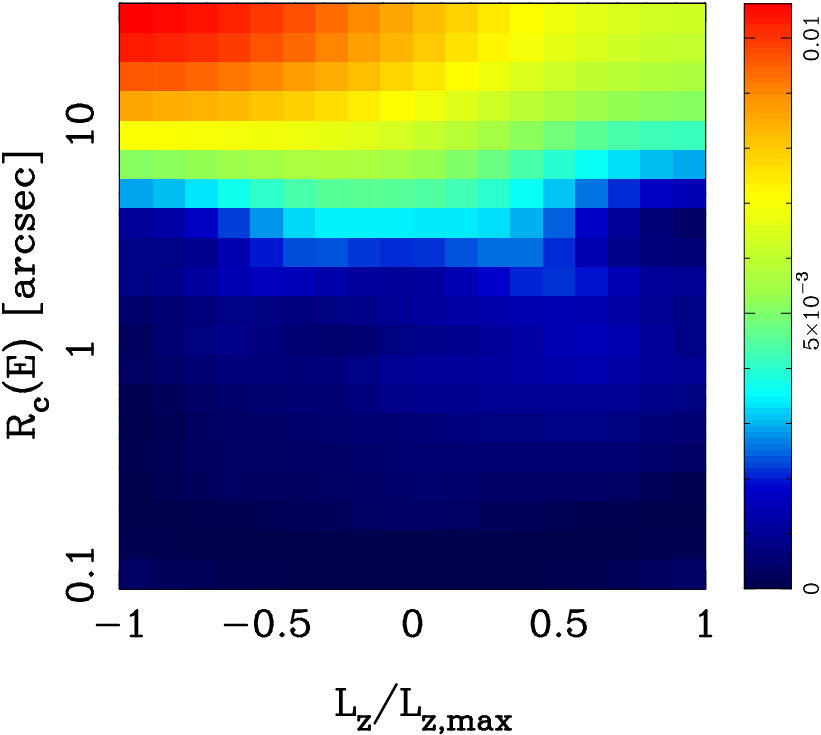}}
    \subfigure[J1451]{\label{fig:DF-1451}\includegraphics[angle=-90,width=0.24\textwidth]{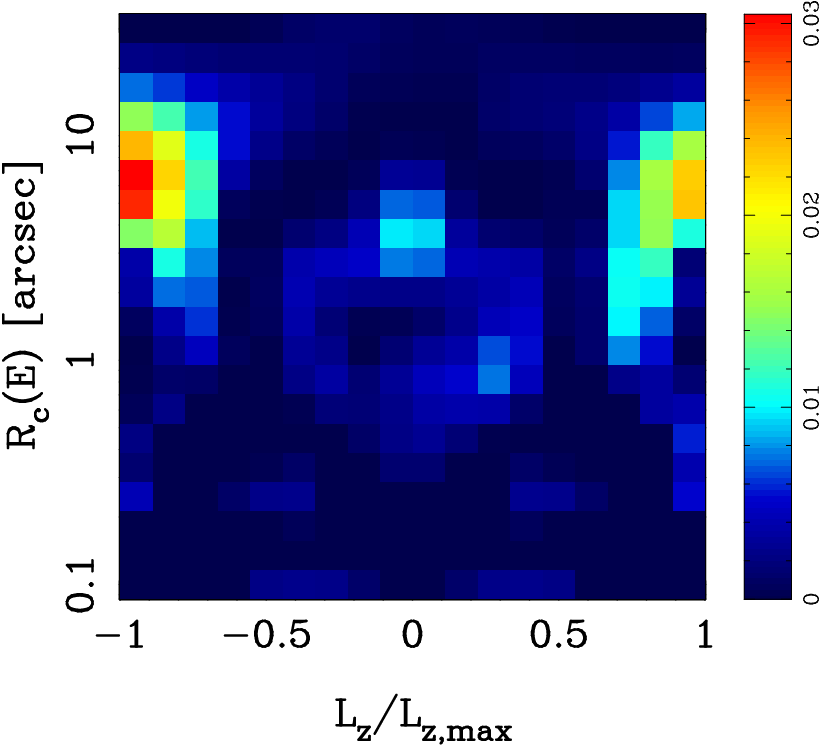}}
    \subfigure[J1330]{\label{fig:DF-1330}\includegraphics[angle=-90,width=0.24\textwidth]{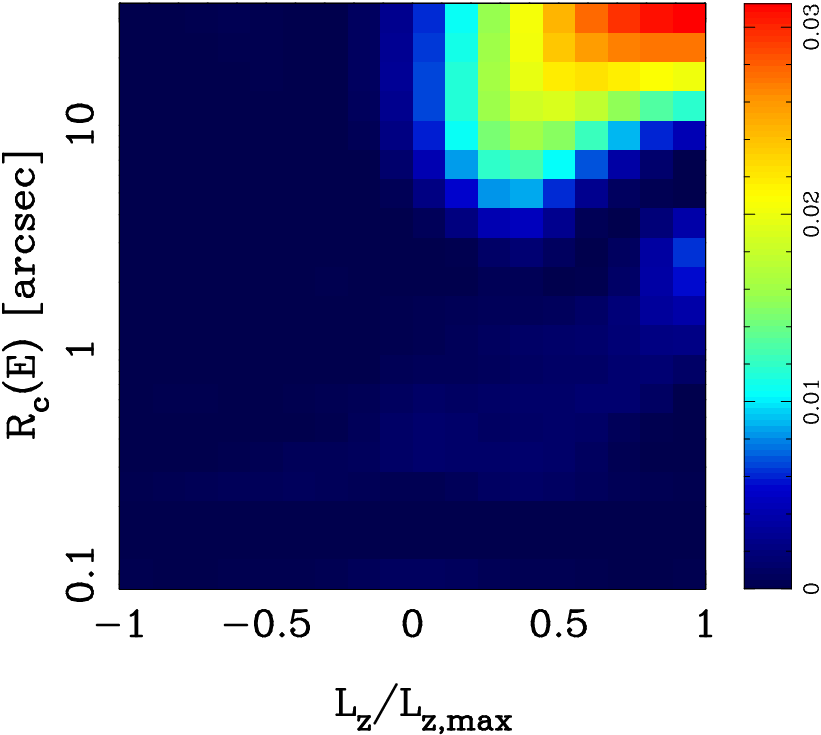}}
  \end{center}
  \vspace{-0.1cm}

  \begin{center}
    \subfigure[J1443]{\label{fig:DF-1443}\includegraphics[angle=-90,width=0.24\textwidth]{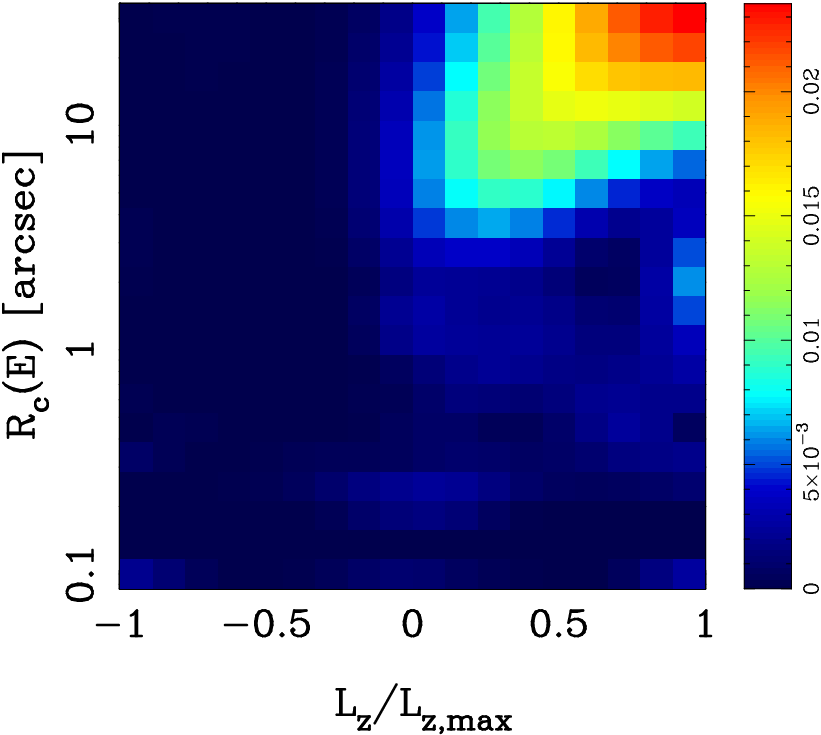}}
    \subfigure[J0959]{\label{fig:DF-0959}\includegraphics[angle=-90,width=0.24\textwidth]{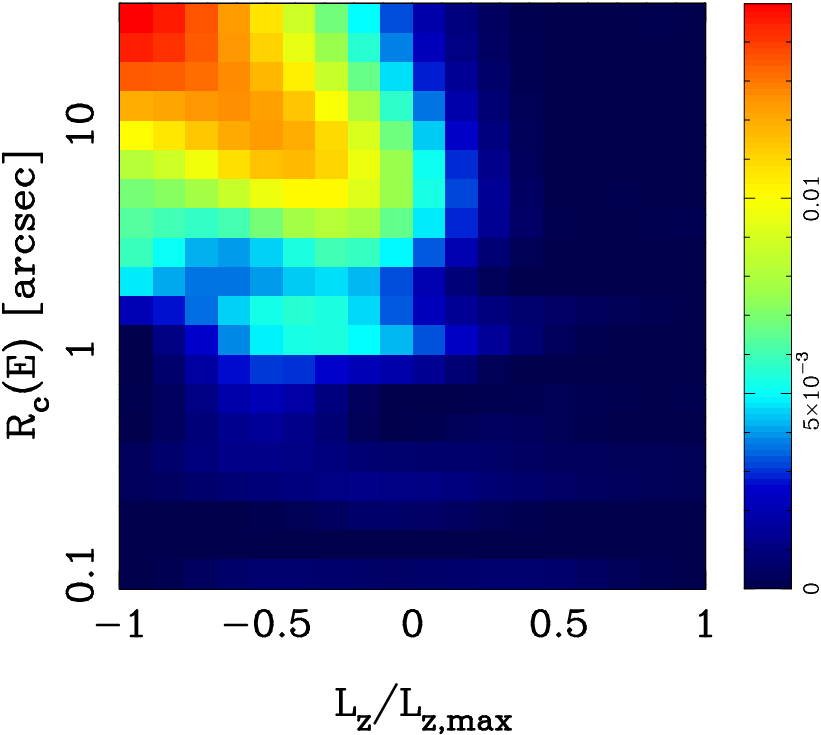}}
    \subfigure[J2238]{\label{fig:DF-2238}\includegraphics[angle=-90,width=0.24\textwidth]{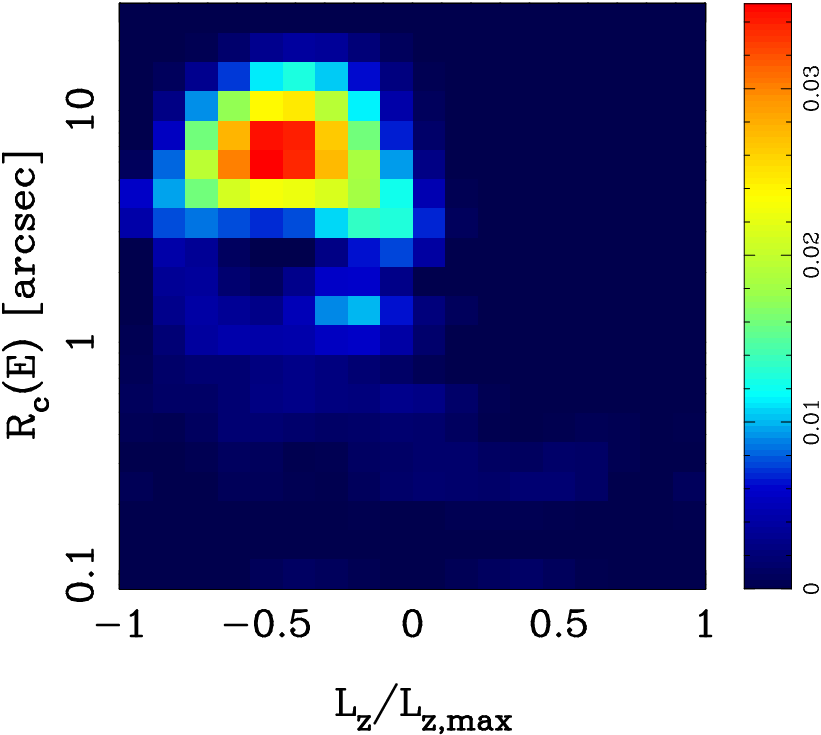}}
    \subfigure[J1251]{\label{fig:DF-1251}\includegraphics[angle=-90,width=0.24\textwidth]{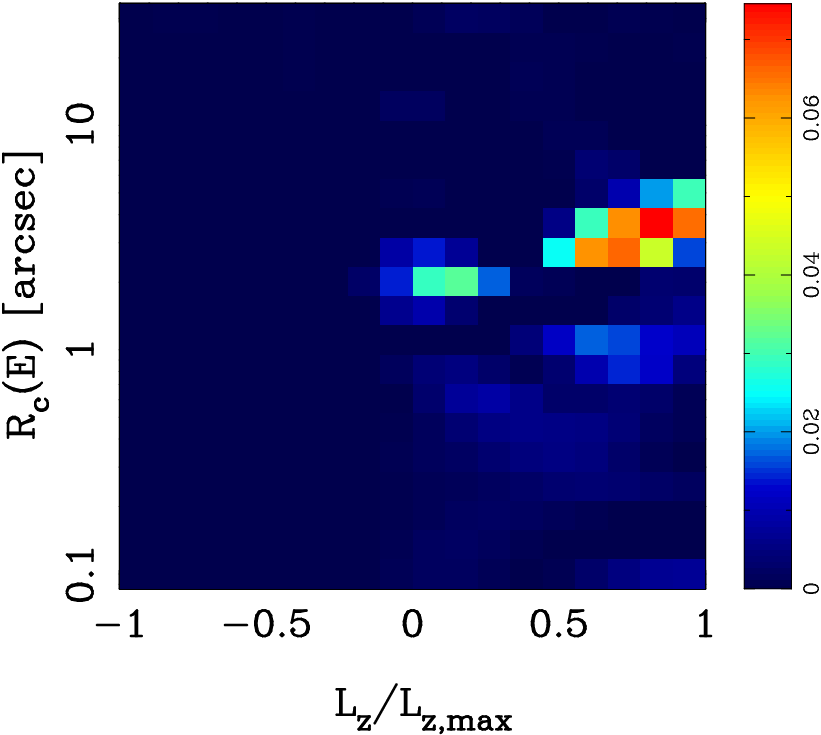}}
  \end{center}

  \caption{Best model reconstruction of the weighted two-integral DFs
    of the sample galaxies. The value of each pixel represents the
    relative contribution of the corresponding TIC to the luminous
    mass of the galaxy. The galaxies are sorted by increasing values
    of the specific angular momentum parameter~$\jz$.}
  \label{fig:DF}
\end{figure*}

\section{Dynamical structure}
\label{sec:dynamics}

A major strength of the method described in Sect.~\ref{sec:analysis}
is that the {\cauldron} algorithm also provides, for each analyzed
galaxy, the associated (weighted) two-integral stellar DF, which
constitutes the most general and complete description of the dynamical
structure of the system within the context of the adopted model. The
DF can equivalently be represented as a bi-dimensional map of the TIC
weights in the integral space $(E, \Lz)$: the value of each pixel
simply indicates the relative stellar-mass contribution due to the TIC
building block of corresponding energy and angular momentum.

Figure~\ref{fig:DF} shows the maps of the TIC weights that are
obtained for the best-reconstructed model (i.e., the MAP model) of
each lens galaxy. In keeping with the analysis of SDSS\,J0728
\citep{Barnabe2010}, the most recent study of a lens galaxy performed
with {\cauldron}, we make use of a library composed of~360 TICS: for
each one of the $\nE = 20$ elements logarithmically sampled in the
circular radius $\Rc$ (corresponding to a grid in energy, see
\citealt{Barnabe-Koopmans2007}) we consider $\nLz = 9$ elements
linearly sampled in the normalized angular momentum $0 \le \Lz/L_{z,
  \mathrm{max}} \le 1$, and the mirrored negative $\Lz/L_{z,
  \mathrm{max}}$ values. Although a grid of $(\nE = 10) \times (\nLz =
5)$ has been shown to be generally sufficient to provide a
satisfactory reconstruction of the observables \citep{Barnabe2009}, a
finer grid does a better job in reducing the undesired discreteness
effect of the TIC superposition (e.g. on the surface brightness,
kinematic and local $\vphi/\bar{\sigma}$ maps).

The DF encapsulates the description of the stellar component of the
galaxy in an very compact way, which can not be straightforwardly
decoded at a glance. Therefore, we now proceed to distill much of that
information into more directly interpreted quantities of astrophysical
interest that characterize the dynamical status of the system on a
global and local level, i.e.\ the ratio between ordered and random
motions, the specific angular momentum and the global anisotropy of
the stellar velocity dispersion tensor.

\subsection{Global and local $v/\sigma$}
\label{ssec:vos}

\begin{figure}
  \centering
  \resizebox{1.00\hsize}{!}{\includegraphics[angle=-90]
            {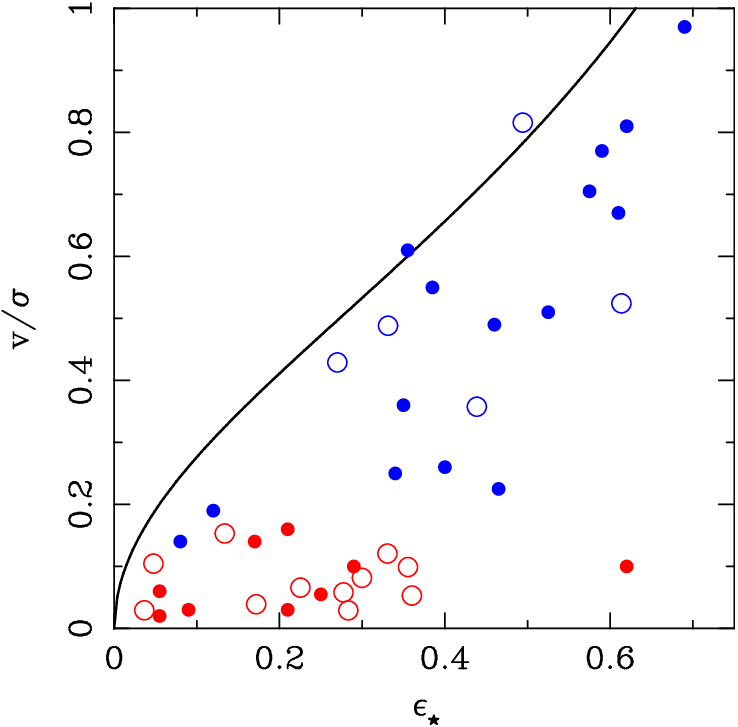}}
  \caption{Model $\vos$ diagram for the sixteen lens galaxies in our sample
    (open circles), divided in fast and slow rotators (blue and red
    points, respectively); $\epstar$ is the intrinsic ellipticity of
    luminous distribution, and $v/\sigma$ is calculated from the best
    model, corrected to an edge-on inclination. As a comparison, the
    plot also shows, as full circles, the corresponding quantities
    (also corrected for inclination) for the 24 nearby SAURON
    ellipticals studied in \citet{Cappellari2007}. The solid line
    represents the location of edge-on isotropic rotators (assuming
    $\alpha = 0.15$, see \citealt{Binney2005}).}
  \label{fig:vos}
\end{figure}

Figure~\ref{fig:vos} shows the location of the sixteen analyzed lens
galaxies on the model version of the classical $\vos$ diagram. The
quantity $v/\sigma$ is commonly used as a global indicator of the
importance of ordered motions (i.e. rotation) with respect to random
motions in a stellar system. For each galaxy we calculate the ratio
$v/\sigma$, corrected to an edge-on view, from the two-dimensional
kinematic maps of the best reconstructed model, following the
prescriptions of \citet{Binney2005} for extended kinematic data
sets. The ratio is plotted against the inclination-corrected
ellipticity of the light distribution $\epstar \equiv 1 - \qstar$,
where $\qstar$ is the axial ratio introduced in
Sect.~\ref{ssec:axialratio}. 

We find that the dynamics of most systems is clearly dominated by
random motions, with $v/\sigma \le 0.15$, while rotation plays a
significant role in the remaining five galaxies. The members of the
latter group have several common characteristics, beyond being above a
$v/\sigma$ ratio of 0.35: (i) they are all fast rotators, according to
both the definition of \citet{Emsellem2007} and their specific angular
momentum content (as discussed below in Sect.~\ref{ssec:jz}); (ii)
they are quite flattened in the light (having $0.38 < \qstar < 0.74$);
(iii) they belong to the lower half of the sample in terms of $\Mtot$,
including, moreover, the three least massive galaxies of the whole
sample (namely, SDSS\,J1330, SDSS\,J1443 and SDSS\,0959).

As a comparison to the SLACS systems, Figure~\ref{fig:vos} also
displays the location on the diagram of a sub-sample of~24 SAURON
early-type galaxies, whose observables have also been corrected for
inclination, analyzed by \citet{Cappellari2007}. The two samples
generally occupy similar positions in the $\vos$ space, although the
SLACS galaxies lack the most extreme cases (such as the very flattened
slow rotator NGC\,4550 and the fast rotator NGC\,3156 with a $v/\sigma
\sim 1$) and show a sharper transition between dispersion- and
rotation-dominated systems. This, however, is likely a consequence of
the limited number of fast rotating galaxies present among the SLACS
sample, mainly comprised of massive systems ($\sigma_{\mathrm{SDSS}}
\gtrsim 200$ km s$^{-1}$) which tend to be slow rotators.

\begin{figure*}
  \begin{center}
    \subfigure[J1204]{\label{fig:vos-1204}\includegraphics[angle=-90,width=0.24\textwidth]{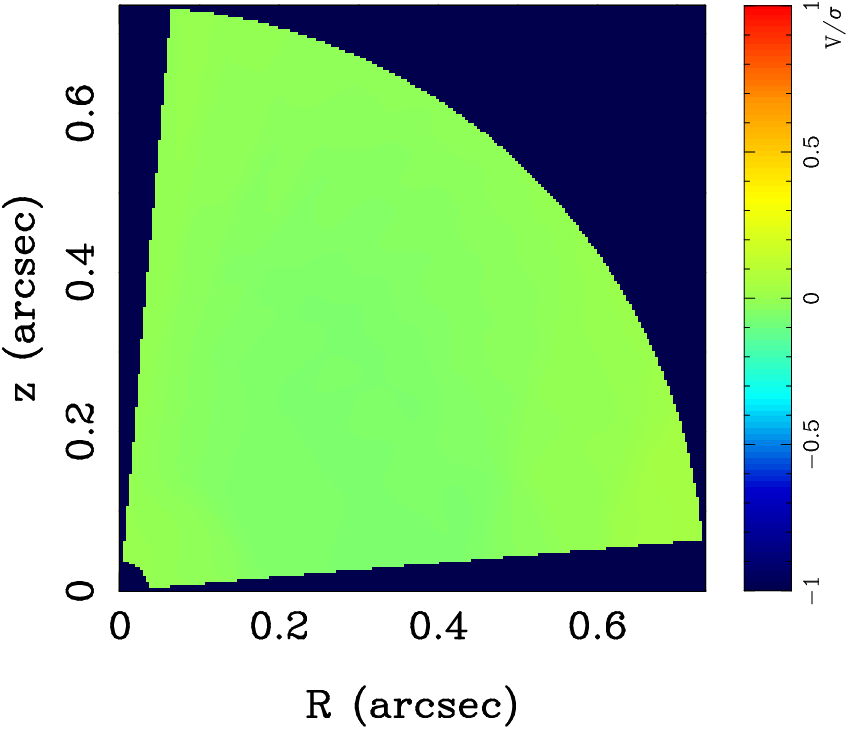}}
    \subfigure[J2321]{\label{fig:vos-2321}\includegraphics[angle=-90,width=0.24\textwidth]{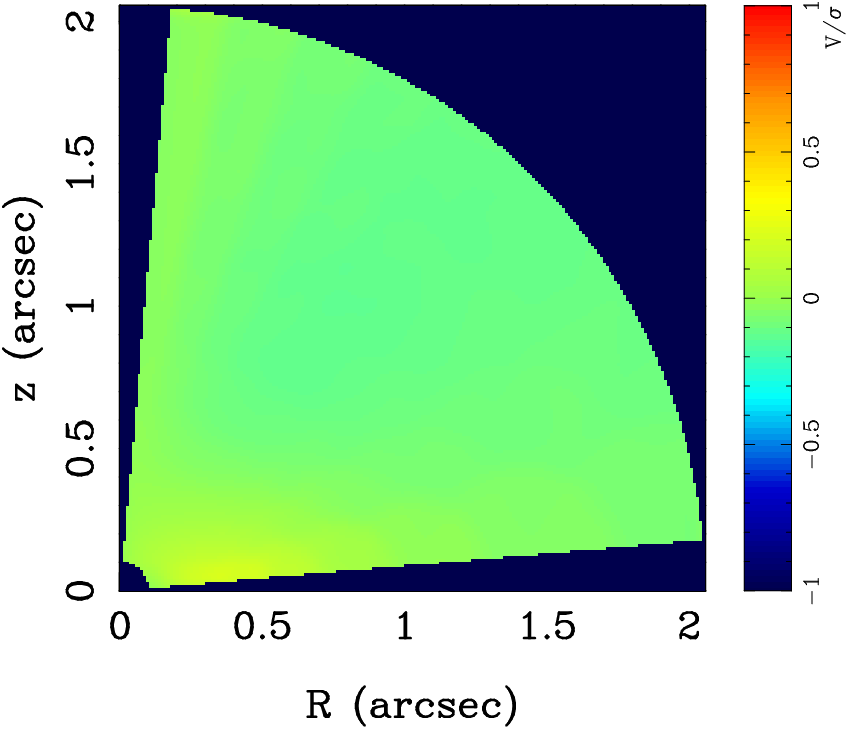}}
    \subfigure[J0216]{\label{fig:vos-0216}\includegraphics[angle=-90,width=0.24\textwidth]{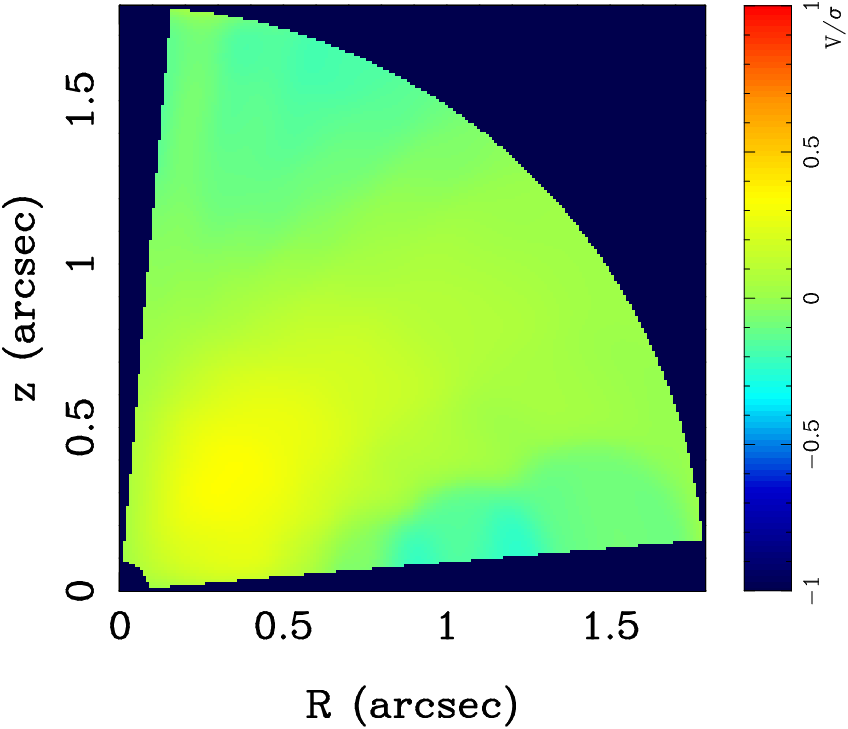}}
    \subfigure[J1250]{\label{fig:vos-1250}\includegraphics[angle=-90,width=0.24\textwidth]{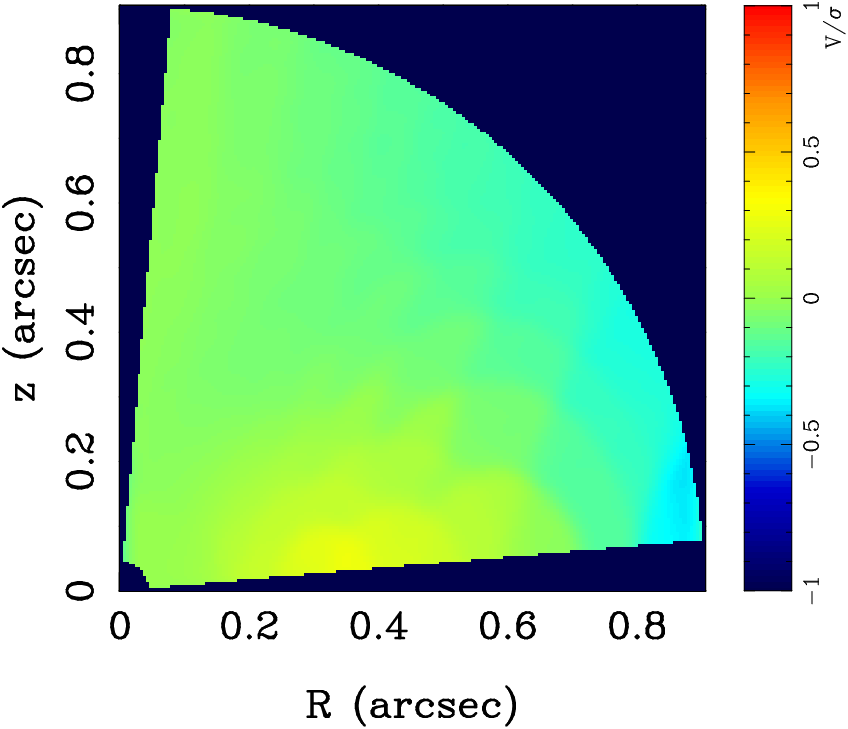}}
  \end{center}
  \vspace{-0.1cm}

  \begin{center}
    \subfigure[J2300]{\label{fig:vos-2300}\includegraphics[angle=-90,width=0.24\textwidth]{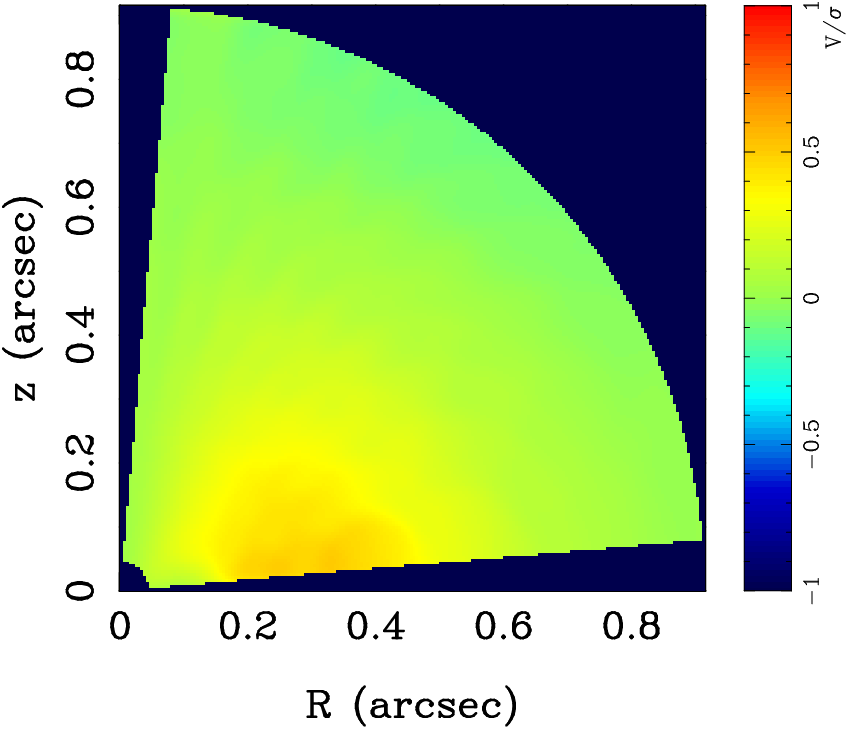}}
    \subfigure[J0935]{\label{fig:vos-0935}\includegraphics[angle=-90,width=0.24\textwidth]{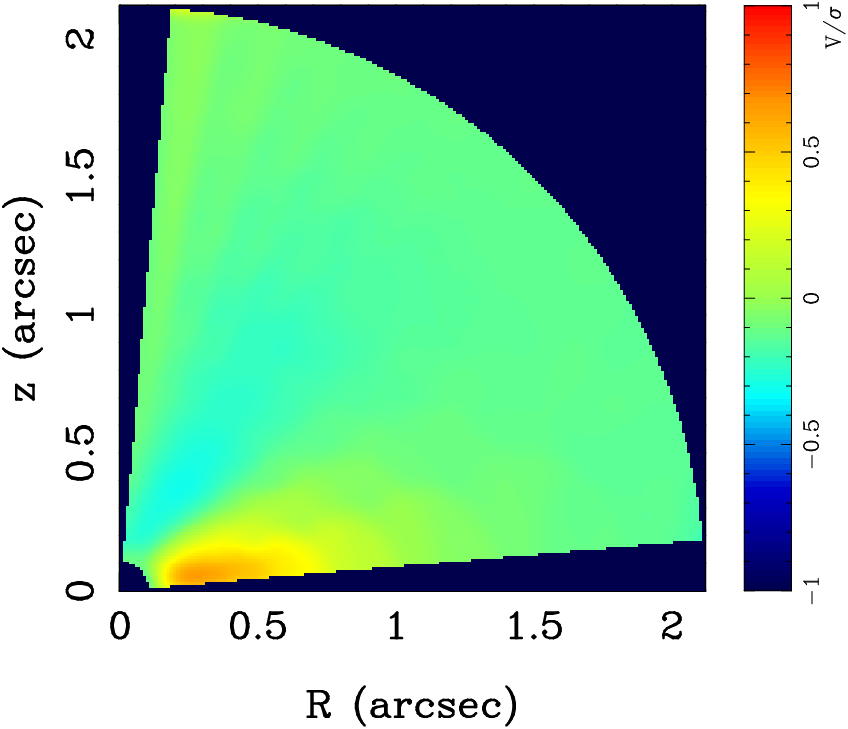}}
    \subfigure[J1627]{\label{fig:vos-1627}\includegraphics[angle=-90,width=0.24\textwidth]{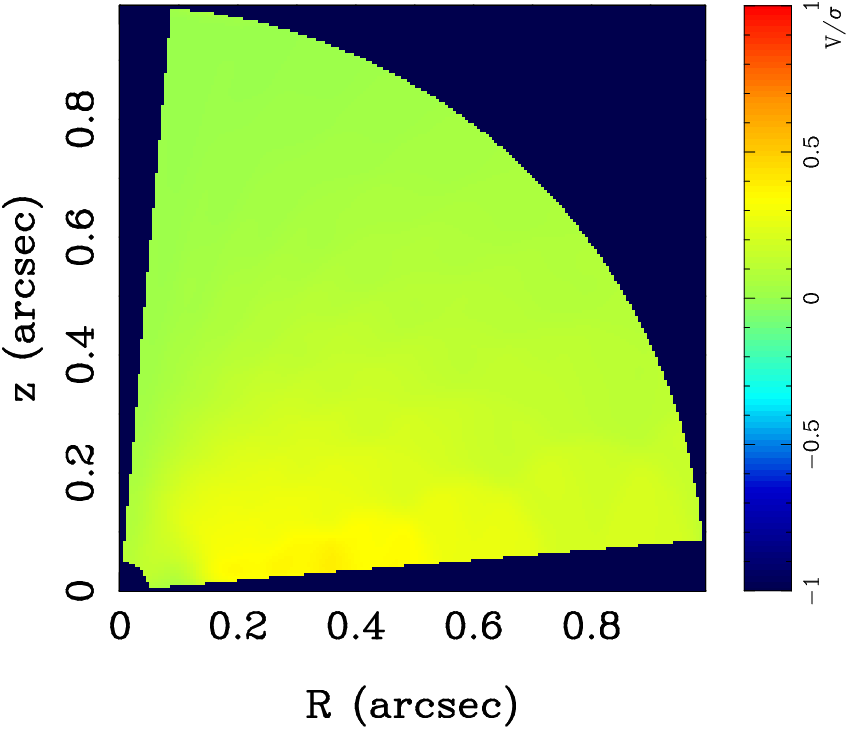}}
    \subfigure[J2303]{\label{fig:vos-2303}\includegraphics[angle=-90,width=0.24\textwidth]{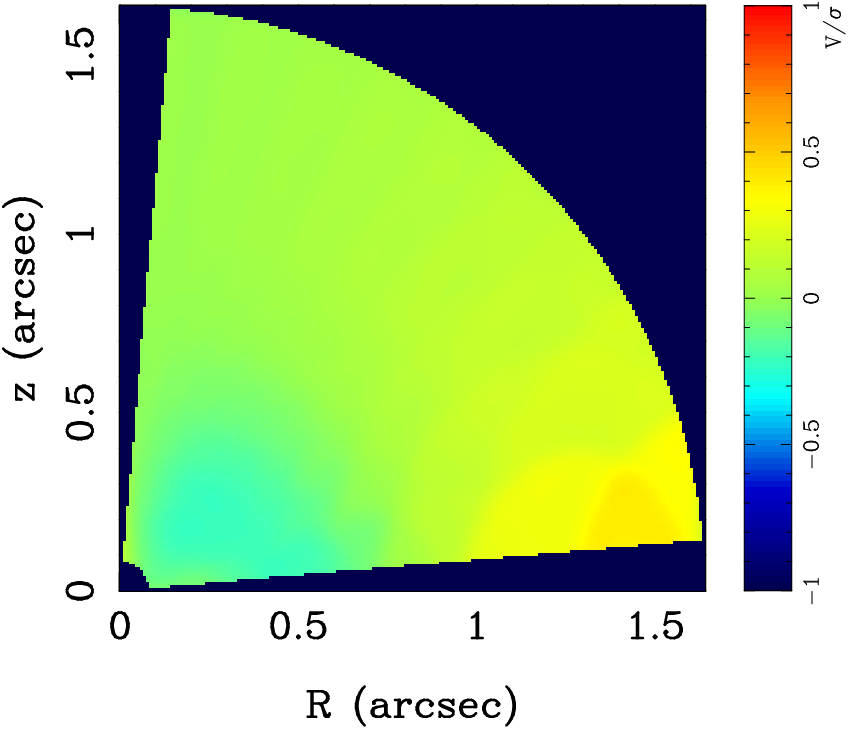}}
  \end{center}
  \vspace{-0.1cm}

  \begin{center}
    \subfigure[J0912]{\label{fig:vos-0912}\includegraphics[angle=-90,width=0.24\textwidth]{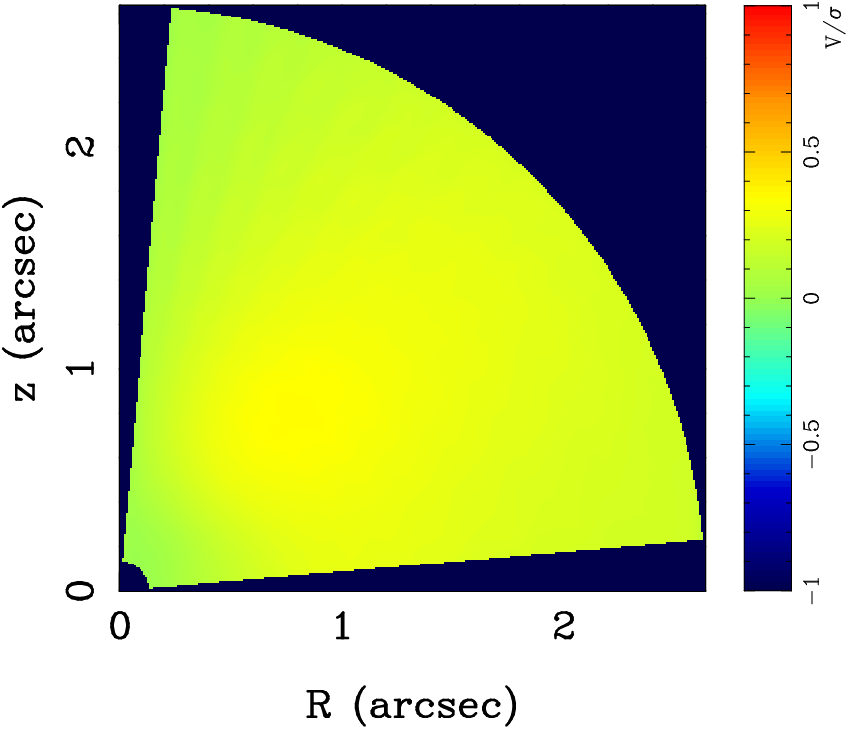}}
    \subfigure[J0037]{\label{fig:vos-0037}\includegraphics[angle=-90,width=0.24\textwidth]{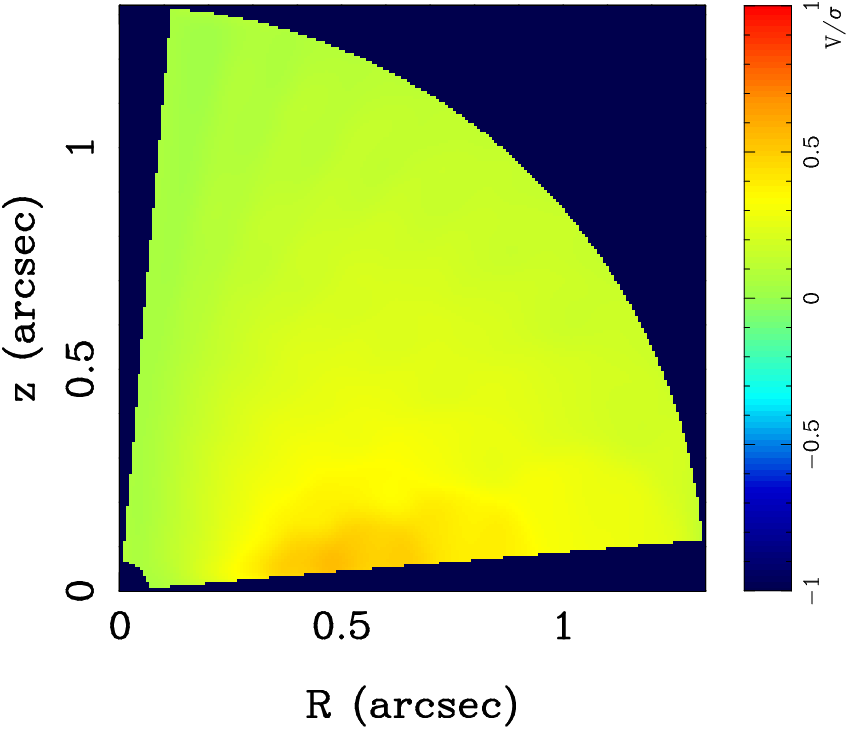}}
    \subfigure[J1451]{\label{fig:vos-1451}\includegraphics[angle=-90,width=0.24\textwidth]{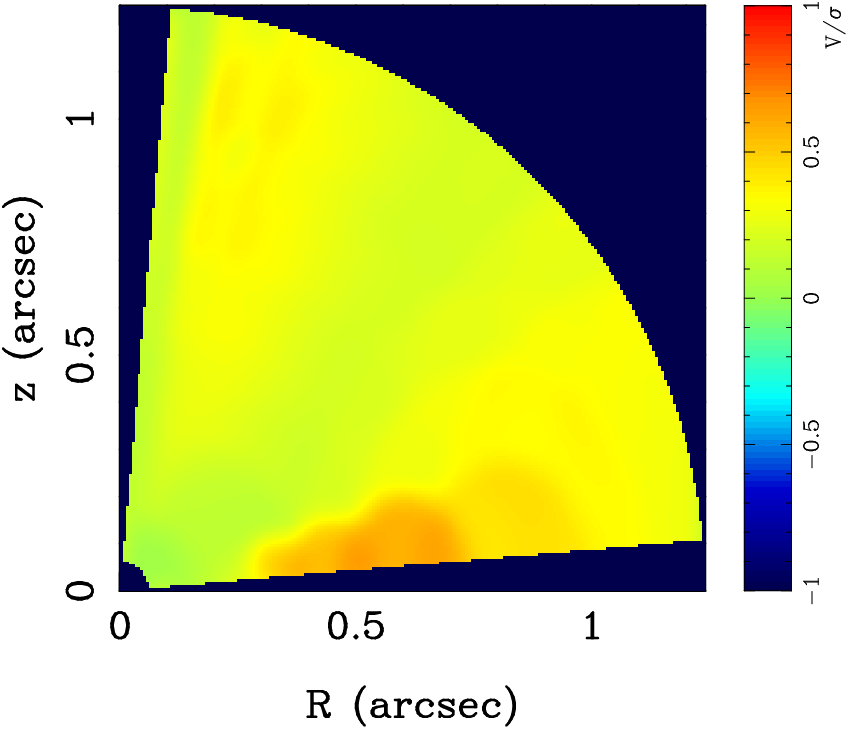}}
    \subfigure[J1330]{\label{fig:vos-1330}\includegraphics[angle=-90,width=0.24\textwidth]{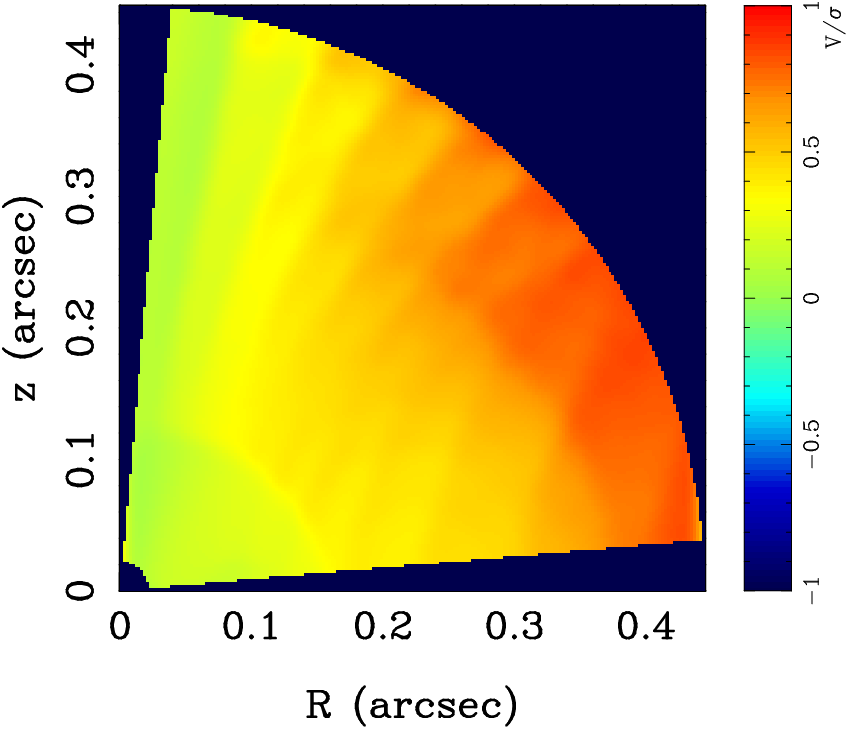}}
  \end{center}
  \vspace{-0.1cm}

  \begin{center}
    \subfigure[J1443]{\label{fig:vos-1443}\includegraphics[angle=-90,width=0.24\textwidth]{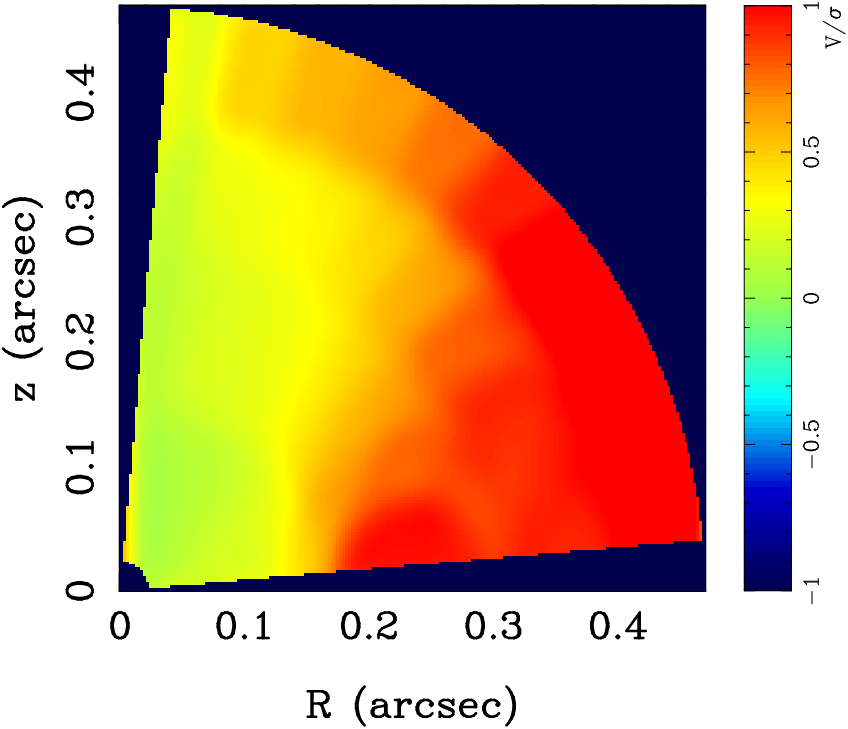}}
    \subfigure[J0959]{\label{fig:vos-0959}\includegraphics[angle=-90,width=0.24\textwidth]{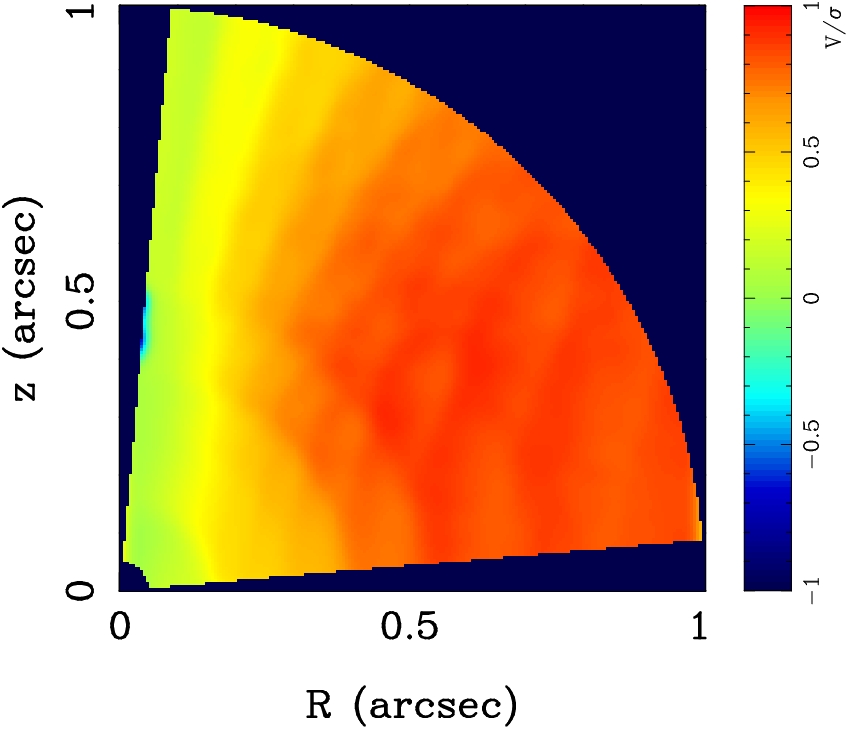}}
    \subfigure[J2238]{\label{fig:vos-2238}\includegraphics[angle=-90,width=0.24\textwidth]{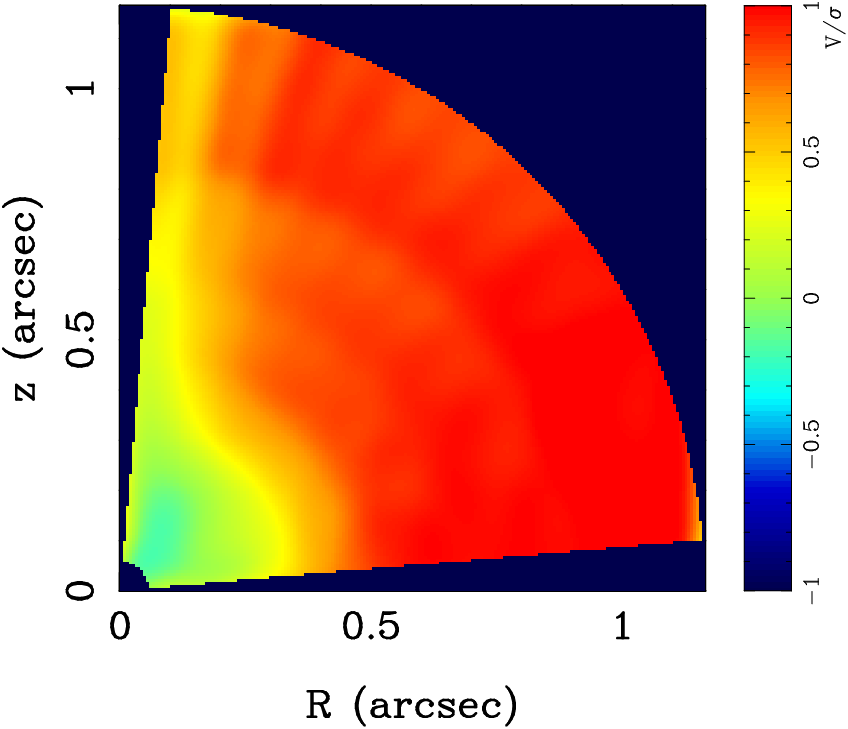}}
    \subfigure[J1251]{\label{fig:vos-1251}\includegraphics[angle=-90,width=0.24\textwidth]{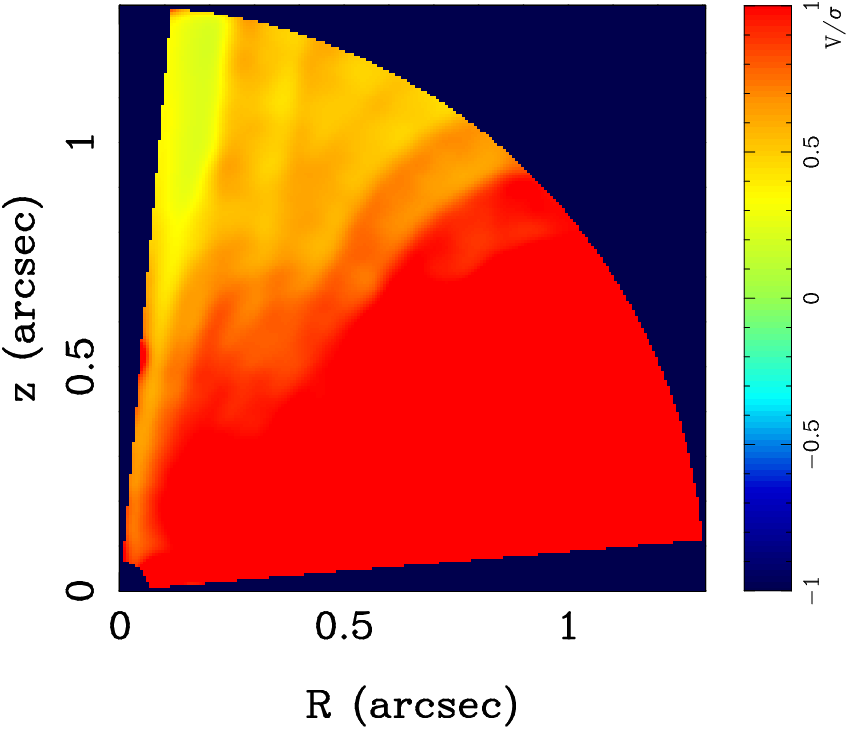}}
  \end{center}

  \caption{Maps of the local $\vphi/\bar{\sigma}$ ratio between the
    mean rotation velocity around the $z$-axis and the mean velocity
    dispersion, plotted up to $\Reff/2$ in the positive quadrant of
    the meridional plane. As in Fig.~\ref{fig:DF}, the galaxies are sorted
    by increasing~$\jz$.}
  \label{fig:local_vos}
\end{figure*}

From the stellar DF it is also possible to derive a description of the
intrinsic kinematics of the galaxy by using the internal velocity
moments to define\,---\,at each point $(R,z)$ in the meridional
plane\,---\,the quantity $\vphi/\bar{\sigma}$, where $\vphi$ is the local
mean azimuthal velocity (i.e.\ around the symmetry axis) and
$\bar{\sigma}^{2} \equiv (\sigma_{R}^{2} + \sigma_{\varphi}^{2} +
\sigma_{z}^{2})/3$ is the mean velocity dispersion. This ratio can be
interpreted as a local and intrinsic analogue of the global indicator
$v/\sigma$, since it characterizes the importance of rotation with
respect to random motions at each position inside the
galaxy. Figure~\ref{fig:local_vos} shows the $\vphi/\bar{\sigma}$ maps
for all the sample galaxies, extending up to $\re/2$ and sorted by
increasing specific angular momentum~$\jz$ (see
Sect.~\ref{ssec:jz}). For visualization purposes, and in particular to
make the comparison among fast rotators easier, the maps are presented
as having the same sense of rotation (where applicable).

By examining the $\vphi/\bar{\sigma}$ maps, the sample galaxies
clearly fall into two broad groups on the basis of their dynamical
status. The first~11 systems are overall dominated by random motions;
in a few systems, such as SDSS\,J0935 and SDSS\,J1451, islands of
moderate rotation ($|\vphi|/\bar{\sigma} \approx 0.5$) are present,
but remain confined to small, spatially limited regions. What sets
apart the remaining five galaxies, on the other hand, is the presence
of strong rotation on a large scale pattern. This structural
difference reflects the classification of the two groups as,
respectively, slow and fast rotators. Guided by the intuition provided
by these maps, in the next Section we provide a more quantitative
criterion, based on the angular momentum content, to specify to which
one of the two groups the modelled galaxies belong.

\subsection{The angular momentum of slow and fast rotators}
\label{ssec:jz}

We recall the definition of the dimensionless specific angular
momentum of the stellar component $\jz$ \citep[see][]{Barnabe2009}:
\begin{equation}
\label{eq:jz}
\jz \equiv \frac{\displaystyle
                 \int \rhostar R \, | \vphi | \, \mathrm{d}^{3} \vec{x}}
                {\displaystyle 
		 \int \rhostar R \sqrt{{\vphi}^{2} + \bar{\sigma}^{2}} 
		 \, \mathrm{d}^{3} \vec{x}} \, ,
\end{equation}
where $\rhostar = \int f \, \mathrm{d}^{3} \vec{v}$ is the density of
the luminous component obtained from the MAP model stellar DF~$f$. For
each galaxy in the sample, we carry out this integral up to $\re/2$,
tabulating the results in Table~\ref{tab:dyn}. We find that the
galaxies can be neatly divided in two groups depending on the value of
the quantity~$\jz$: on one side~11 low angular momentum systems, with
$\jz \lesssim 0.3$, and on the other side five objects with $\jz
\simeq 0.5$ or greater. By comparing with the local
$\vphi/\bar{\sigma}$ maps, we see that these five systems are also the
only ones that clearly exhibit large scale rotation. Therefore, we
choose to classify the galaxies on the basis of their specific angular
momentum content by labelling as slow and fast rotators the systems
that are, respectively, below and above $\jz = 0.4$.

\begin{figure}
  \centering
  \resizebox{1.00\hsize}{!}{\includegraphics[angle=-90]
            {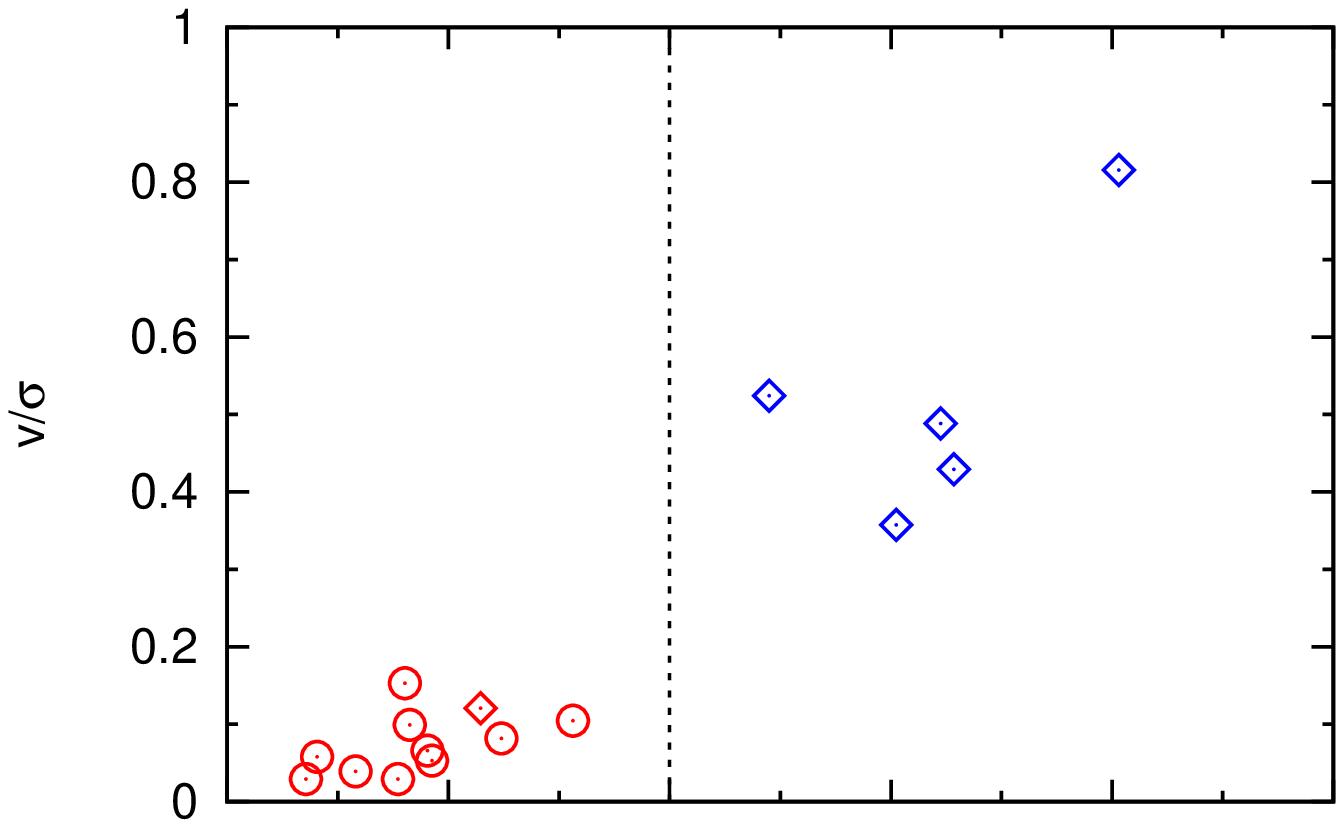}}

  \vspace{-1.0cm}

  \resizebox{1.00\hsize}{!}{\includegraphics[angle=-90]
            {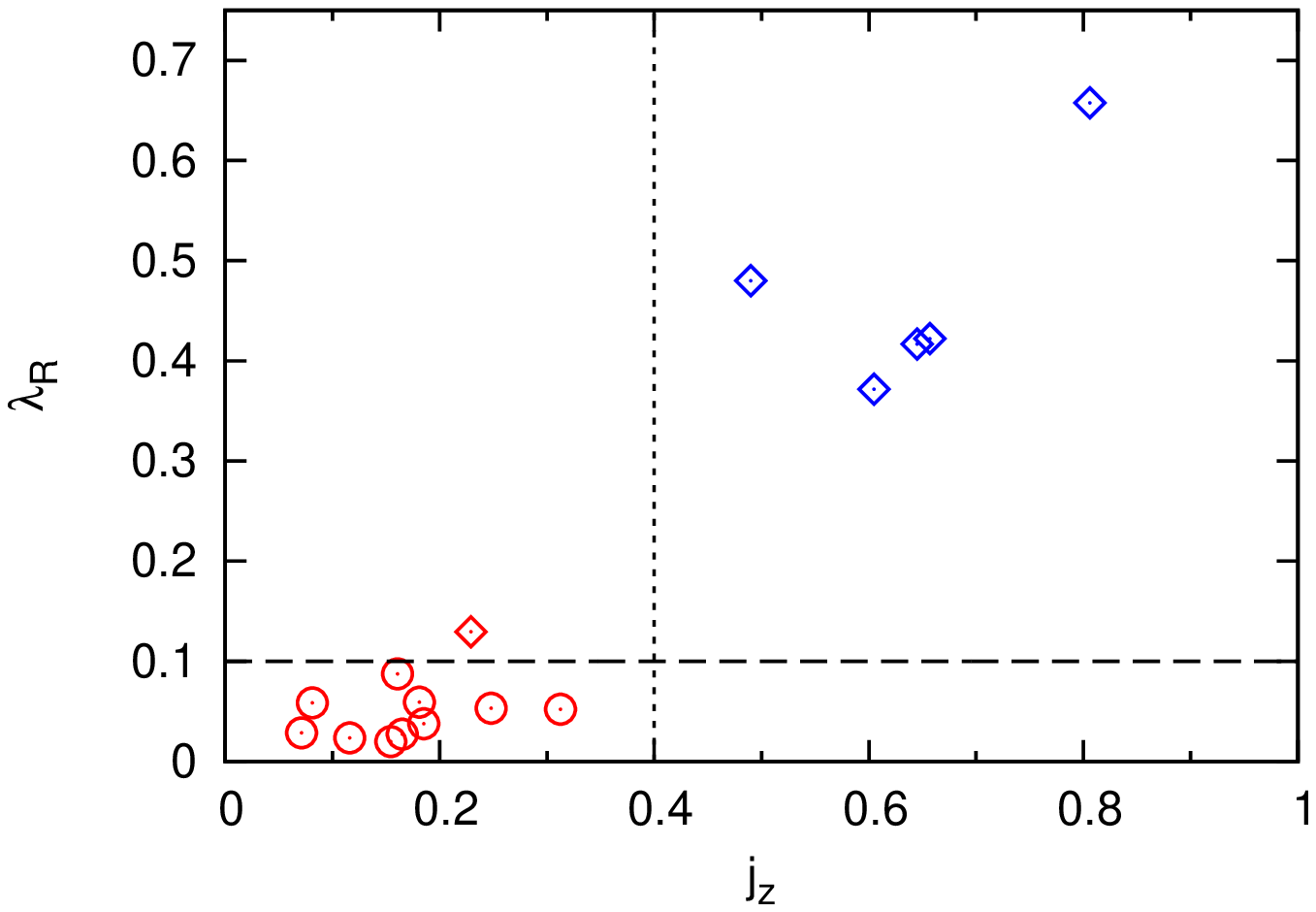}}

  \caption{The intrinsic dynamical parameter~$\jz$ versus the
    inclination-corrected $v/\sigma$ ratio (top panel) and the
    Emsellem kinematic parameter~$\lamR$ (bottom panel) for the
    sixteen SLACS lens galaxies in the sample. The circles and
    diamonds indicate, respectively, the slow and fast rotating
    systems classified according to the \citet{Emsellem2007}
    definition of the threshold value that divides the two types,
    i.e.\ $\lamR = 0.1$ (horizontal dashed line). The red and blue
    symbols denote the slow and fast rotators when the galaxies are
    classified on the basis of the dimensionless angular momentum
    (threshold value $\jz = 0.4$, vertical dotted line). The two
    classifications produce fully equivalent results for all systems
    barring SDSS\,J0912 (indicated with the red diamond), which is the
    galaxy with the most spatially limited kinematic maps (see text).}
  \label{fig:jz}
\end{figure}

In Figure~\ref{fig:jz} we plot, for each system, $\jz$ versus the
quantities $v/\sigma$ (inclination-corrected, Sect.~\ref{ssec:vos})
and $\lamR$, computed on the kinematic data set of the best
reconstructed model. The observational parameter $\lamR$ was first
introduced by \citet{Emsellem2007} as a consistent way to quantify the
specific stellar angular momentum (in projection) of an early-type
galaxy for which two-dimensional kinematic maps are available, and its
value is employed as the criterion to discriminate between slow
($\lamR < 0.1$) and fast ($\lamR > 0.1$) rotating systems. The model
quantity~$\jz$ is, by construction, the three-dimensional and
intrinsic analogue of~$\lamR$.  Figure~\ref{fig:jz} shows that
classifying the rotators on the basis of either of these two
parameters yields fully equivalent results for all galaxies but one,
SDSS\,J0912. We have already singled out this system (in
Sect.~\ref{sec:mass}) as the only one for which the kinematic maps do
not reach half of the effective radius. The misclassification of
SDSS\,J0912 can be ascribed to the spatially limited data set, since
slow rotators commonly show decreasing $\lamR$ profiles for $R/\Reff
\lesssim 0.5$, and thus the parameter can exceed the threshold value
of~0.1 in the inner regions (see \citealt{Emsellem2007} and in
particular their Figure~2).

It is worth noting that, for slow rotators, the spread in $\lamR$ is
much smaller than the spread in $\jz$, indicating that the latter
constitutes a better discriminator of the intrinsic properties of
these systems.

\subsection{Global velocity dispersion tensor}
\label{ssec:anisotropy}

The global stellar velocity dispersion tensor constitutes a useful and
concise characterization of the overall dynamical structure of a
stellar system. It is customary to describe its shape by means of the
three global anisotropy parameters \citep{Cappellari2007,BT08}
\begin{equation}
  \label{eq:AP}
  \beta \equiv 1 - \frac{\Pi_{zz}}{\Pi_{RR}}, \quad
  \gamma \equiv 1 - \frac{\Pi_{\varphi\varphi}}{\Pi_{RR}} 
  \quad \textrm{and} \quad
  \delta \equiv 1 - \frac{2 \Pi_{zz}}{\Pi_{RR} + \Pi_{\varphi\varphi}},
\end{equation}
where we denote the total unordered kinetic energy along the
coordinate direction~$k$ as
\begin{equation}
  \label{eq:AP:PI}
  \Pi_{kk} = \int \rhostar \sigma^{2}_{k}\, \mathrm{d}^{3} \vec{x} \, ,
\end{equation}
and $\sigma^{2}_{k}(\vec{x}) \equiv \langle v^{2}_{k} \rangle -
{\langle v_{k} \rangle}^{2}$ is the local stellar velocity dispersion
in the $k$~direction. For a fully isotropic object the values of all
three parameters become zero. If the galaxy is only isotropic in the
meridional plane, then $\beta = 0$ and the remaining parameters are
related by $\gamma = 2 \delta/(\delta - 1)$. The latter scenario is
the one that applies to the galaxy models considered in this analysis,
since, for any axisymmetric collisionless system supported by a
two-integral DF, $\langle v^{2}_{R} \rangle = \langle v^{2}_{z}
\rangle$ everywhere, the meridional velocities $\langle v_{R} \rangle$
and $\langle v_{z} \rangle$ are null, and therefore $\Pi_{RR} =
\Pi_{zz}$.

The values of the global parameters~$\delta$ and~$\gamma$ for the
SLACS galaxies, calculated within $\re/2$, are reported in
Table~\ref{tab:dyn}. The slow rotators are all mildly anisotropic,
quite similarly to what is found for their nearby counterparts
\citep{Cappellari2007}: the values of~$\delta$ fall between~0.05
and~0.25, and $\delta \lesssim 0.15$ for most systems. The fast
rotating galaxies, with the exception of SDSS\,J1443, exhibit instead
negative $\delta$ values. This is a consequence of the adopted
dynamical model: due to the isotropy in the meridional plane, it
follows from Eq.~(\ref{eq:AP}) that every system for which
$\sigma^{2}_{\varphi} < \sigma^{2}_{R}$ over most of the luminosity
density-weighted volume (so that $\Pi_{\varphi\varphi} < \Pi_{RR}$) is
characterized by $\delta < 0$, i.e. is anisotropic in the sense of
having a smaller pressure along the equatorial plane than
perpendicular to it. Moreover\,---\,within the approach of modelling
based on TIC superposition\,---\,very fast rotating systems tend to
have small azimuthal velocity dispersions~$\sigma^{2}_{\varphi}$,
since their dynamical description is dominated by co-rotating TIC
building blocks of similar angular momentum. This can be more easily
illustrated by considering the limiting case of a hypothetical system
described by a single TIC: for this simple object one has $\langle
v^{2}_{\varphi} \rangle = {\langle v_{\varphi} \rangle}^{2}$, and thus
$\sigma^{2}_{\varphi} = 0$, everywhere in the meridional plane within
the zero-velocity curve, so that $\delta_{\mathrm{TIC}} = -1$ and
$\gamma_{\mathrm{TIC}} = 1$. It is not surprising, therefore, that the
three fastest rotating galaxies in the sample (SDSS\,J0959,
SDSS\,J2238 and SDSS\,J1251) are all found to have clearly negative
values of~$\delta$.

\section{Discussion}
\label{sec:discussion}

The study presented in this paper represents the most detailed
analysis conducted so far of the total mass density profile of a
relatively large sample of E/S0 galaxies beyond the local Universe,
making consistent use of axially symmetric density distributions for
both the lensing and dynamical modelling, and taking advantage of the
constraints provided by two-dimensional kinematic information.

\subsection{Total density profile}
\label{ssec:disc:rho}

The results of this work corroborate the mounting evidence (see
Sect.~\ref{sec:introduction}) that the inner regions (on the scale of
the effective radius) of massive early-type galaxies are characterized
by an approximately isothermal total density distribution (i.e.,
$\rho_{\mathrm{tot}} \propto r^{-\slope}$, with $\slope \approx 2$), a
property often referred to as the `bulge--halo conspiracy'. This name
owes to the fact that there appears to be no obvious physical
explanation of why the luminous and dark matter components of each
system, despite not obeying a power-law distribution, should often
`conspire' to produce a nearly $1/r^{2}$ total profile.

\begin{figure}
  \centering
  \resizebox{1.00\hsize}{!}{\includegraphics[angle=-90]
            {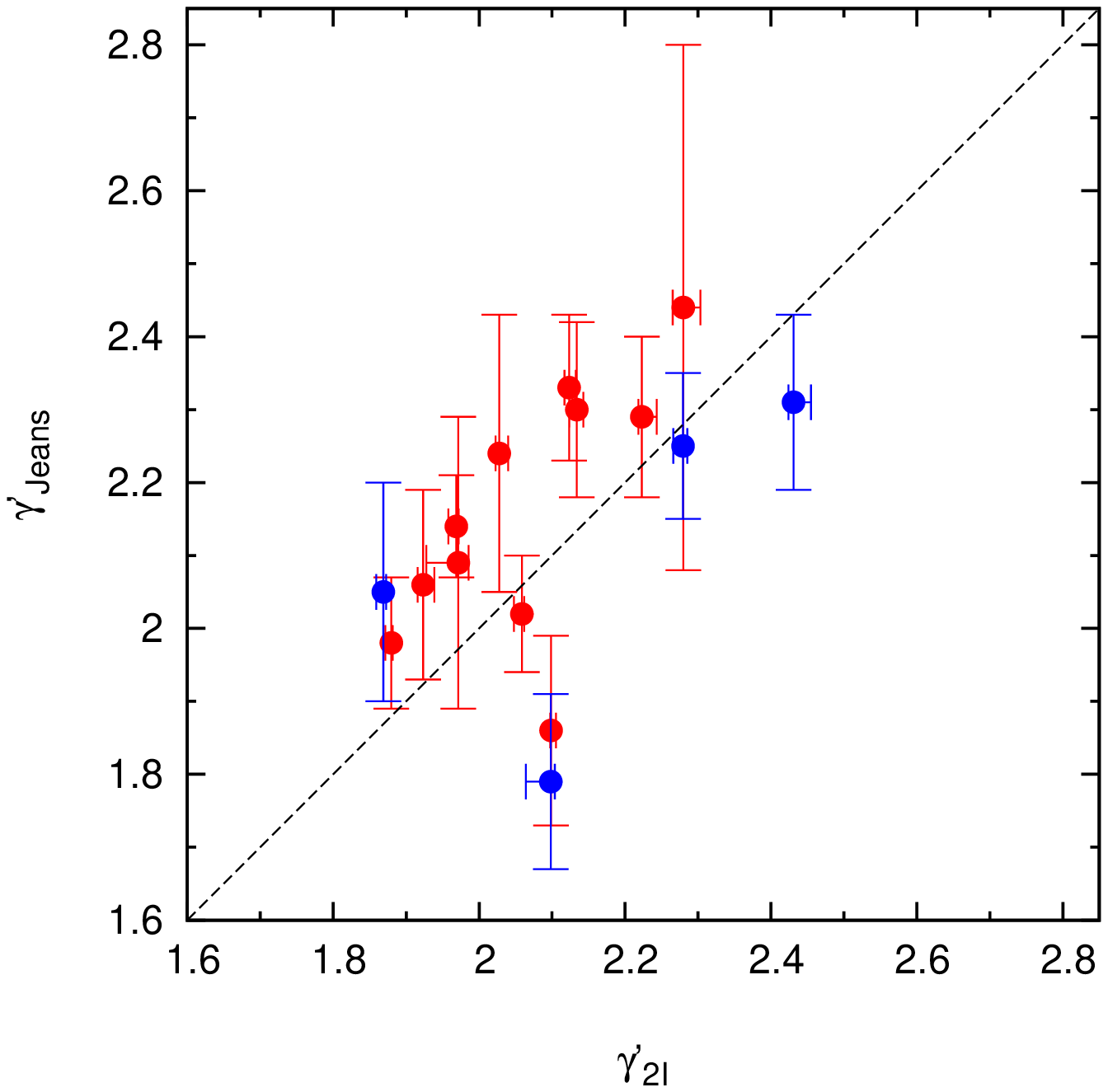}}
  \caption{Comparison between the total density logarithmic slope
    $\slope_{\mathrm{2I}}$ of the sample lens galaxies obtained from
    the analysis conducted in this paper, making use of two-integral
    dynamical models and VLT VIMOS integral field spectroscopy, and
    the slope $\slope_{\mathrm{Jeans}}$ derived by \citet{Auger2010},
    using spherical Jeans models and a single kinematic
    measurement. Red and blue points denote, respectively, slow and
    fast rotators. The dashed line traces the one-to-one
    relationship. For both $\slope_{\mathrm{2I}}$ and
    $\slope_{\mathrm{Jeans}}$, the uncertainties represent the 68 per
    cent confidence interval. The slope of galaxy SDSS\,1251 is not
    listed, as it was not included in the \citet{Auger2010} sample of
    early-type systems.}
  \label{fig:gamma-gamma}
\end{figure}

By examining individual lens galaxies in great detail, however, we can
conclude that the purported conspiracy only holds in first
approximation: even though we provide confirmation that all systems,
quite remarkably, can be effectively modelled by a simple and smooth
single power-law profile for $\rho_{\mathrm{tot}}$, the average
logarithmic slope is found to be $\mslope = 2.074^{+0.043}_{-0.041}$,
i.e. slightly (but significantly) super-isothermal. Moreover, and
perhaps more importantly, different galaxies do exhibit noticeably
different slopes, with an intrinsic scatter in $\mslope$ of about
7~per cent. These results are in agreement with the findings of
\citet{Koopmans2009} and \citet{Auger2010} for the full sample of
SLACS lenses, obtained by making use of spherical Jeans models for the
dynamics, constrained by a single kinematic measurement, i.e. the
stellar velocity dispersion $\sigma_{\mathrm{SDSS}}$ inside the
3~arcsec diameter SDSS fiber.  This indicates that the combined
lensing and dynamics approach is a generally robust method to
determine the total density distribution of a sample of lens galaxies,
even when a very simple (and not fully consistent) dynamical model is
employed. The values of the logarithmic slope obtained for
\emph{individual} systems with the two methods\,---\,that we denote
here as $\slope_{\mathrm{2I}}$ and $\slope_{\mathrm{Jeans}}$\,---\,can
differ by up to about~10-15 per cent, as illustrated in
Figure~\ref{fig:gamma-gamma}. This discrepancy is typically of the
order of the 1-$\sigma$ errors in $\slope_{\mathrm{Jeans}}$ but, in
most cases, much larger than the few per cent uncertainties on
$\slope$ obtained with this study, stressing the importance of having
access to two-dimensional kinematic data sets and more sophisticated
dynamical models when a precise measurement of a lens galaxy density
profile is called for. It is interesting to note that another joint
lensing and dynamics study conducted by \citet{Ruff2011} on a
different sample of~11 massive early-type galaxies from the Strong
Lenses in the Legacy Survey (SL2S) at a higher redshift (median
$z_{\mathrm{lens}} = 0.5$) also finds an average slope and an
intrinsic scatter consistent with the results presented here.

It has been well established by N-body cosmological simulations that
dissipationless processes produce inner density distributions with
logarithmic slopes $\slope \approx 1$ \citep[e.g.][]{Navarro1996} or
even shallower \citep[e.g.][]{Graham2006, Navarro2010}. Moreover,
subsequent dissipationless merging (i.e., dry merging) preserves the
steepness of the inner density profile \citep{Dehnen2005,
  Kazantzidis2006}. It follows, therefore, that the interplay between
dark matter and the baryonic component, with its complex dissipative
processes, plays a key role and must be taken into account in order to
explain the emergence of the isothermal profile in the inner regions
of galaxies. Numerical work (see e.g. \citealt{Gnedin2004},
\citealt{Abadi2010} and references therein) on the response of dark
matter to the condensation of baryons does indeed show that the
assembly of a central galaxy makes the halo significantly more
concentrated, producing a steeper total density slope.  The
high-resolution simulations of moderately massive systems (with halo
masses comparable to that of the Milky Way) carried out by
\citet{Tissera2010}, which include a detailed treatment of baryonic
processes, reveal a nearly isothermal profile over a relatively large
radial range extending up to the baryonic radius (defined as the
radius containing 83 per cent of the stellar and gaseous component of
the galaxy, i.e. the region within which the baryons play an important
role). Aside from this general global result, however, the details of
the density distribution are largely determined by the specific
assembly history of each system. This suggests that the observed
scatter in the $\slope$ values of the SLACS lenses is a consequence of
the idiosyncratic formation histories of the individual galaxies, and
can potentially provide information on the processes involved in their
build-up. Once the isothermal structure is in place, dissipationless
mergers\,---\,both major and minor\,---\,preserve it not only in the
inner regions, but throughout the systems \citep*{Nipoti2009}. We
stress, however, that, even in the case of a nearly perfectly
$1/r^{2}$ seed distribution, this process produces a non-negligible
scatter of about 10 per cent in the logarithmic slope (i.e. comparable
to the observed level), which overlaps with the scatter that,
potentially, might have been present in the parent (pre- dry mergers)
density distribution, further complicating the task of reconstructing
the galaxy formation history.

\subsection{Mass and dynamical structure}
\label{ssec:disc:mass}

Another defining property of early-type galaxies that we have aimed to
quantify in this study is the fractional amount of dark matter that is
present in the systems' central regions. If the assumption of a
constant stellar mass-to-light ratio holds, at least within the inner
$\Reff$ or so, then most of the analyzed systems do require a
non-negligible fraction of dark matter ($\fDM$ greater than 10 per
cent) even when the contribution of the luminous component is
maximized \emph{ad hoc}, without any concerns for the plausibility of
the $\Mstar/L_{\sun, B}$ ratios involved in the rescaling. In other
words, what is shown by this `maximum bulge' approach is that the
mass-follows-light hypothesis is necessarily violated in the majority
of cases.

Adopting a physically motivated rescaling of the luminous
profile\,---\,based on the stellar masses inferred from SPS
models\,---\,leads to interesting results.

First of all, in agreement with previous work on the SLACS lenses
\citep{Koopmans2006, Bolton2008b, Auger2010, Barnabe2010}, we find a
significant amount of dark matter within the galaxy inner regions: in
the case of a Salpeter IMF, the average $\fDM$ is~31 per cent, with a
wide variation within individual objects, ranging from galaxies
consistent with having no dark matter, to systems where the baryonic
component accounts for less than half of the total mass already
at~$\re$. These values are broadly consistent with the results of
earlier dynamical studies of nearby early-type galaxies employing a
variety of different techniques and assumptions \citep{Gerhard2001,
  Cappellari2006, Thomas2007b, Weijmans2008, Tortora2009}, which find
dark matter fractions inside one half-light radius varying from~0 to
about~50 per cent of the total mass, with a typical average value
close to~30 per cent. If we prescribe, instead, an IMF lighter than
Salpeter\,---\,as advocated for example by \citealt{Cappellari2006}
based on their analysis of the SAURON ellipticals, which have lower
velocity dispersions than the SLACS lenses considered here\,---\,then
the dark component becomes the dominant one in the central regions for
almost all systems (average $\fDM$ is $61$ per cent with a Chabrier
IMF), in striking contrast with the traditional picture of early-type
galaxies. It is worth noting that similarly abundant dark matter
contents within the galaxy inner regions are commonly found by studies
of E/S0 galaxies that adopt a Chabrier-like IMF: \citet{Grillo2010b},
e.g., calculates a median value of~64 per cent (in projection) within
1~$\Reff$ for a large sample of $1.7 \times 10^{5}$ massive SDSS
early-type galaxies. \citet{Tortora2009}, applying spherical Jeans
models to a homogeneous sample of 335 local E/S0 systems, derive a
$\fDM \sim 60$ per cent. \citet{Weijmans2009T} performs a
sophisticated dynamical analysis of the nearby galaxy NGC\,2549 using
high-quality photometry and extended kinematic maps and finds,
likewise, a dark matter fraction of~65 per cent within the half-light
radius by adopting a \citet{Kroupa2001} IMF.

The second relevant result is the non-negligible positive correlation
between the derived dark matter fraction and the total mass enclosed
within~$\re$. This relation is in agreement with the results of the
\citet{Padmanabhan2004} study of a sample of about 30,000 SDSS
ellipticals, and with the findings of \citet{Tortora2009} and
\citet{Graves-Faber2010} at lower redshift. Conversely,
\citet{Grillo2010b}\,---\,by determining the stellar masses from SDSS
multicolor photometry and the total mass from a simple dynamical
relation, under the assumption that the galaxies can be described as
spherical and isothermal systems\,---\,comes to a different
conclusion, i.e.\ that the (projected) dark matter fraction within the
half-light radius remains almost constant for values of the total mass
between a few $10^{10}$ and $\sim 10^{12}$ $M_{\sun}$.

If the IMF is truly universal, then the trend for $\fDM$ implies a
genuine increase of the dark matter contribution with $\Mtot$, and for
the most massive ellipticals the stars represent a minority component,
even in the central regions (cf.\ Sect.~\ref{ssec:stellar-pop}).
However, an alternative explanation is possible if the IMF varies with
galaxy mass from a Chabrier/Kroupa-like functional form to a
Salpeter-like one (\citealt{Treu2010}; see also
\citealt{Auger2010imf}), or even to a steeper low-mass end slope, as
argued by \citet{vanDokkum-Conroy2010} for the most massive systems
(i.e., $\sigma \gtrsim 250$ km s$^{-1}$, which would include many of
the objects studied here).  We note, however, that the results from
the SLACS survey only imply a mass-to-light ratio that is higher than
expected for a Chabrier or Kroupa IMF. The excess mass can be provided
by low mass stars (as in the case of a Salpeter IMF, as advocated by,
e.g., \citealt{vanDokkum-Conroy2010}), or by high mass stars remnants,
in the form of neutron star and black holes \citep{Treu2010,
  Auger2010imf}.

Finally, in terms of dynamical structure, the SLACS lenses present no
surprises, and are found to be very similar to their local
counterparts as observed, e.g., by SAURON \citep{Emsellem2007,
  Cappellari2007}. More than two-thirds of the galaxies are
slow-rotating objects, as a consequence of the sample being skewed
towards the high velocity dispersion end. The five fast-rotators
largely coincide with the low mass tail of the sample, and all of them
are high angular momentum systems, with $\lamR$ well above $0.1$, and
very clearly identified as such when their $\vphi/\bar{\sigma}$
meridional plane maps are examined. The dark matter fraction of the
fast-rotators is consistent with the general trend, and they do not
appear as a different population in terms of $\fDM$, as suggested by
\citet{Cappellari2006}.

There is tentative evidence that fast-rotating galaxies are
systematically more flattened in the luminous distribution than in the
total density profile, despite the fact that these systems\,---\,as
discussed above\,---\,are among the least massive in the sample and
thus likely to be the ones where the baryonic component is most
important. This would then imply that these objects lie in dark haloes
whose central regions are significantly rounder than the light we
observe. Such a scenario appears to be in general agreement with the
results of numerical simulations \citep[e.g.][]{Debattista2008,
  Abadi2010}, which show that the assembly of massive high angular
momentum galaxies changes the dark halo into nearly axially symmetric,
slightly oblate systems over a wide radial range. The question of
whether there is an actual structural difference between the dark
haloes of slow and fast rotators, however, remains pending.


\section{Summary}
\label{sec:conclusions}

In this work we have conducted a combined, self-consistent
gravitational lensing and stellar dynamics analysis of the full sample
of SLACS lens systems with available VLT VIMOS integral-field
spectroscopic data, employing axially symmetric dynamical models
supported by two-integral DFs. This is integrated with the constraints
from stellar mass estimates derived from SPS analysis of multiband HST
imaging. The application of these modelling tools on this remarkable
data set has enabled us to investigate in detail the properties and
three-dimensional structure of the considered sample of sixteen
early-type galaxies in the redshift range $z = 0.08 - 0.35$. 

This study represents an improvement over previous joint lensing and
dynamics analyses (which adopt dynamical models based on Jeans
equations) and constitutes, in a sense, an extension beyond the local
Universe of SAURON-type studies of the inner regions of E/S0
galaxies. Although the detailed kinematic maps available for nearby
systems are clearly superior to the corresponding data sets currently
obtainable for objects at $z \gtrsim 0.1$, the additional information
derived from gravitational lensing provides an invaluable asset for
the study of distant galaxies, which allows us to put robust
constraints on their mass distribution and amount of dark matter
within the probed regions.

The main conclusions from this analysis are summarized as follows.

\begin{enumerate}

\item The total mass density distribution of all the sample galaxies
  is well described, within the inner regions probed by the data sets
  (of order one half-light radius), by an axially symmetric power-law
  model. The average logarithmic slope is slightly super-isothermal,
  with $\mslope = 2.074^{+0.043}_{-0.041}$; there is an intrinsic
  scatter $\sigmams = 0.144^{+0.055}_{-0.014}$ (corresponding to less
  than~10 per cent) around this value.

\item We find that the lens galaxies have a fairly round total mass
  distribution: the average axial ratio is $\langle q \rangle = 0.77
  \pm 0.04$ and\,---\,while there are significant differences between
  individual systems, with an intrinsic scatter of nearly~20 per
  cent\,---\,all objects, barring one, are rounder than $q =
  0.6$. Most galaxies are about as flattened in the total distribution
  as they are in the luminous one.

\item The \emph{lower limit} for the fraction of dark over total mass
  within the three-dimensional radius~$\re$ (calculated with the
  maximum bulge approach under the assumption of constant stellar
  mass-to-light ratio inside the considered region) varies
  significantly between individual systems from nearly zero to
  almost~50 per cent, with an average value $\fDM = 16$ per cent
  (median value: $\fDM = 12$ per cent). The $\Mstar/L_{\sun, B}$
  ratios corresponding to these maximal light profiles range from~3 to
  almost~10.

\item When the normalization of the luminous profile is set using
  stellar mass determinations obtained from SPS models, under the
  assumption of a universal IMF of the Salpeter type, we derive an
  average dark matter fraction $f^{\mathrm{Salp}}_{\mathrm{DM}} = 31$
  per cent inside $\re$ (median value:
  $f^{\mathrm{Salp}}_{\mathrm{DM}} = 37$ per cent).  None of the
  analyzed galaxies has a luminous mass distribution strongly at odds
  with the assumption of a Salpeter IMF (the luminous profile does
  exceed unphysically the circularized total mass profile for two
  systems, but in both cases the discrepancy is just a few per cent
  above the 1-$\sigma$ uncertainty).

  If one adopts a Chabrier IMF the recovered dark matter fraction
  becomes much higher, i.e. $f^{\mathrm{Chab}}_{\mathrm{DM}} = 61$ per
  cent (median value: $f^{\mathrm{Chab}}_{\mathrm{DM}} = 64$ per
  cent), which would imply that the stars represent a minority mass
  component even in the inner regions of early-type galaxies.

\item If the stellar masses are determined in this way, i.e. assuming
  a universal IMF, then the derived $\fDM$ shows a clear correlation
  (\mbox{3-sigma}) with $\Mtot$, in the sense that the dark matter
  contribution increases with the total mass of the galaxy,
  becoming\,---\,even in the case of Salpeter IMF\,---\,the dominant
  component within~$\re$ for the most massive ellipticals (i.e. the
  systems with $\log [\Mtot/M_{\sun}] \gtrsim 11.5$). If the IMF is
  not universal, than a progressive steepening of its profile with the
  galaxy total mass could explain, at least partially, the observed
  trend.

\item The SLACS lenses can be divided into two dynamically distinct
  groups \emph{based on the value of their specific angular momentum
    parameter}~$\jz$ (which we can introduce based on our
  three-dimensional models): about two-thirds of the galaxies are
  dominated by random motions and show little large-scale ordered
  rotation ($0 \lesssim \jz \lesssim 0.3$), whereas the remaining
  five, all of which are among the least massive systems in the
  sample, are high angular momentum objects ($\jz \simeq 0.5$ and
  above). These two groups correspond to the standard classes of slow
  and fast rotators, as revealed by several other properties
  (e.g.\ mass range, flattening, the Emsellem parameter $\lamR$), and
  in particular by the characteristics of the respective
  $\vphi/\bar{\sigma}$ maps, which illustrate the local distribution
  of ordered-to-random motions ratios in the meridional plane. For
  slow rotators, the spread in $\jz$ is much larger than the spread in
  $\lamR$, suggesting that the specific angular momentum parameter is
  a better discriminator of the intrinsic properties of these systems.

\item Our analysis shows that the SLACS lens systems are overall
  analogous to their local Universe counterparts in terms of density
  profile and global structural and dynamical properties. This in turn
  indicates that massive early-type galaxies have experienced at most
  limited structural evolution, at least as far as their inner regions
  are concerned, within the last four billion years.

\end{enumerate}


\section*{Acknowledgments}

We are grateful to Matt Auger, Rapha\"{e}l Gavazzi, Phil Marshall,
Leonidas Moustakas and Simona Vegetti for their substantial
contributions to the SLACS project.  We thank Roger Blandford, Brendon
Brewer, Claudio Grillo and Sherry Suyu for useful discussion. We also
thank the anonymous referee for providing useful comments.
M.B. acknowledges support from the Department of Energy contract
DE-AC02-76SF00515.  L.K. is supported through an NWO-VIDI program
subsidy (project number 639.042.505).  T.T. acknowledges support from
the NSF through CAREER award NSF-0642621, and from the Packard
Foundation through a Packard Fellowship. Support for programs \#10174,
\#10494, \#10588, \#10798 and \#11202 was provided by NASA through a
grant from the Space Telescope Science Institute, which is operated by
the Association of Universities for Research in Astronomy, Inc., under
NASA contract NAS 5-26555.


\bibliography{my_bibliography}

\label{lastpage}

\clearpage
\newpage


\appendix

\section{Best model reconstructions for the sample galaxies}
\label{app:models}

In this Appendix we present the data sets (i.e., galaxy-subtracted
lensed image, surface brightness maps and kinematics maps), the
corresponding best model reconstructed observables, and the residuals
for eleven systems in the considered sample. The corresponding panels
for the remaining five galaxies (namely J0037, J0216, J0912,
J0959 and J1627) were presented in a previous publication in this
series \citep{Barnabe2009} and therefore will not be duplicated
here. The results for galaxy J2321 were first described in
\citep{Czoske2008}; however, since an improved analysis of the J2321
kinematic data set has become available, we have re-analyzed that
system and we have included it in the selection presented here.

A thorough description of the data sets will be provided in a
forthcoming publication (Czoske et al. 2011, in preparation).


\begin{figure}
  \centering
  \resizebox{0.99\hsize}{!}{\includegraphics[angle=-90]
            {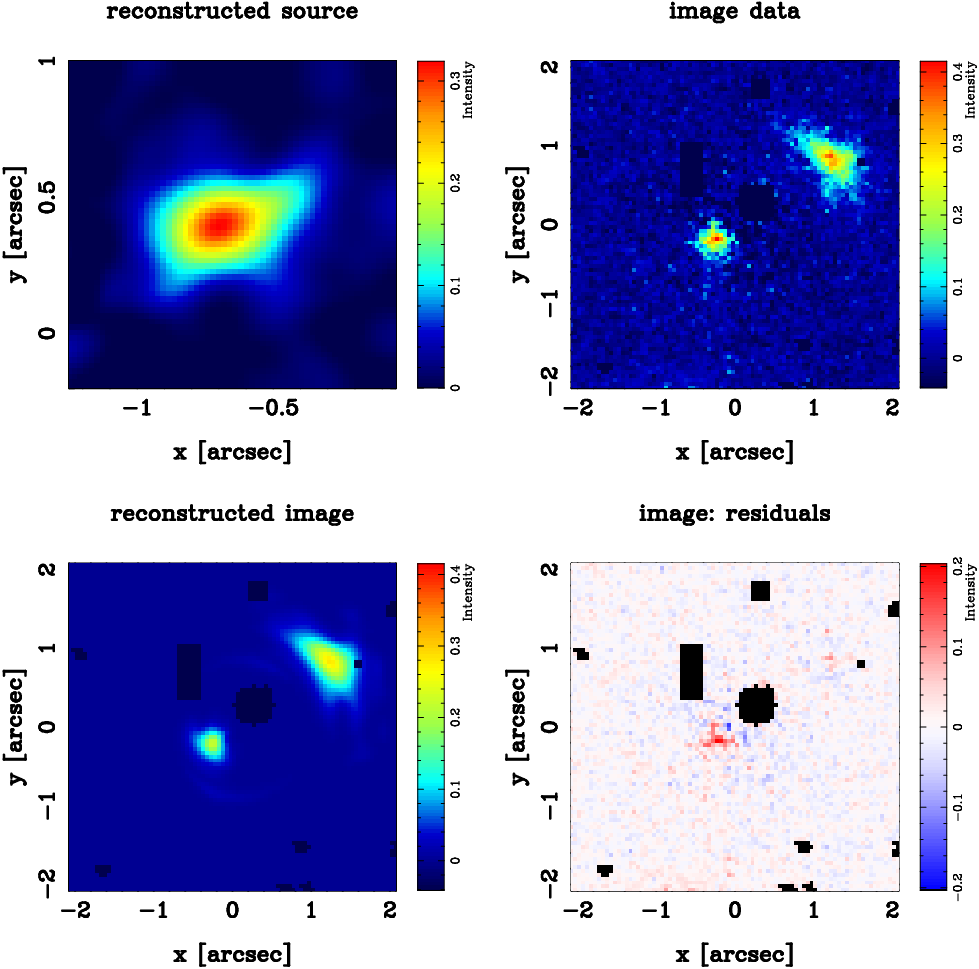}}
  \caption{Best model lens image reconstruction for the system
    SDSS\,J0935. From the top left-hand to bottom right-hand panel:
    reconstructed source model; \textit{HST}/ACS data showing the lens
    image after subtraction of the lens galaxy; lens image
    reconstruction; residuals. In the panels, North is up and East is
    to the right.}
  \label{fig:J0935_LEN}

  \resizebox{0.99\hsize}{!}{\includegraphics[angle=-90]
            {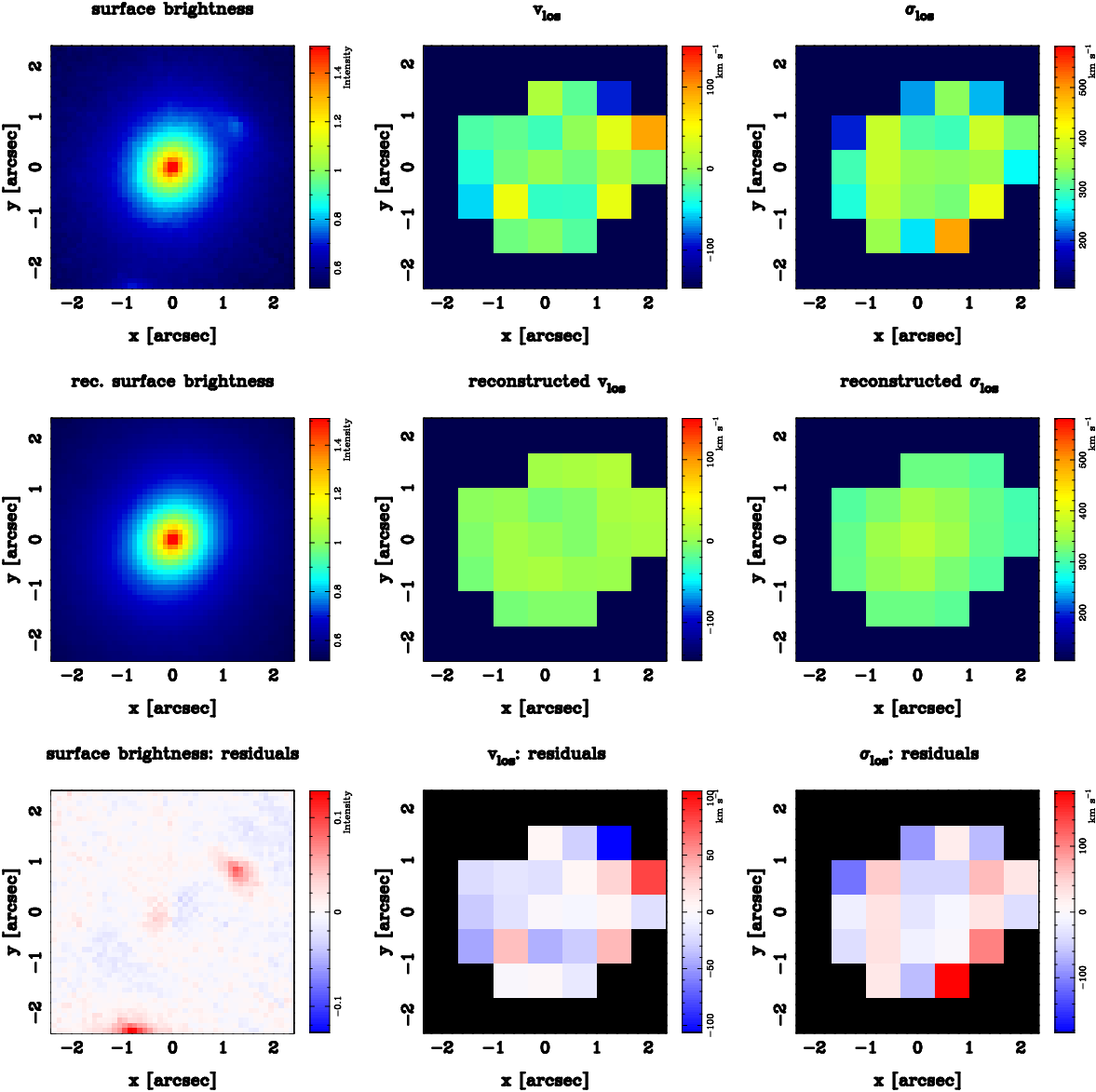}}
  \caption{Best dynamical model for the galaxy SDSS\,J0935. First row:
    observed surface brightness distribution, projected line-of-sight
    velocity and line-of-sight velocity dispersion. Second row:
    corresponding reconstructed quantities for the best model. Third
    row: residuals. In the panels, North is up and East is
    to the right.}
  \label{fig:J0935_DYN}
\end{figure}

\begin{figure}
  \centering
  \resizebox{0.99\hsize}{!}{\includegraphics[angle=-90]
            {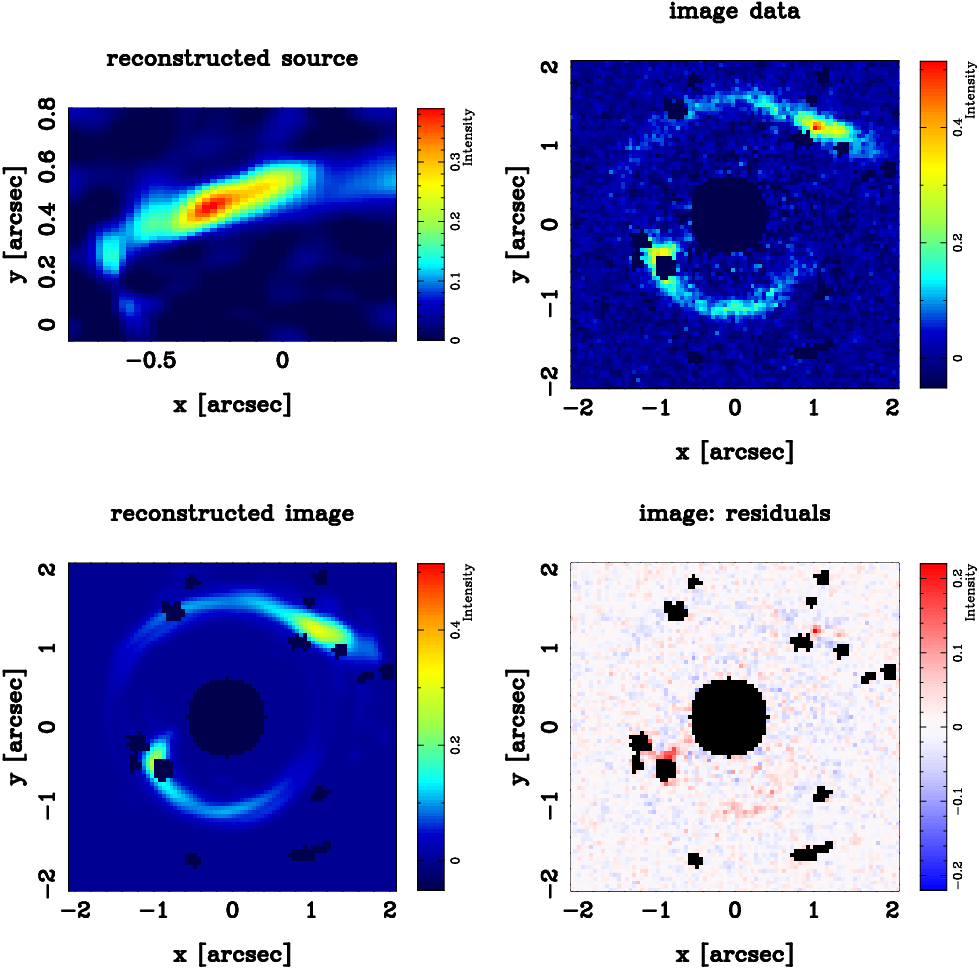}}
  \caption{Best model lens image reconstruction for the galaxy
    SDSS\,J1204. Panels meaning as in Fig.~\ref{fig:J0935_LEN}.}
  \label{fig:J1204_LEN}

  \resizebox{0.99\hsize}{!}{\includegraphics[angle=-90]
            {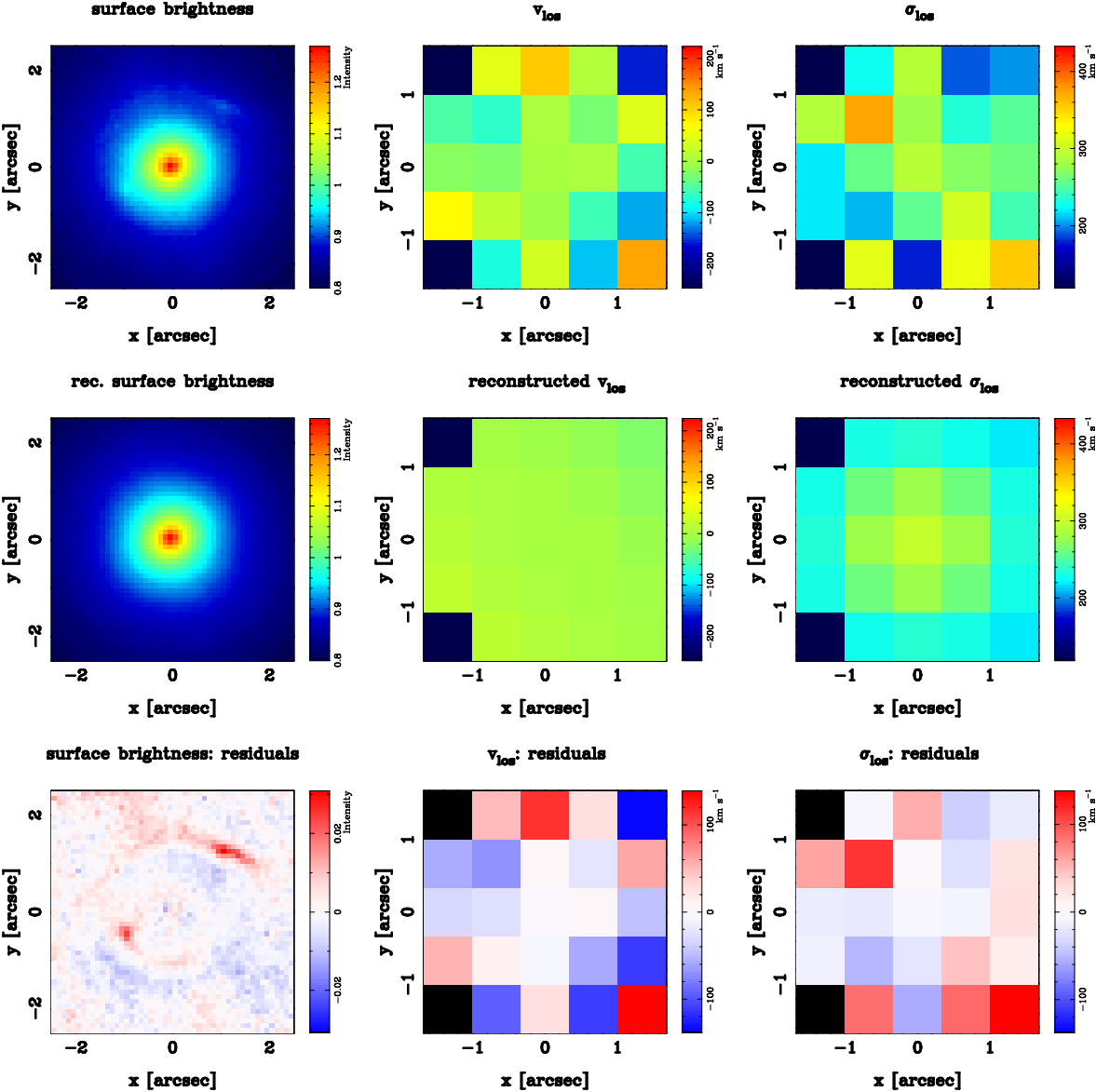}}
  \caption{Best dynamical model for the galaxy SDSS\,J1204. Panels
    meaning as in Fig.~\ref{fig:J0935_DYN}.}
  \label{fig:J1204_DYN}
\end{figure}

\begin{figure}
  \centering
  \resizebox{0.99\hsize}{!}{\includegraphics[angle=-90]
            {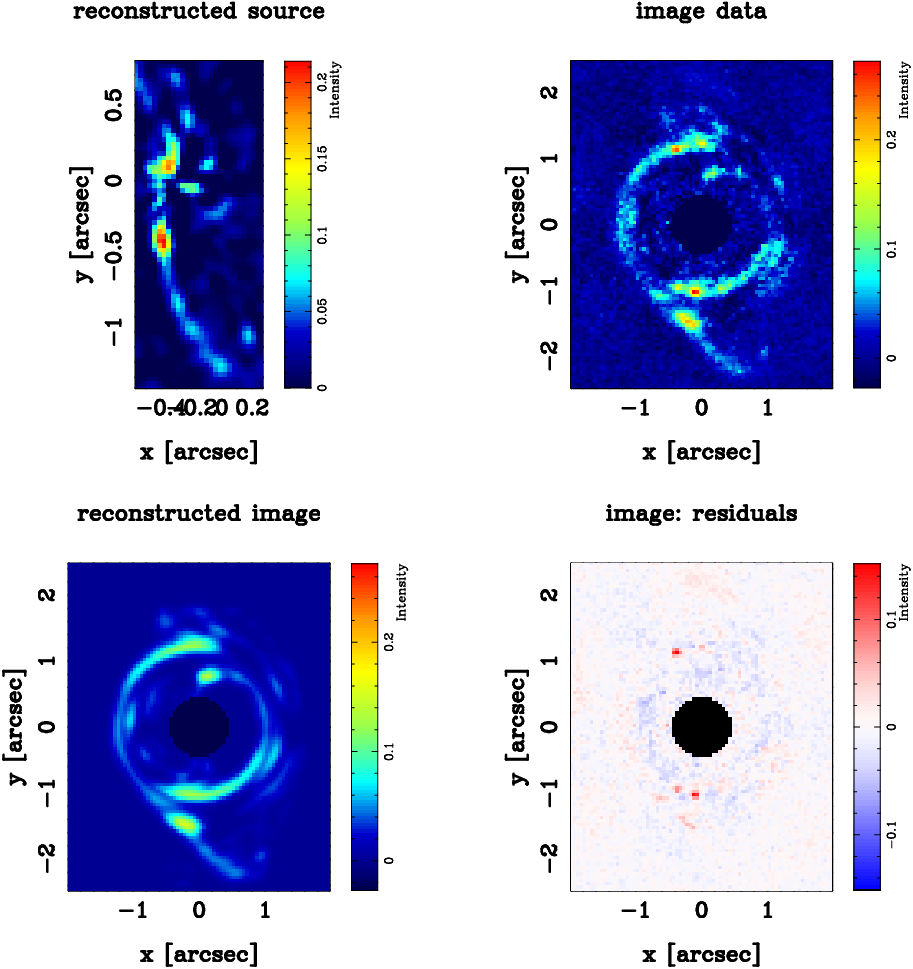}}
  \caption{Best model lens image reconstruction for the galaxy
    SDSS\,J1250. Panels meaning as in Fig.~\ref{fig:J0935_LEN}.}
  \label{fig:J1250_LEN}

  \resizebox{0.99\hsize}{!}{\includegraphics[angle=-90]
            {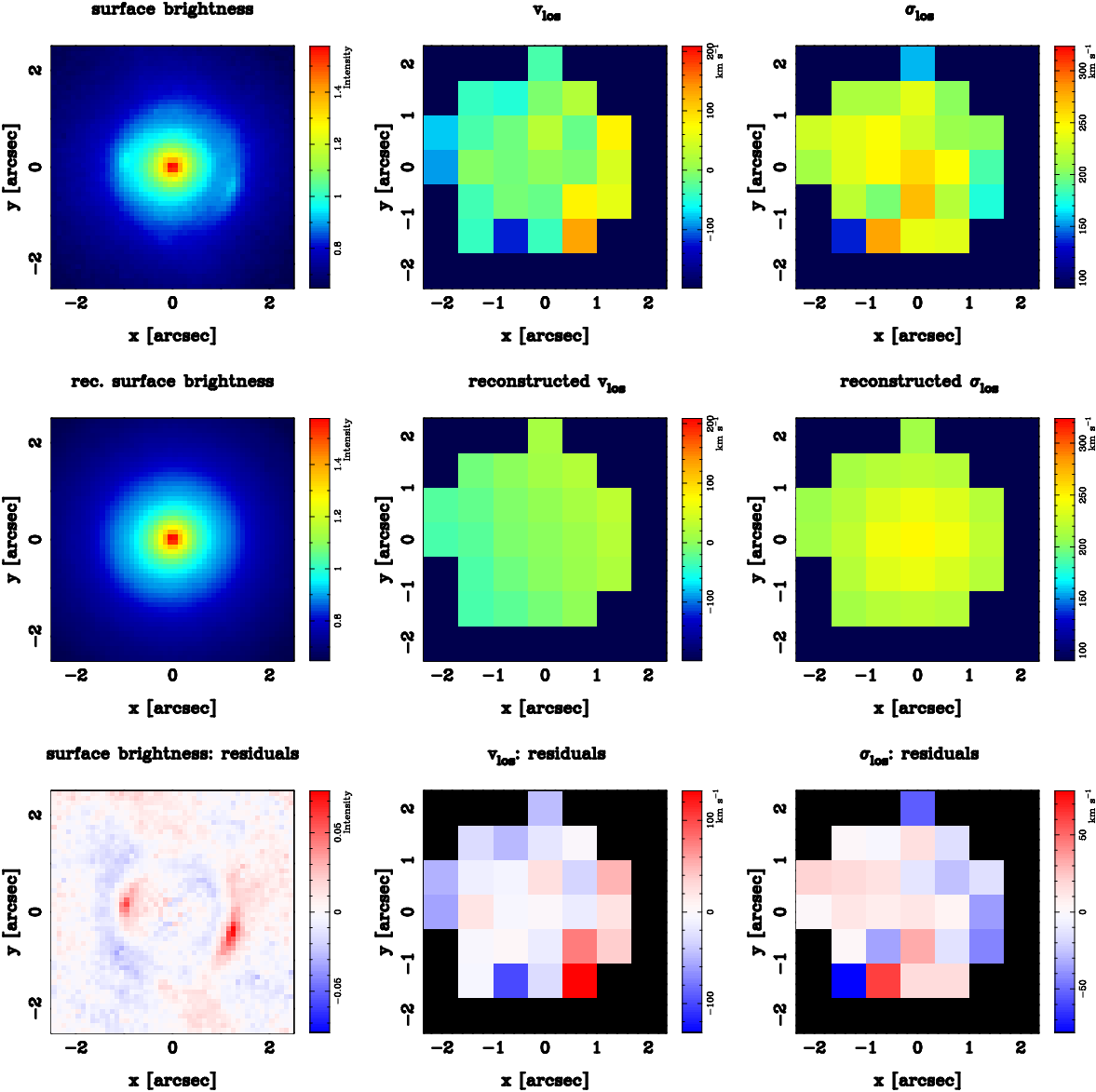}}
  \caption{Best dynamical model for the galaxy SDSS\,J1250. Panels
    meaning as in Fig.~\ref{fig:J0935_DYN}.}
  \label{fig:J1250_DYN}
\end{figure}

\begin{figure}
  \centering
  \resizebox{0.99\hsize}{!}{\includegraphics[angle=-90]
            {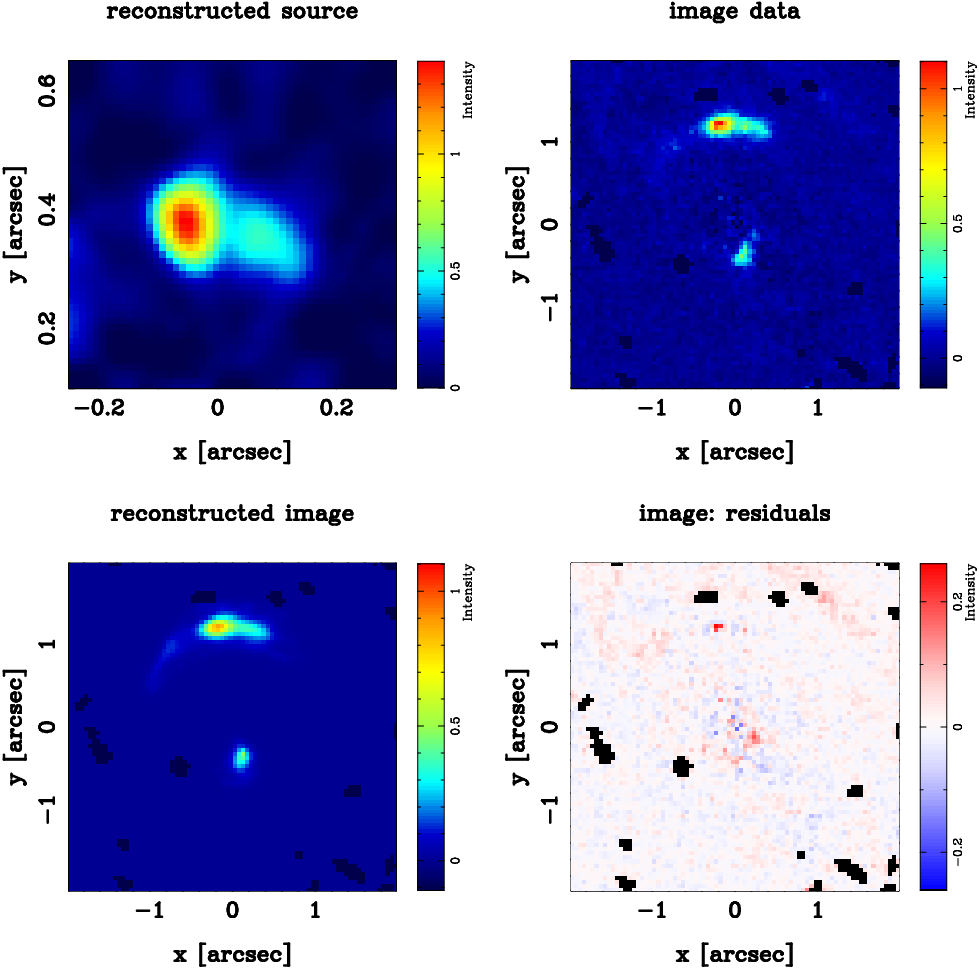}}
  \caption{Best model lens image reconstruction for the galaxy
    SDSS\,J1251. Panels meaning as in Fig.~\ref{fig:J0935_LEN}.}
  \label{fig:J1251_LEN}

  \resizebox{0.99\hsize}{!}{\includegraphics[angle=-90]
            {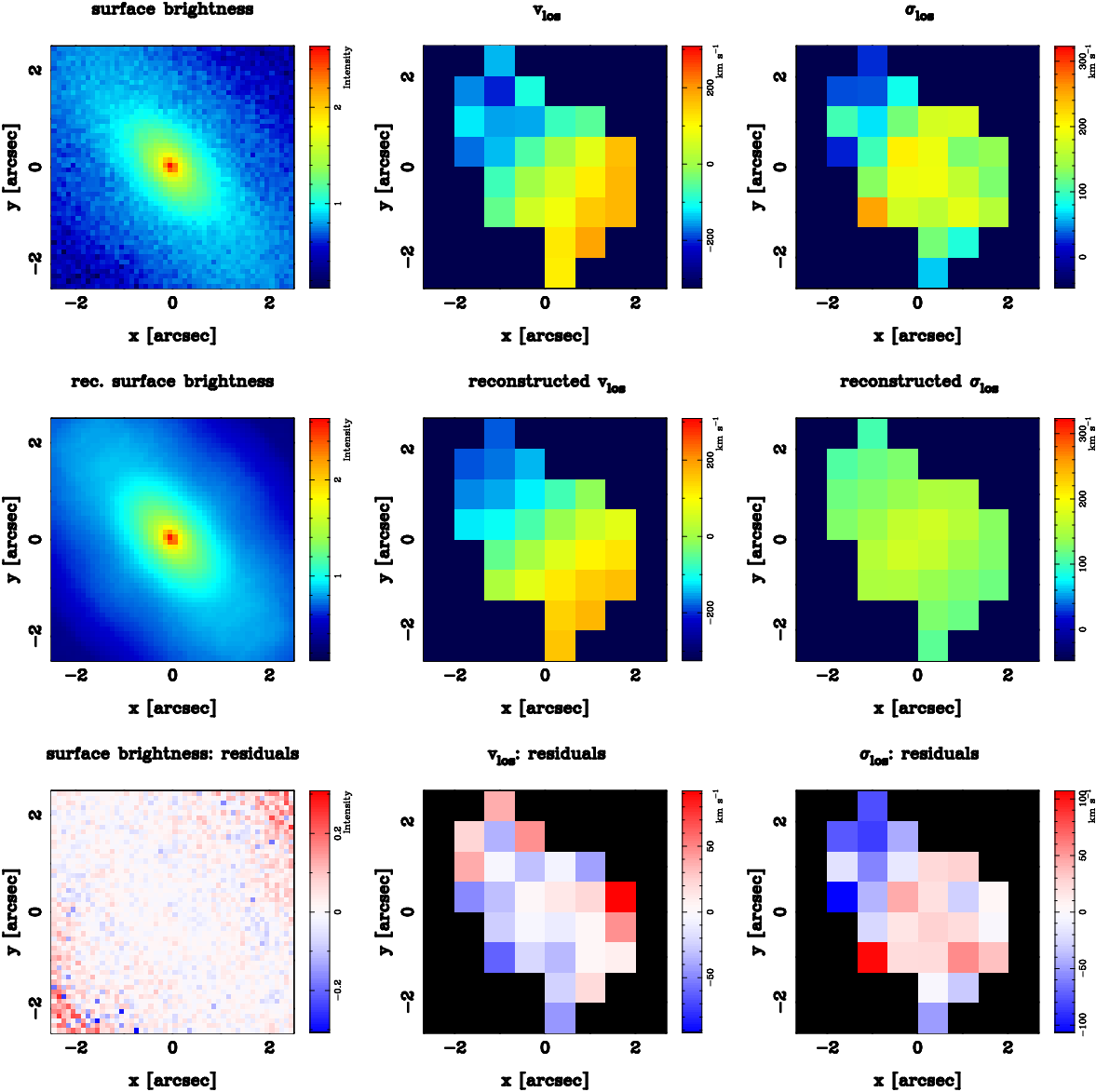}}
  \caption{Best dynamical model for the galaxy SDSS\,J1251. Panels
    meaning as in Fig.~\ref{fig:J0935_DYN}.}
  \label{fig:J1251_DYN}
\end{figure}

\begin{figure}
  \centering
  \resizebox{0.99\hsize}{!}{\includegraphics[angle=-90]
            {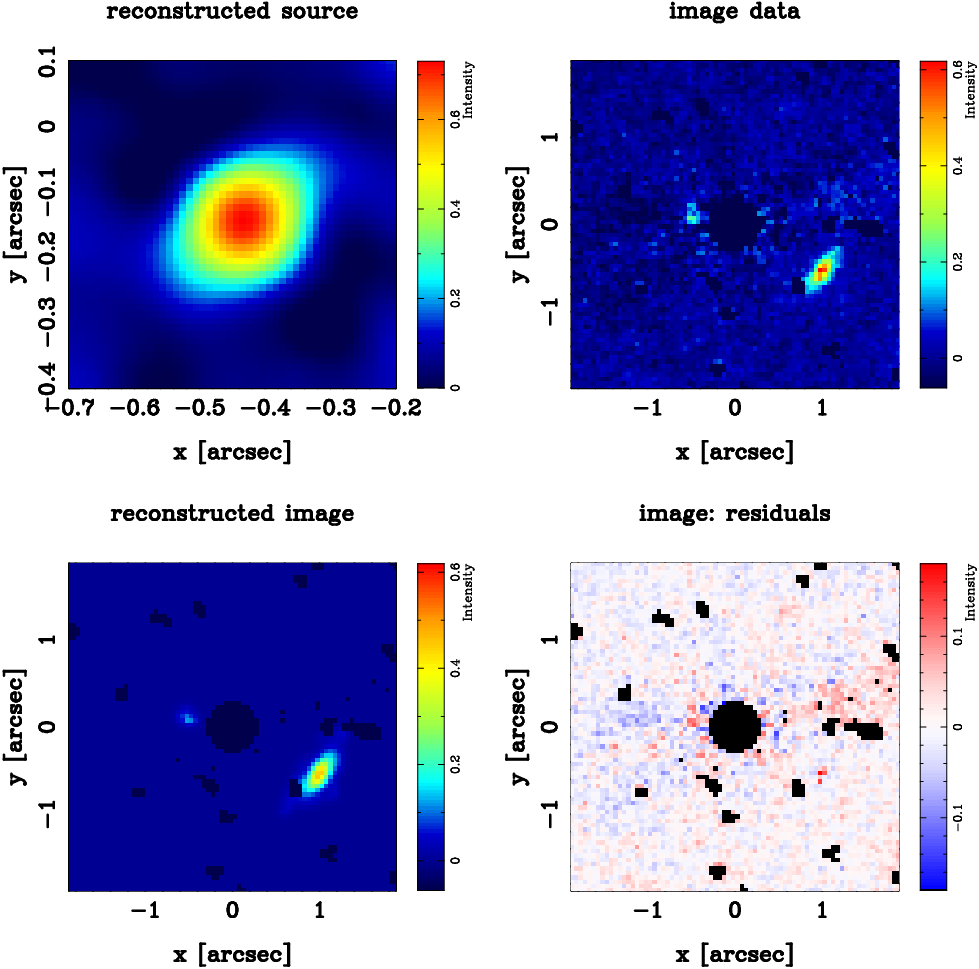}}
  \caption{Best model lens image reconstruction for the galaxy
    SDSS\,J1330. Panels meaning as in Fig.~\ref{fig:J0935_LEN}.}
  \label{fig:J1330_LEN}

  \resizebox{0.99\hsize}{!}{\includegraphics[angle=-90]
            {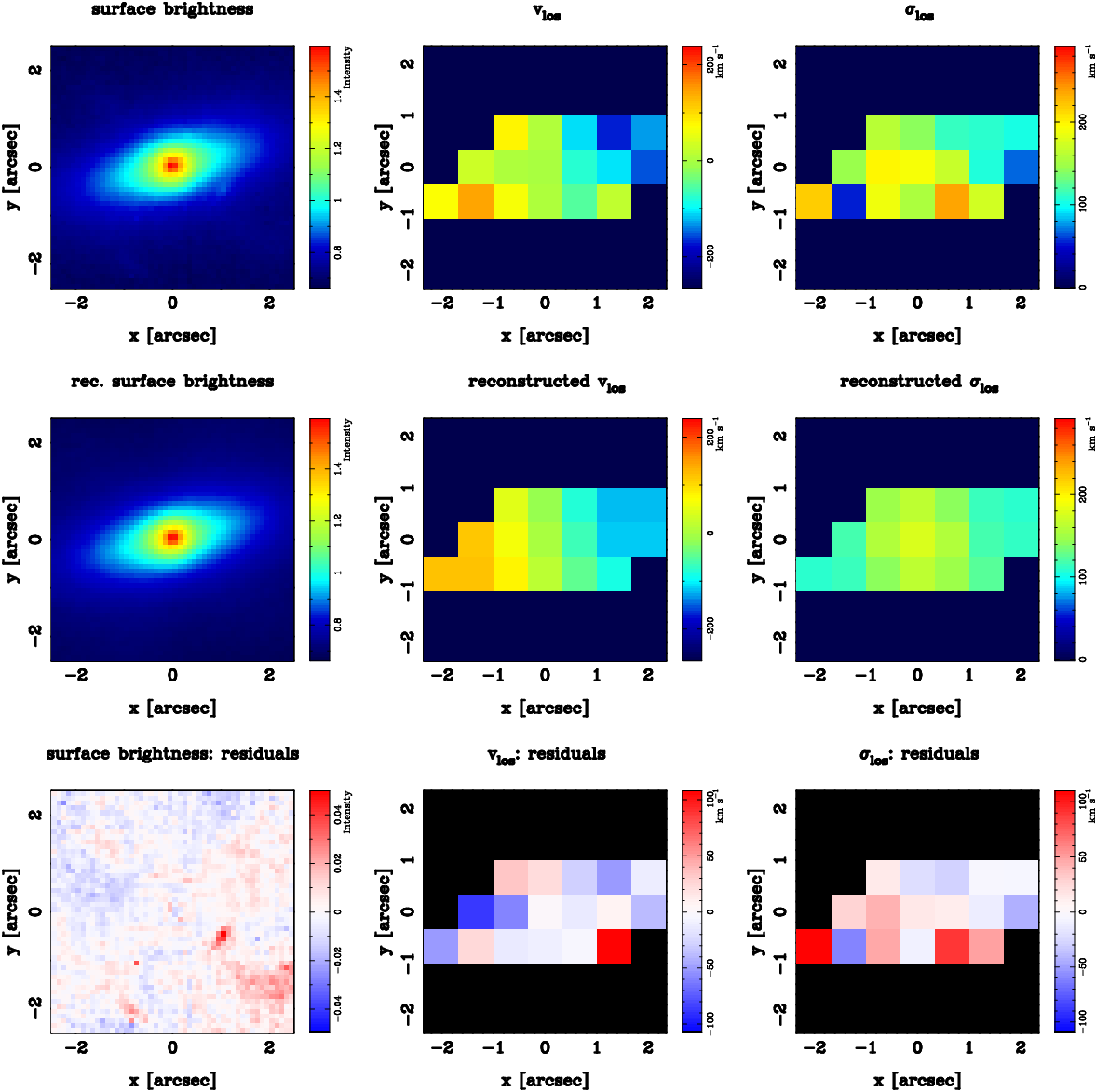}}
  \caption{Best dynamical model for the galaxy SDSS\,J1330. Panels
    meaning as in Fig.~\ref{fig:J0935_DYN}.}
  \label{fig:J1330_DYN}
\end{figure}

\begin{figure}
  \centering
  \resizebox{0.99\hsize}{!}{\includegraphics[angle=-90]
            {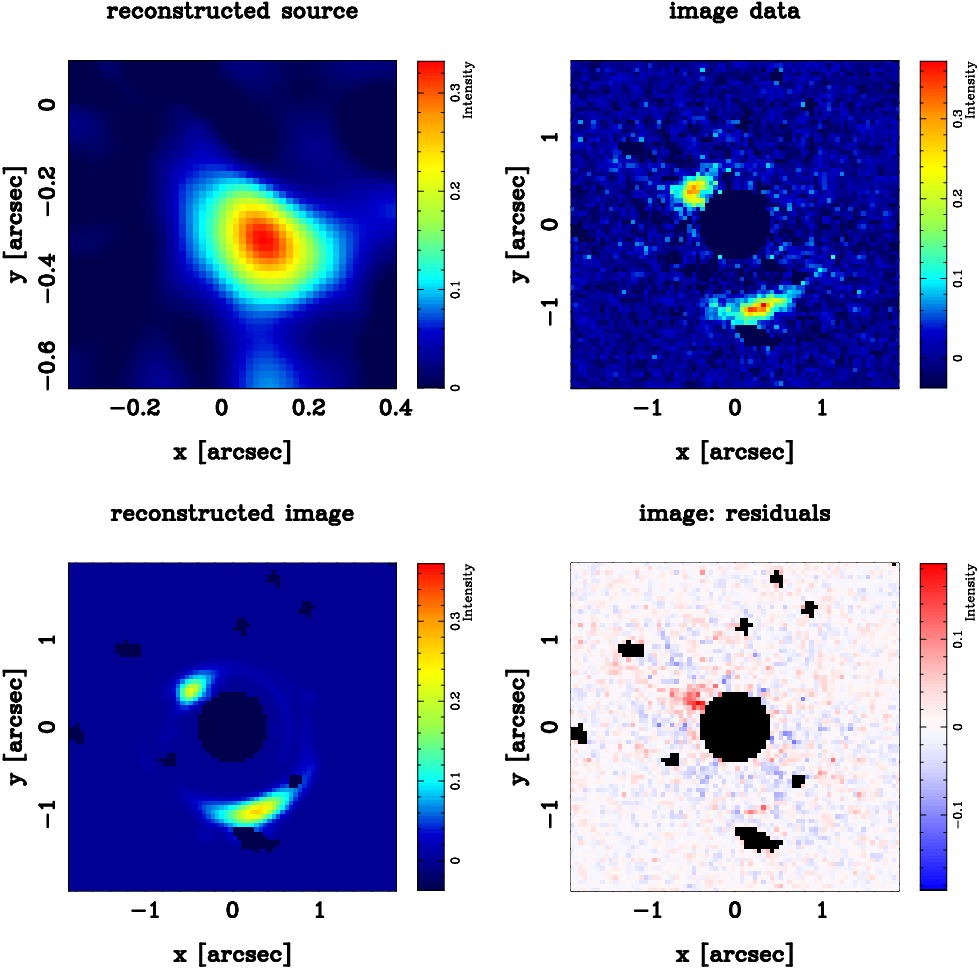}}
  \caption{Best model lens image reconstruction for the galaxy
    SDSS\,J1443. Panels meaning as in Fig.~\ref{fig:J0935_LEN}.}
  \label{fig:J1443_LEN}

  \resizebox{0.99\hsize}{!}{\includegraphics[angle=-90]
            {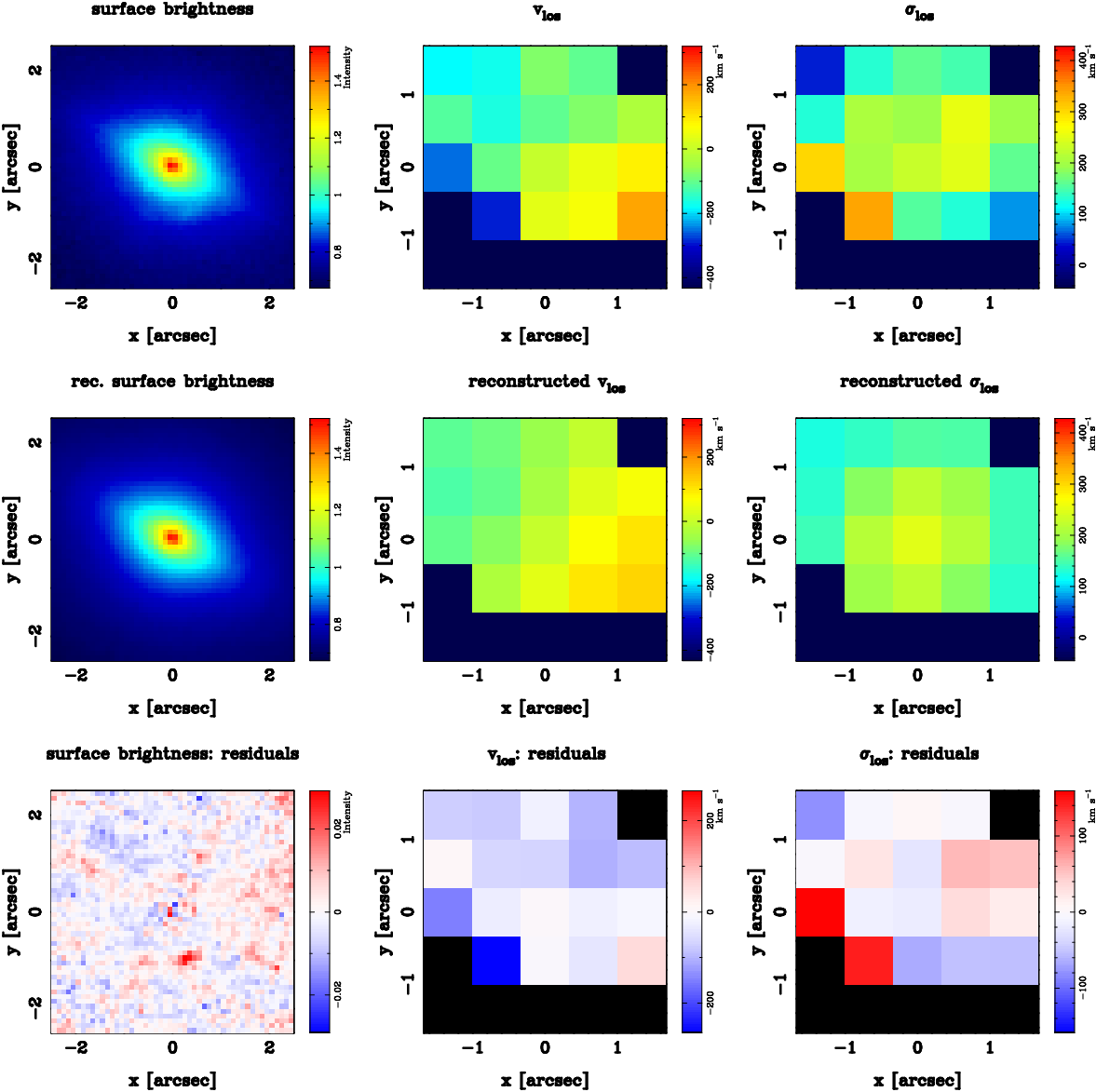}}
  \caption{Best dynamical model for the galaxy SDSS\,J1443. Panels
    meaning as in Fig.~\ref{fig:J0935_DYN}.}
  \label{fig:J1443_DYN}
\end{figure}

\begin{figure}
  \centering
  \resizebox{0.99\hsize}{!}{\includegraphics[angle=-90]
            {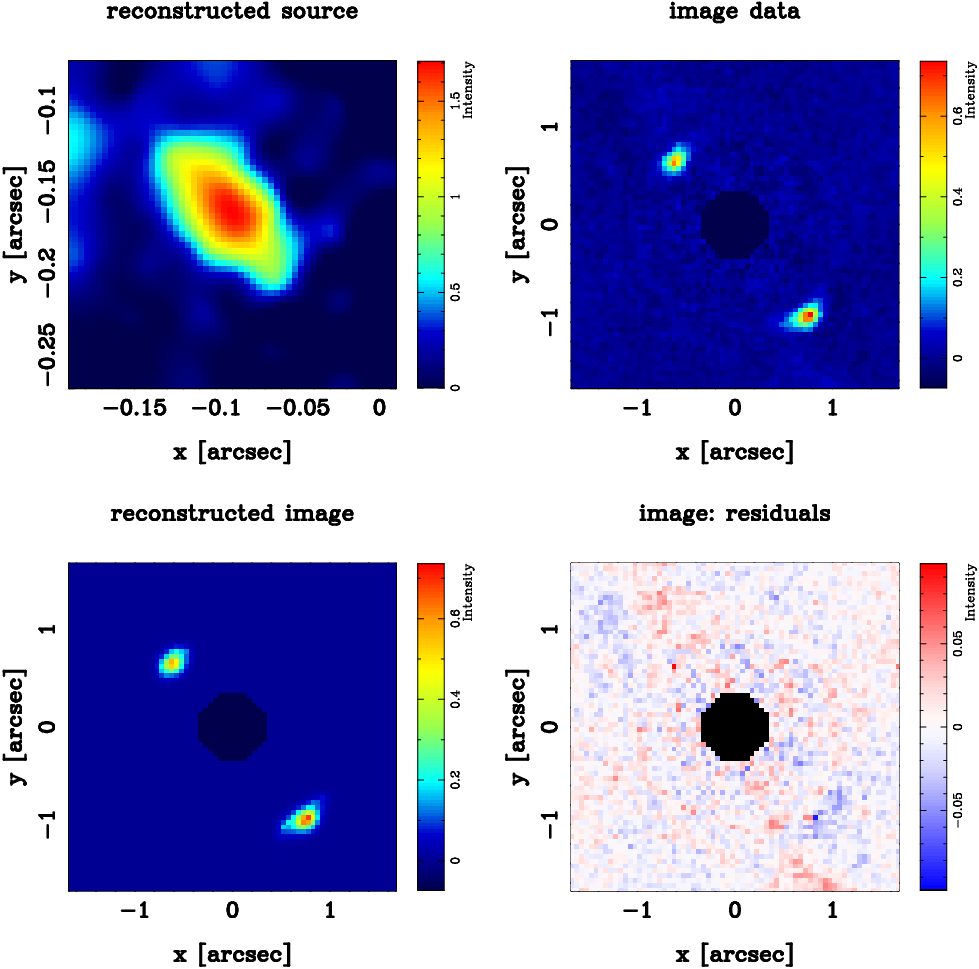}}
  \caption{Best model lens image reconstruction for the galaxy
    SDSS\,J1451. Panels meaning as in Fig.~\ref{fig:J0935_LEN}.}
  \label{fig:J1451_LEN}

  \resizebox{0.99\hsize}{!}{\includegraphics[angle=-90]
            {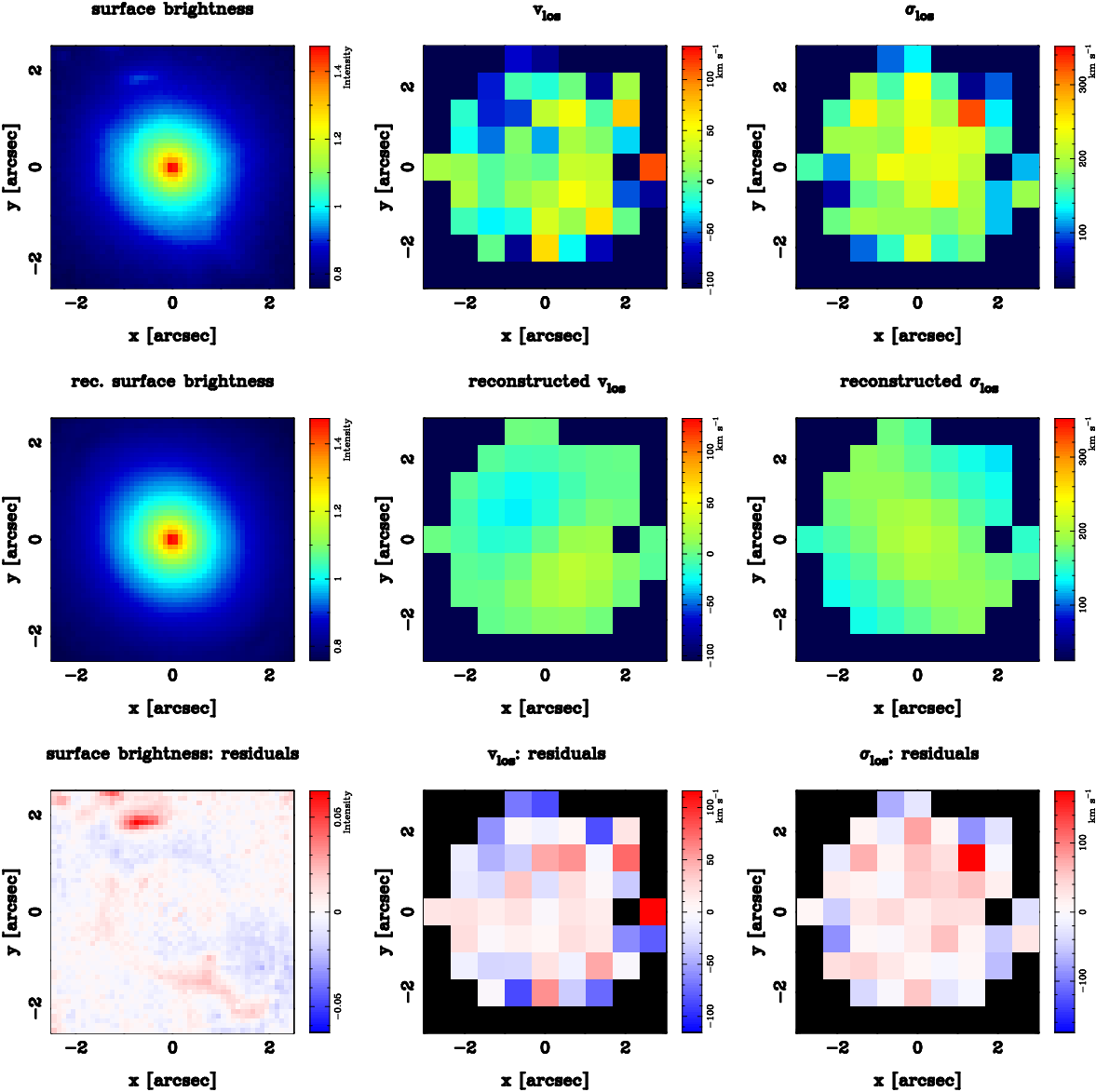}}
  \caption{Best dynamical model for the galaxy SDSS\,J1451. Panels
    meaning as in Fig.~\ref{fig:J0935_DYN}.}
  \label{fig:J1451_DYN}
\end{figure}

\begin{figure}
  \centering
  \resizebox{0.99\hsize}{!}{\includegraphics[angle=-90]
            {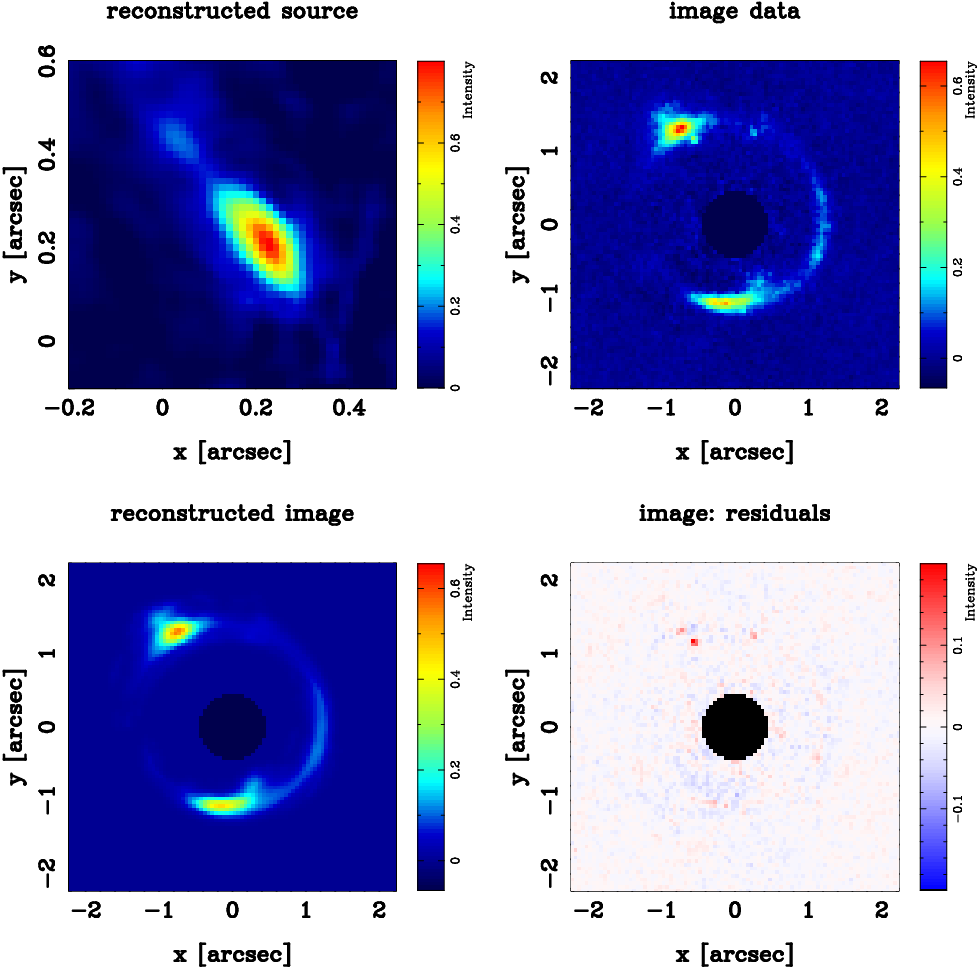}}
  \caption{Best model lens image reconstruction for the galaxy
    SDSS\,J2238. Panels meaning as in Fig.~\ref{fig:J0935_LEN}.}
  \label{fig:J2238_LEN}

  \resizebox{0.99\hsize}{!}{\includegraphics[angle=-90]
            {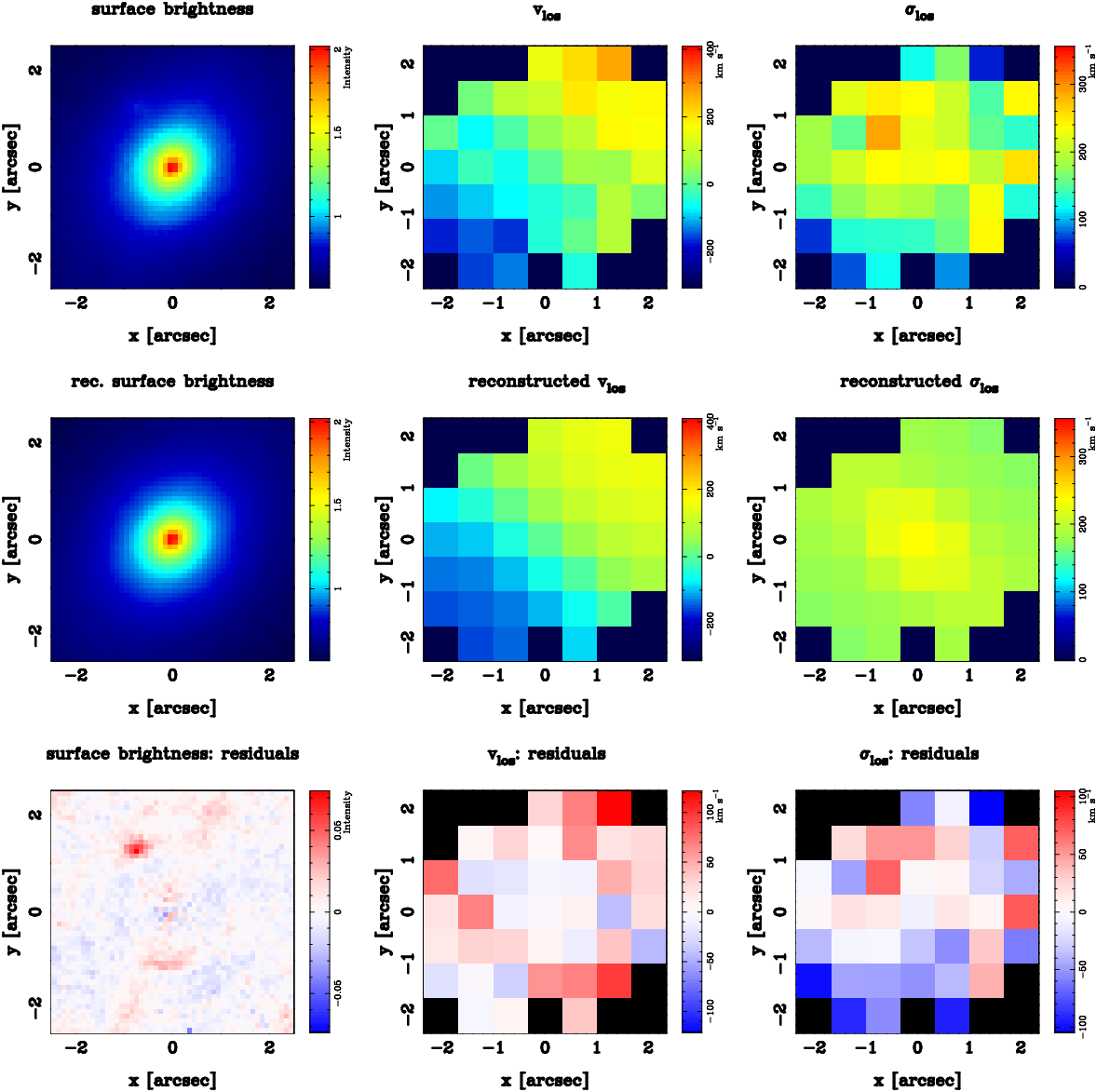}}
  \caption{Best dynamical model for the galaxy SDSS\,J2238. Panels
    meaning as in Fig.~\ref{fig:J0935_DYN}.}
  \label{fig:J2238_DYN}
\end{figure}

\begin{figure}
  \centering
  \resizebox{0.99\hsize}{!}{\includegraphics[angle=-90]
            {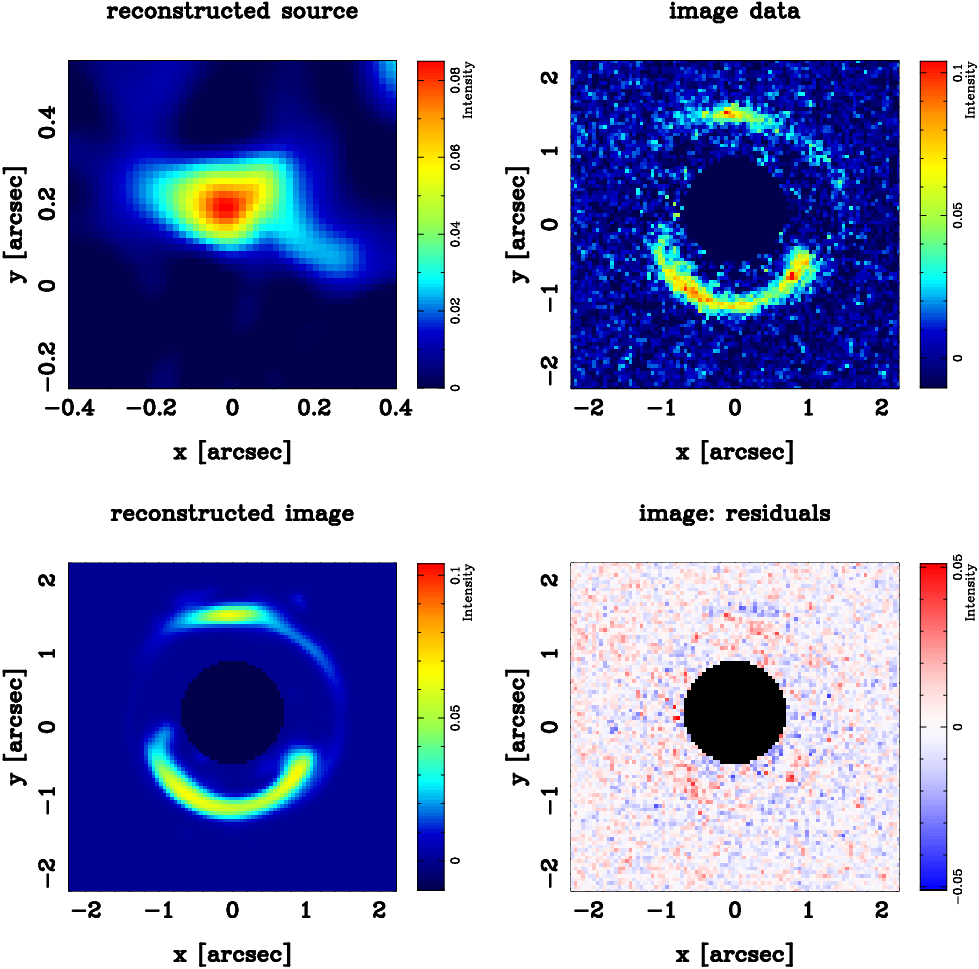}}
  \caption{Best model lens image reconstruction for the galaxy
    SDSS\,J2300. Panels meaning as in Fig.~\ref{fig:J0935_LEN}.}
  \label{fig:J2300_LEN}

  \resizebox{0.99\hsize}{!}{\includegraphics[angle=-90]
            {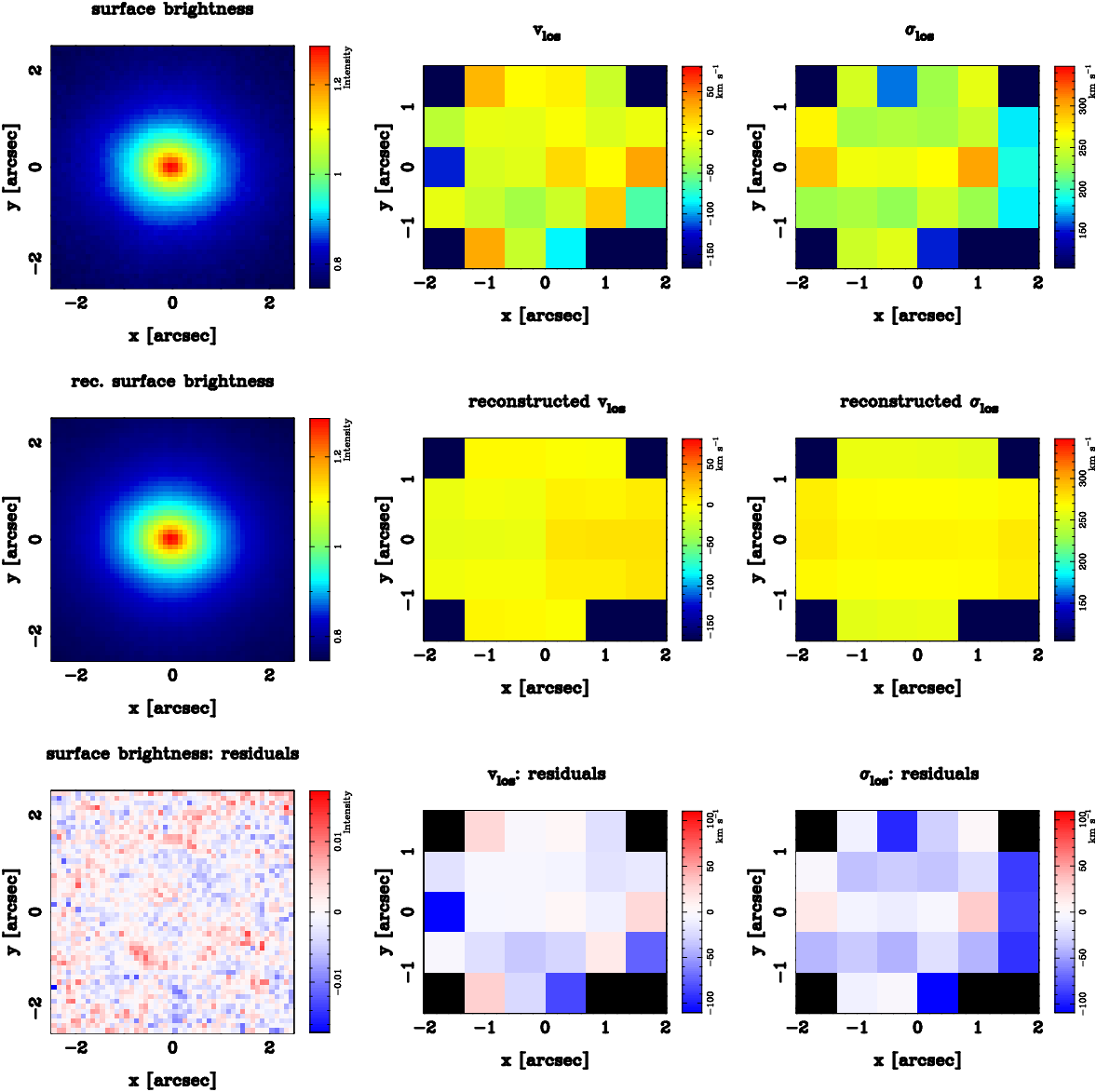}}
  \caption{Best dynamical model for the galaxy SDSS\,J2300. Panels
    meaning as in Fig.~\ref{fig:J0935_DYN}.}
  \label{fig:J2300_DYN}
\end{figure}

\begin{figure}
  \centering
  \resizebox{0.99\hsize}{!}{\includegraphics[angle=-90]
            {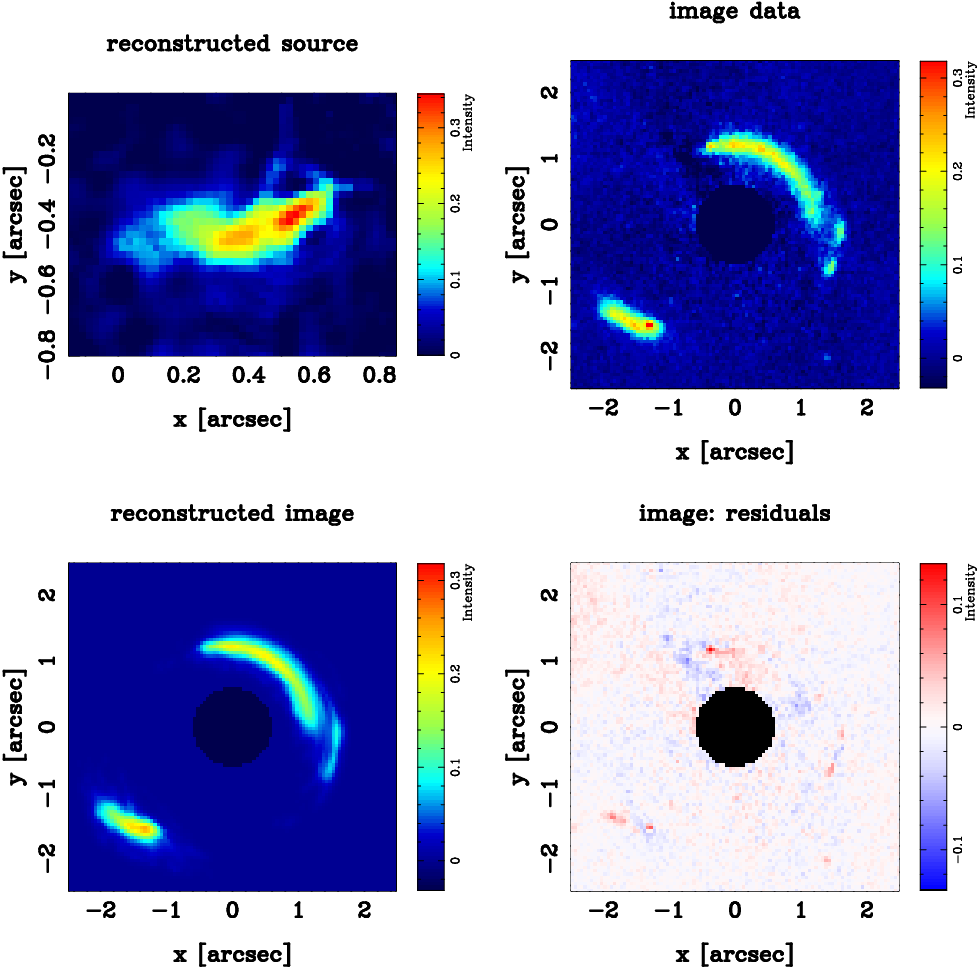}}
  \caption{Best model lens image reconstruction for the galaxy
    SDSS\,J2303. Panels meaning as in Fig.~\ref{fig:J0935_LEN}.}
  \label{fig:J2303_LEN}

  \resizebox{0.99\hsize}{!}{\includegraphics[angle=-90]
            {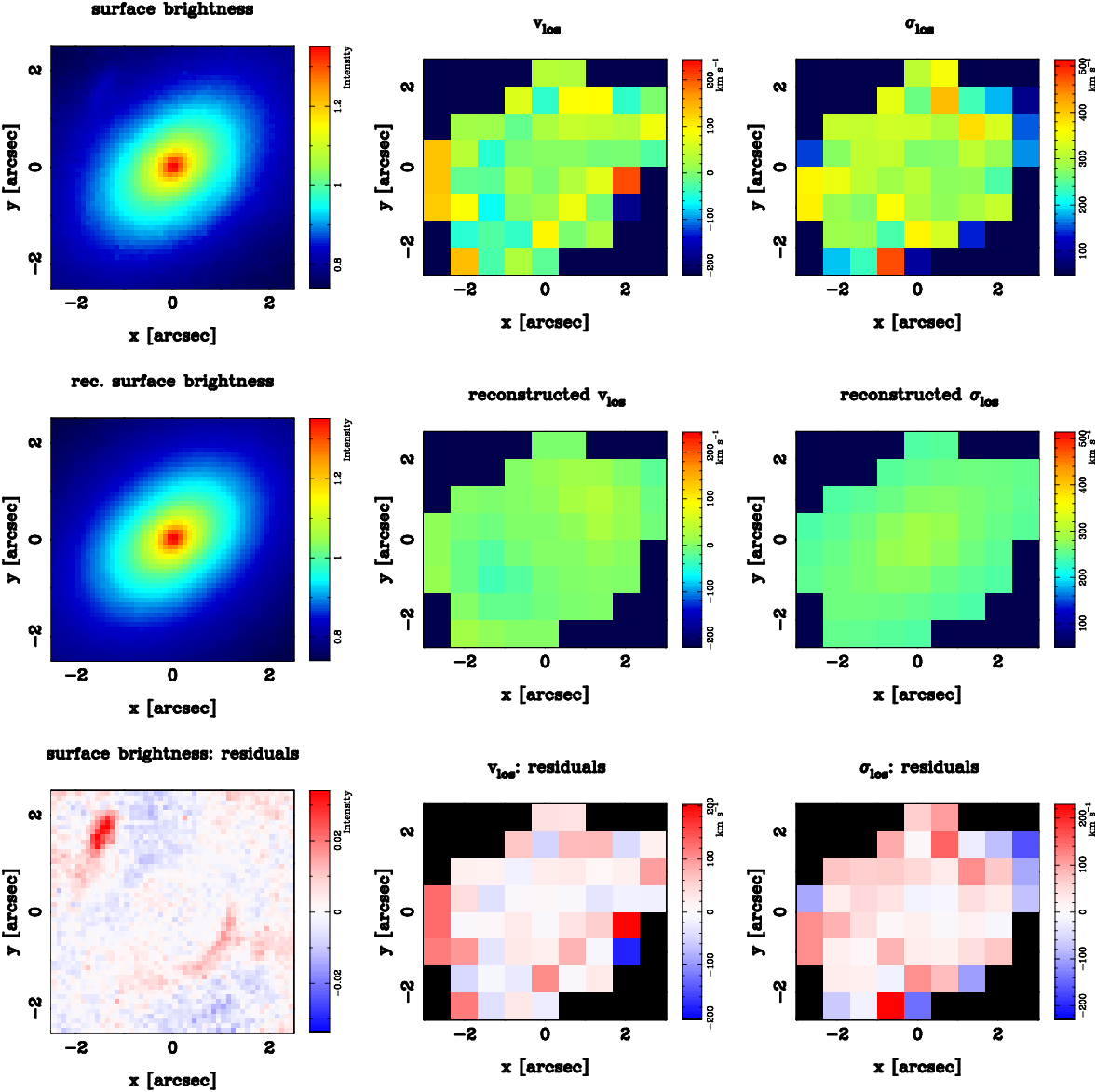}}
  \caption{Best dynamical model for the galaxy SDSS\,J2303. Panels
    meaning as in Fig.~\ref{fig:J0935_DYN}.}
  \label{fig:J2303_DYN}
\end{figure}

\begin{figure}
  \centering
  \resizebox{0.99\hsize}{!}{\includegraphics[angle=-90]
            {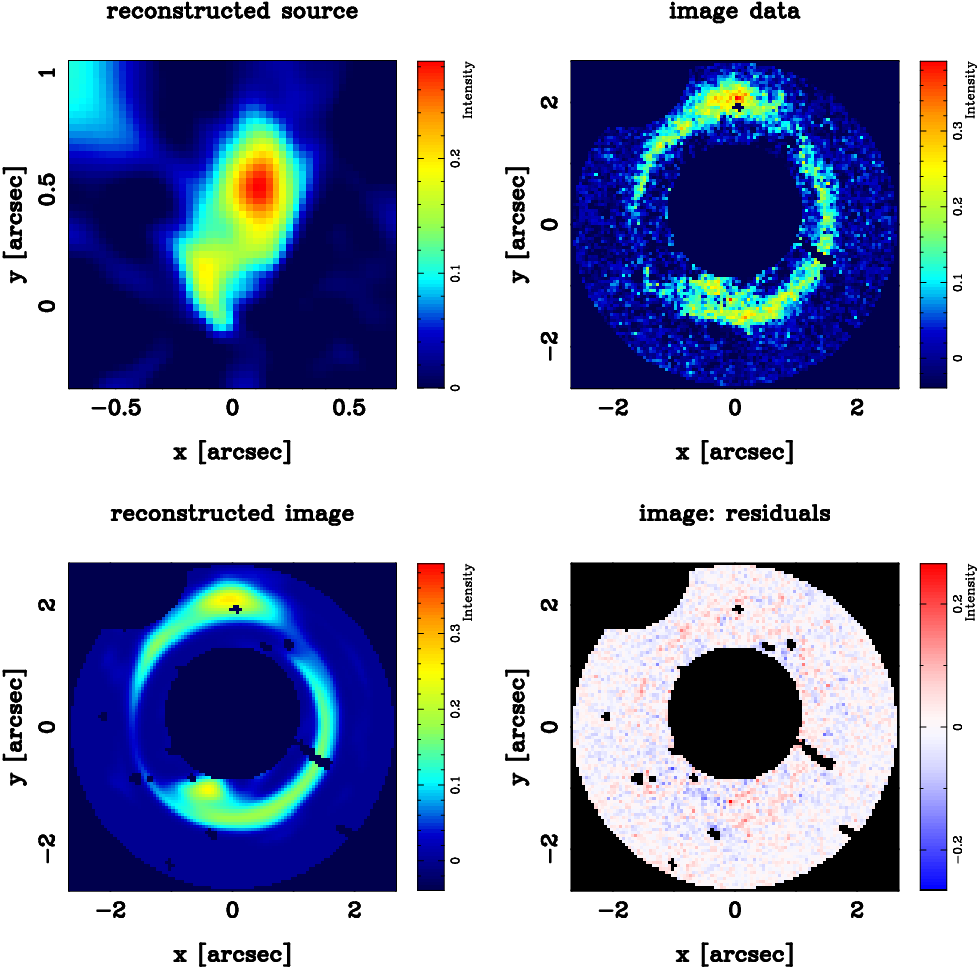}}
  \caption{Best model lens image reconstruction for the galaxy
    SDSS\,J2321. Panels meaning as in Fig.~\ref{fig:J0935_LEN}.}
  \label{fig:J2321_LEN}

  \resizebox{0.99\hsize}{!}{\includegraphics[angle=-90]
            {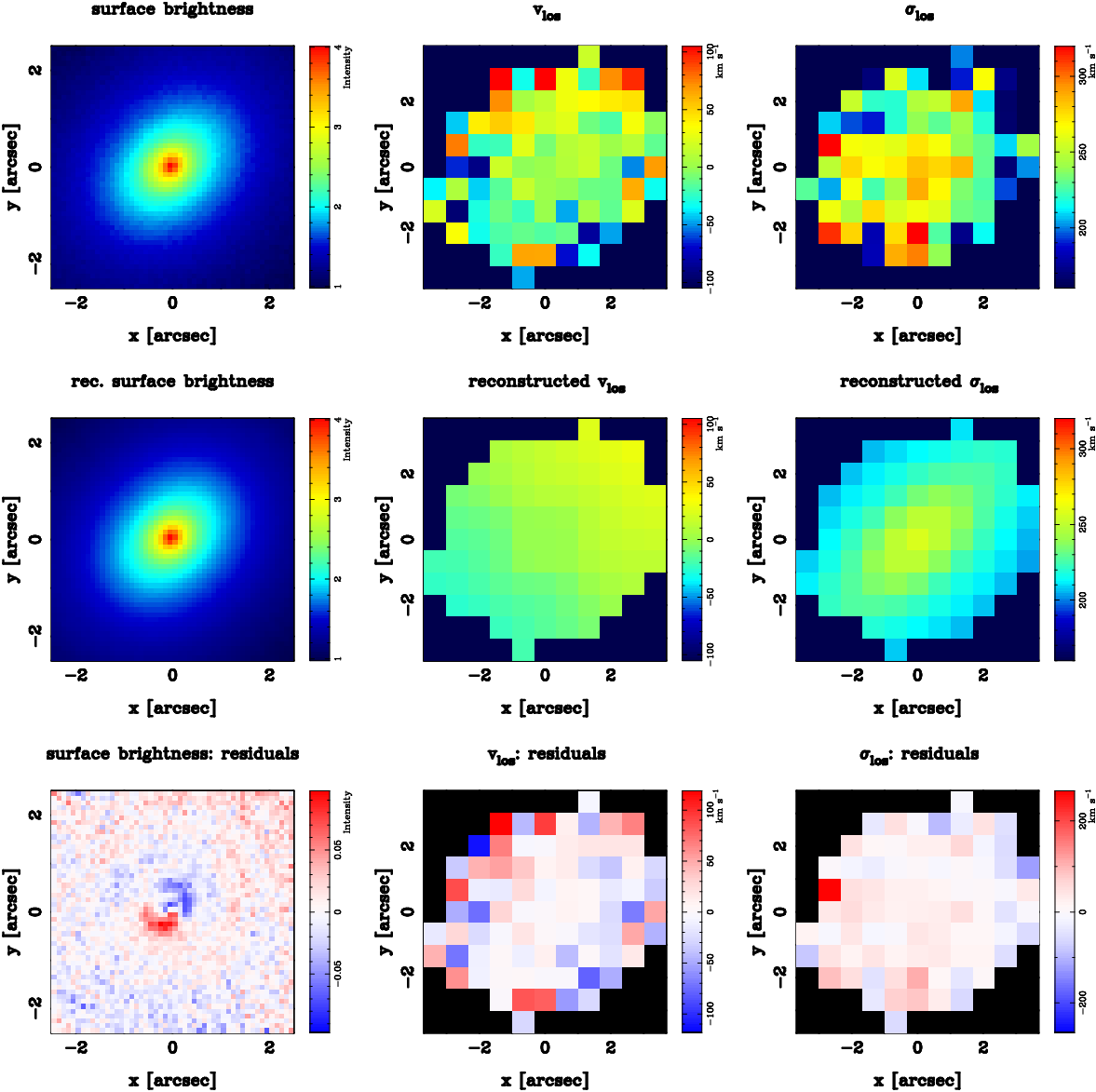}}
  \caption{Best dynamical model for the galaxy SDSS\,J2321. Panels
    meaning as in Fig.~\ref{fig:J0935_DYN}.}
  \label{fig:J2321_DYN}
\end{figure}

\clearpage 

\section{Uncertainties}
\label{app:unc}

In this Appendix we present, for all the analyzed systems, the
one-dimensional marginalized posterior PDFs for the considered model
parameters ($i$, $\slope$, $\talp$, $q$), from which the corresponding
uncertainties (quoted in the Table~\ref{tab:eta}) are derived.

\begin{figure*}
  \begin{center}
    \subfigure[J0037]{\label{fig:NS-0037}\includegraphics[angle=-90,width=0.47\textwidth]{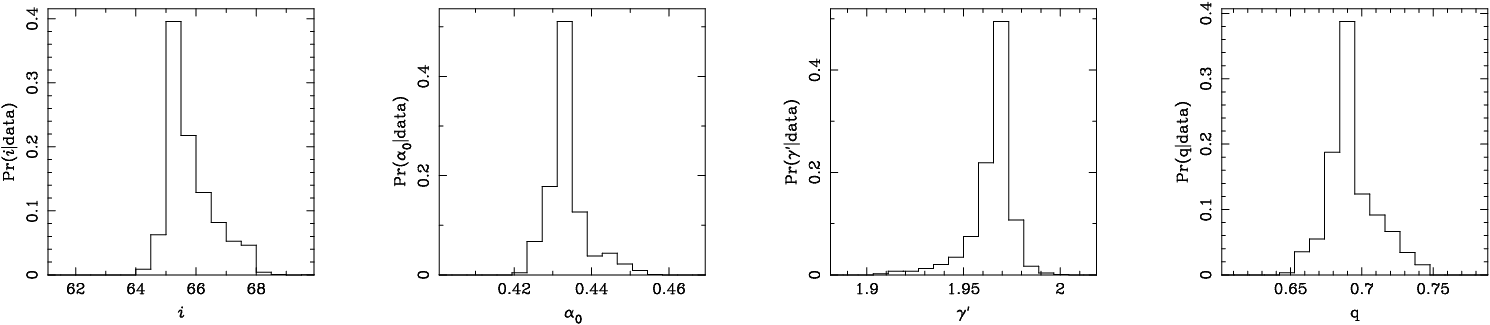}} \qquad $\;$
    \subfigure[J0216]{\label{fig:NS-0216}\includegraphics[angle=-90,width=0.47\textwidth]{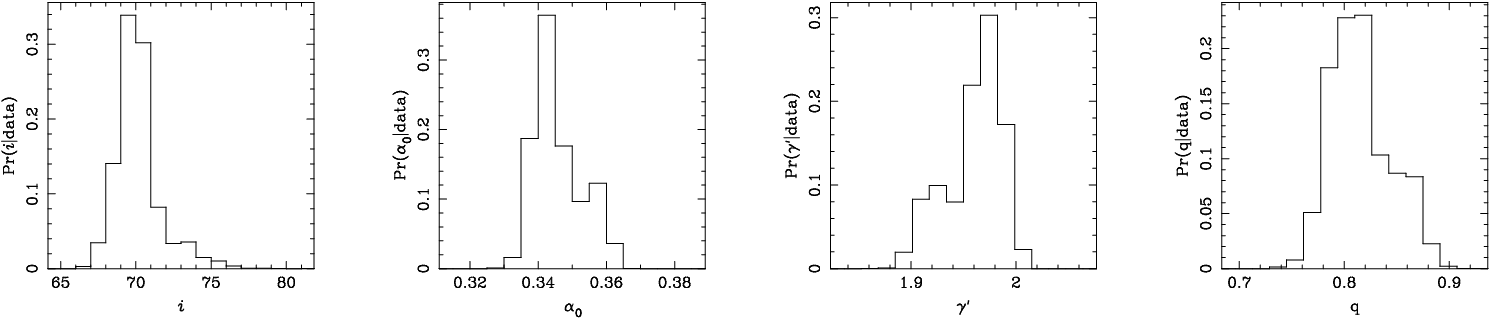}}
  \end{center}
  \vspace{-0.5cm}

  \begin{center}
    \subfigure[J0912]{\label{fig:NS-0912}\includegraphics[angle=-90,width=0.47\textwidth]{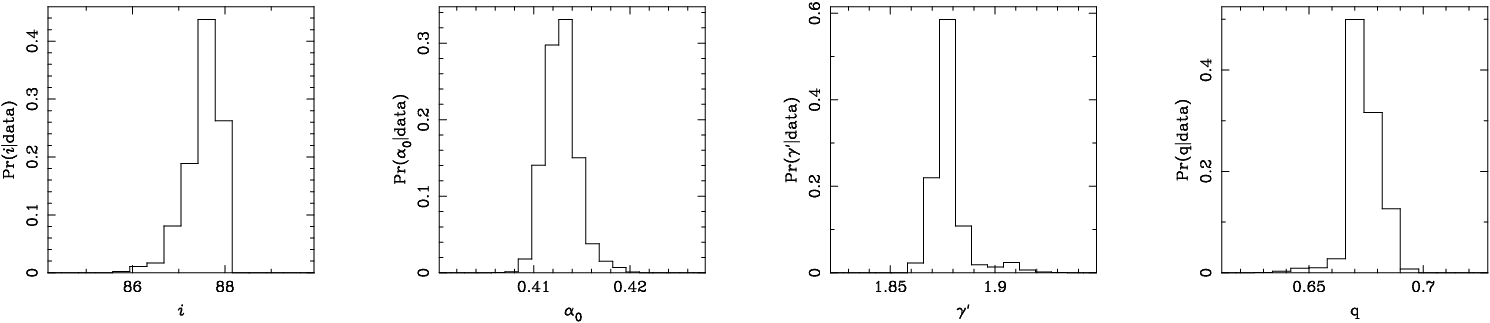}} \qquad $\;$
    \subfigure[J0935]{\label{fig:NS-0935}\includegraphics[angle=-90,width=0.47\textwidth]{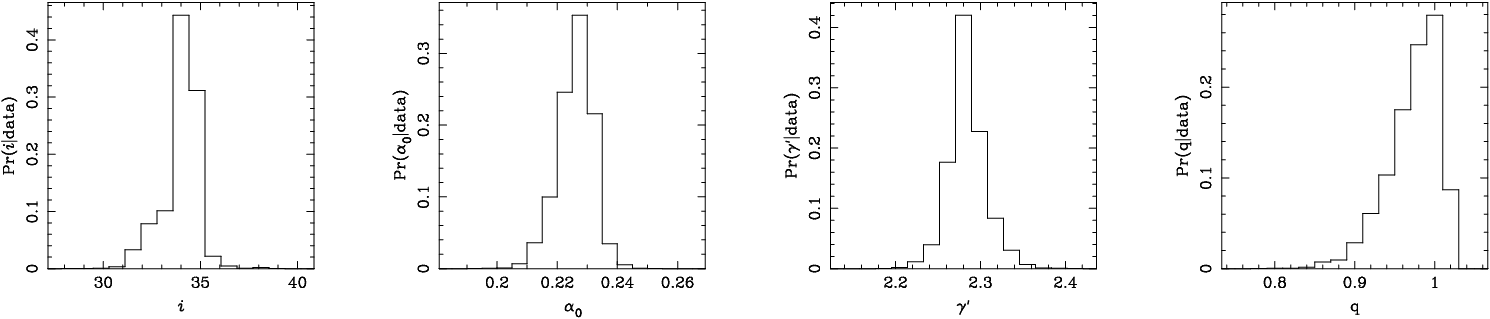}}
  \end{center}
  \vspace{-0.5cm}

  \begin{center}
    \subfigure[J0959]{\label{fig:NS-0959}\includegraphics[angle=-90,width=0.47\textwidth]{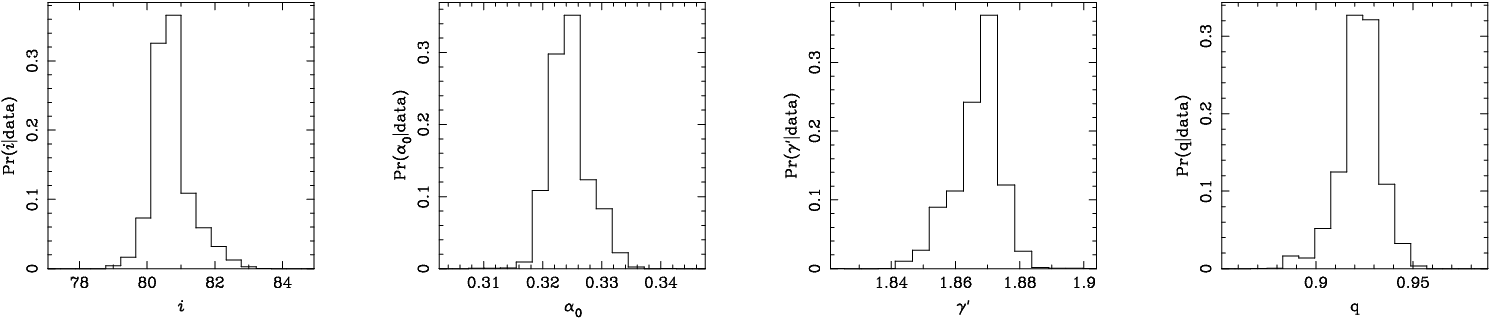}} \qquad $\;$
    \subfigure[J1204]{\label{fig:NS-1204}\includegraphics[angle=-90,width=0.47\textwidth]{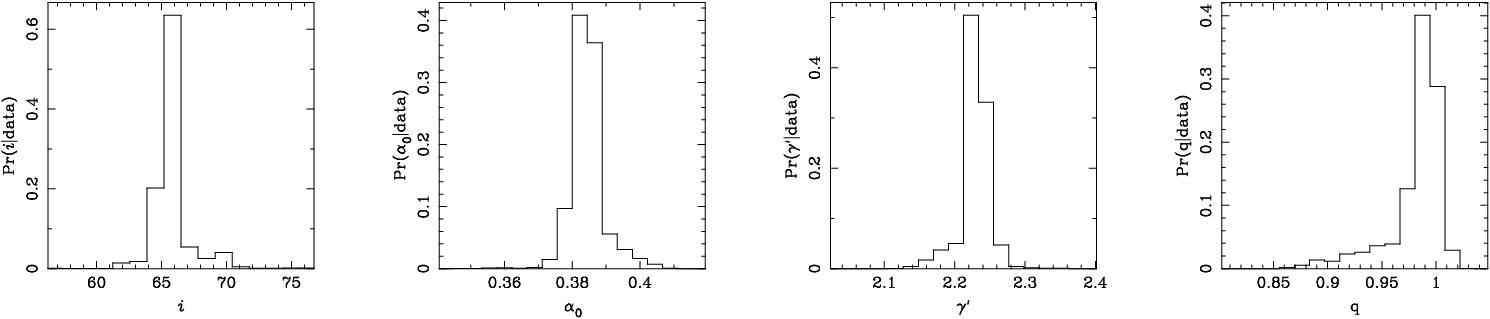}}
  \end{center}
  \vspace{-0.5cm}

  \begin{center}
    \subfigure[J1250]{\label{fig:NS-1250}\includegraphics[angle=-90,width=0.47\textwidth]{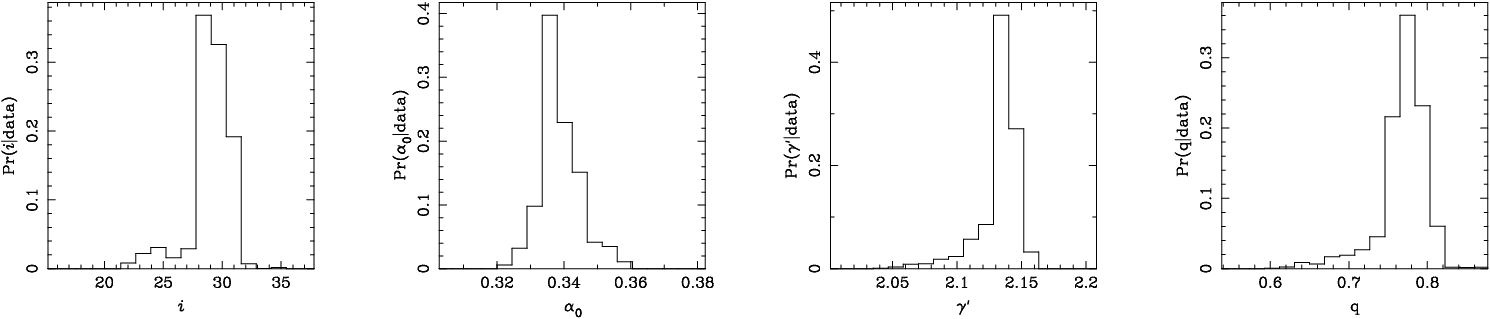}} \qquad $\;$
    \subfigure[J1251]{\label{fig:NS-1251}\includegraphics[angle=-90,width=0.47\textwidth]{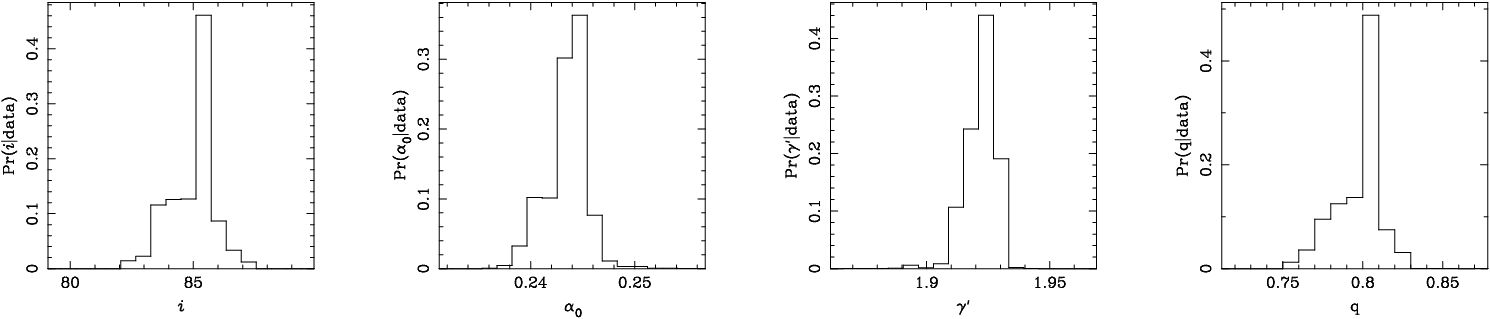}}
  \end{center}
  \vspace{-0.5cm}

  \begin{center}
    \subfigure[J1330]{\label{fig:NS-1330}\includegraphics[angle=-90,width=0.47\textwidth]{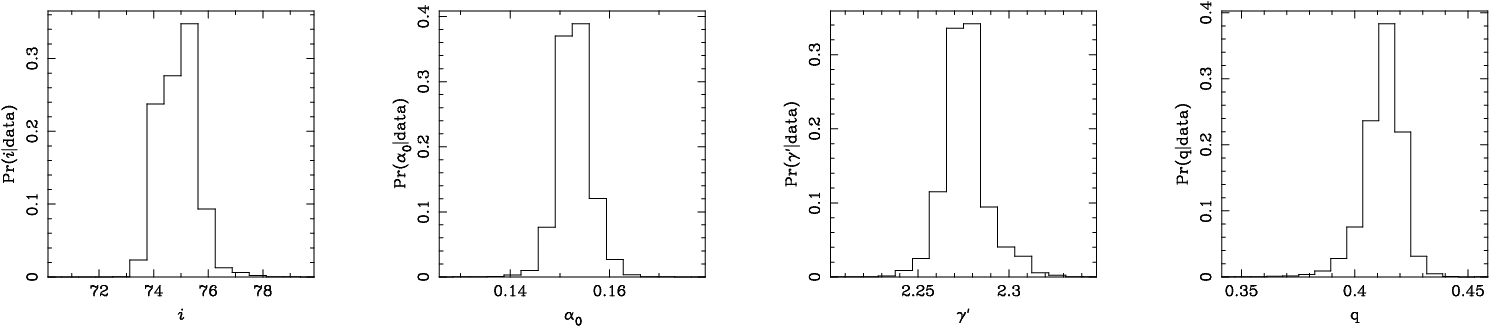}} \qquad $\;$
    \subfigure[J1443]{\label{fig:NS-1443}\includegraphics[angle=-90,width=0.47\textwidth]{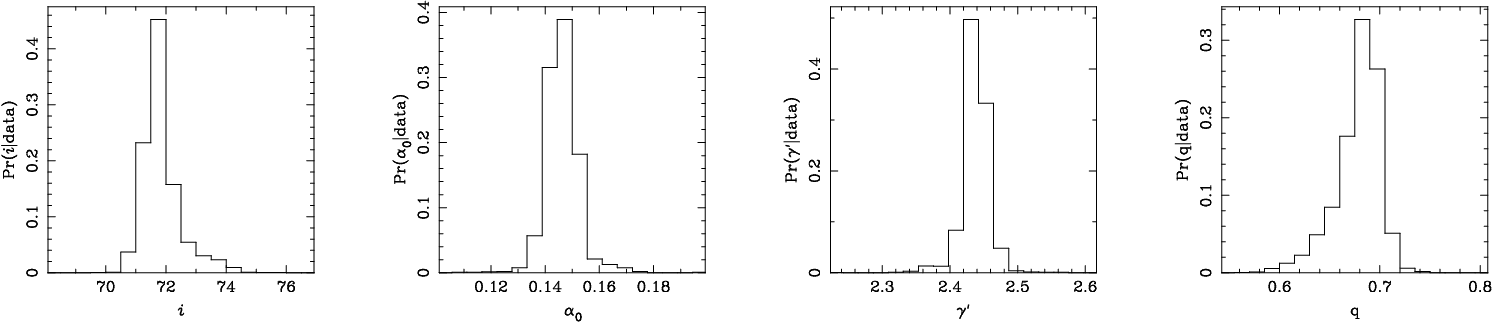}}
  \end{center}
  \vspace{-0.5cm}

  \begin{center}
    \subfigure[J1451]{\label{fig:NS-1451}\includegraphics[angle=-90,width=0.47\textwidth]{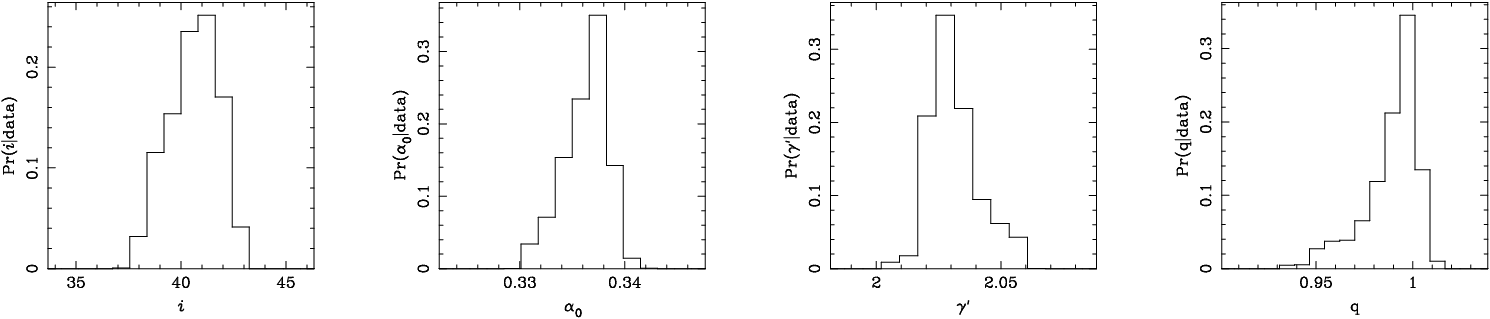}} \qquad $\;$
    \subfigure[J1627]{\label{fig:NS-1627}\includegraphics[angle=-90,width=0.47\textwidth]{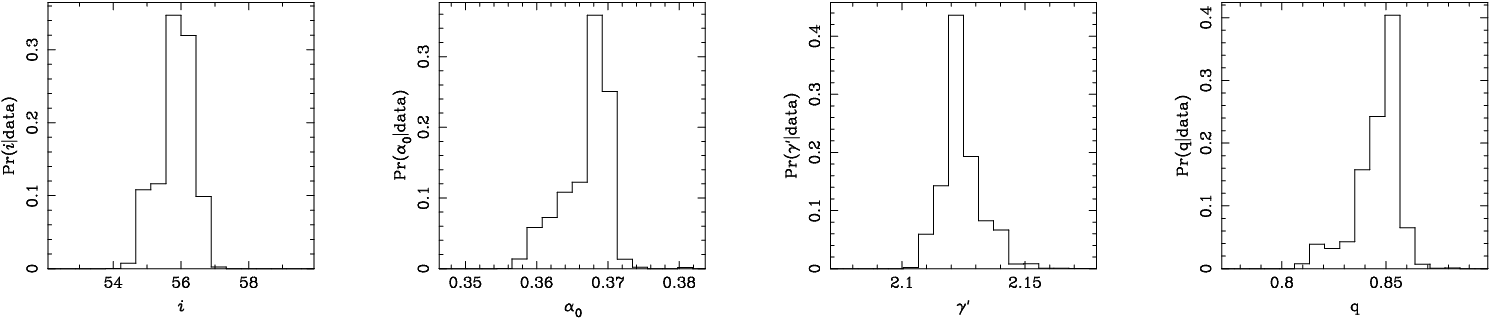}}
  \end{center}
  \vspace{-0.5cm}

  \begin{center}
    \subfigure[J2238]{\label{fig:NS-2238}\includegraphics[angle=-90,width=0.47\textwidth]{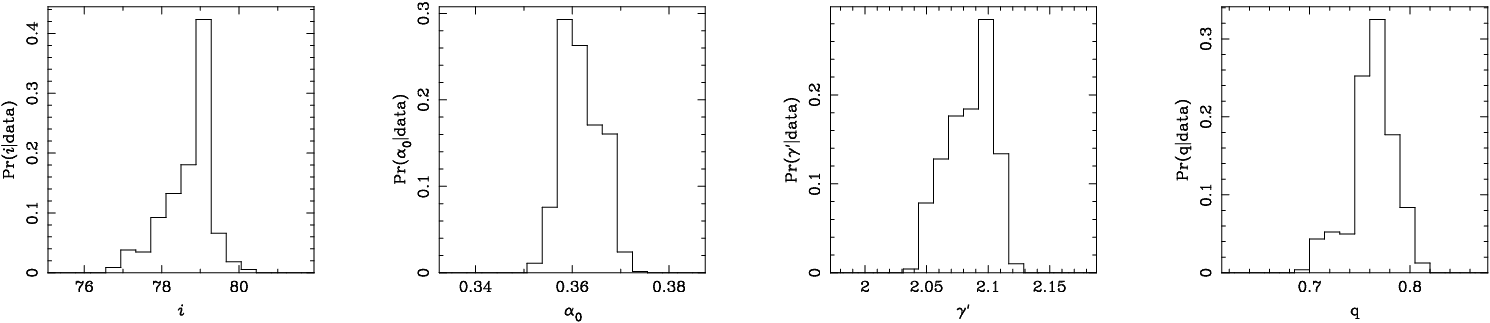}} \qquad $\;$
    \subfigure[J2300]{\label{fig:NS-2300}\includegraphics[angle=-90,width=0.47\textwidth]{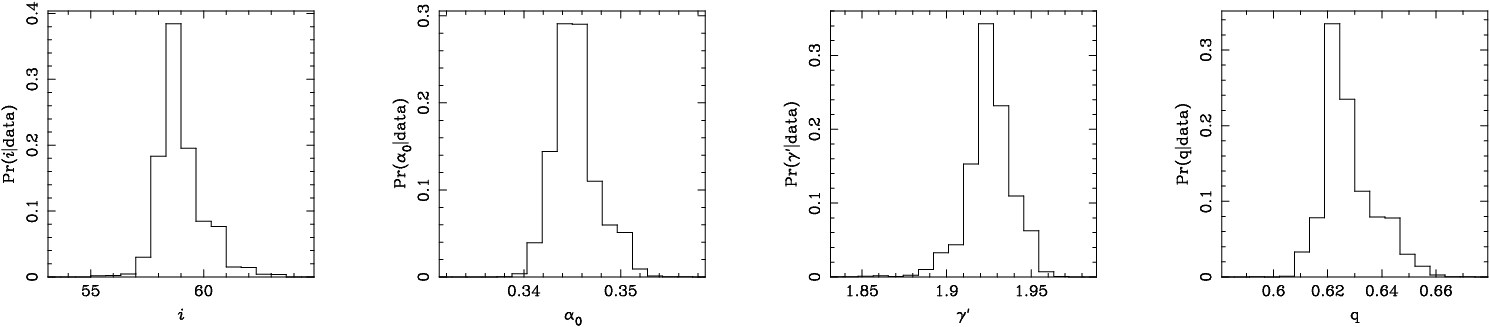}}
  \end{center}
  \vspace{-0.5cm}

  \begin{center}
    \subfigure[J2303]{\label{fig:NS-2303}\includegraphics[angle=-90,width=0.47\textwidth]{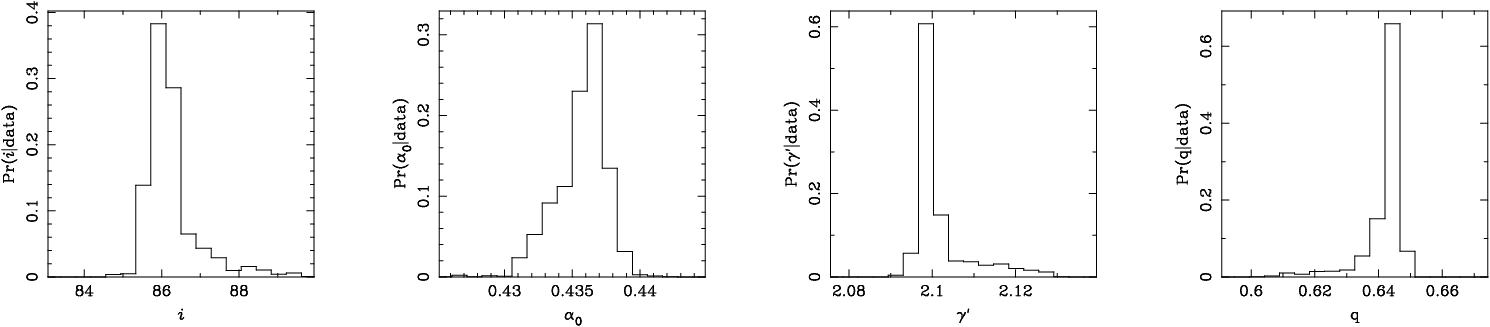}} \qquad $\;$
    \subfigure[J2321]{\label{fig:NS-2321}\includegraphics[angle=-90,width=0.47\textwidth]{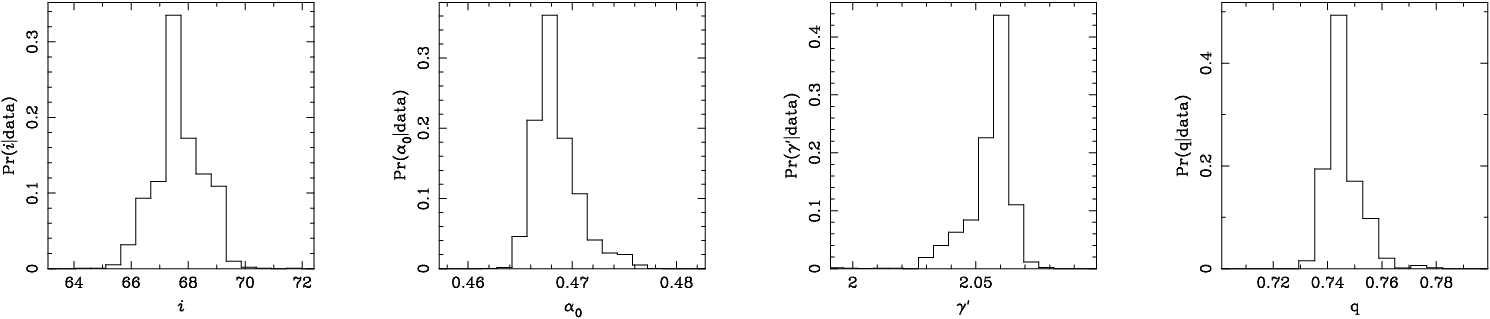}}
  \end{center}

  \caption{Marginalized posterior probability distribution functions
    of the power-law model parameters $i$ (inclination), $\slope$
    (logarithmic slope), $\talp$ (lens strength) and $q$ (axial ratio)
    for each of the analyzed systems, obtained from the nested
    sampling evidence exploration (see text, Sect.~\ref{sec:unc}).
    From left to right, top to bottom, the galaxy are presented in the
    same order as in Tables~\ref{tab:eta} and~\ref{tab:dyn}.}
  \label{fig:NSerrors}
\end{figure*}

\end{document}